\documentclass[10pt,aip,amsmath,amssymb]{revtex4-1}
\usepackage{graphicx}

\newcommand{\Order}{\mbox{$\mathcal{O}$}} 
\newcommand\Alfven{Alfv\'en }
\newcommand\Alfvenic{Alfv\'enic }
\newcommand{\V}[1]{\mathbf{#1}}

\newcommand{\zhat}{\mbox{$\hat{\mathbf{z}}$}} 
\newcommand{\kphat}{\mbox{$\hat{\mathbf{k}}_\perp$}} 
\newcommand{\xhat}{\mbox{$\hat{\mathbf{x}}$}} 
\newcommand{\yhat}{\mbox{$\hat{\mathbf{y}}$}} 
\newcommand{\figref}[1]{Figure~\ref{#1}}
\newcommand{\secref}[1]{\S\ref{#1}}

\begin{document}


\title{\Alfven Wave Collisions, The Fundamental 
Building Block of Plasma Turbulence I: \\Asymptotic Solution} 



\author{G.~G. Howes}
\email[]{gregory-howes@uiowa.edu}
\author{K.~D. Nielson}
\affiliation{Department of Physics and Astronomy, University of Iowa, Iowa City, 
Iowa 52242, USA.}

\date{\today}

\begin{abstract}
The nonlinear interaction between counterpropagating \Alfven waves is
the physical mechanism underlying the cascade of energy to small
scales in astrophysical plasma turbulence. Beginning with the
equations for incompressible MHD, an asymptotic analytical solution
for the nonlinear evolution of these \Alfven wave collisions is
derived in the weakly nonlinear limit. The resulting qualitative
picture of nonlinear energy transfer due to this mechanism involves
two steps: first, the primary counterpropagating \Alfven waves
interact to generate an inherently nonlinear, purely magnetic
secondary fluctuation with no parallel variation; second, the two
primary waves each interact with this secondary fluctuation to
transfer energy secularly to two tertiary \Alfven waves.  These
tertiary modes are linear \Alfven waves with the same parallel
wavenumber as the primary waves, indicating the lack of a parallel
cascade.  The amplitude of these tertiary modes increases linearly
with time due to the coherent nature of the resonant four-wave
interaction responsible for the nonlinear energy transfer. The
implications of this analytical solution for turbulence in
astrophysical plasmas is discussed. The solution presented here
provides valuable intuition about the nonlinear interactions
underlying magnetized plasma turbulence, in support of an experimental
program to verify in the laboratory the nature of this fundamental
building block of astrophysical plasma turbulence.
\end{abstract}


\maketitle 


\section{Introduction}
Turbulence plays a crucial role in the transport of mass, momentum,
and energy in a wide variety of plasma environments, from distant
astrophysical systems, such as galaxy clusters and accretion disks
around black holes, to the solar corona and solar wind in our own
heliosphere, to the laboratory plasmas of the magnetic confinement
fusion program. Of particular importance in space and astrophysical
plasmas is the governing role that turbulence plays in transferring
energy from large scales, where the turbulent motions are typically
driven by violent events or instabilities, down to sufficiently small
scales that dissipative mechanisms can damp the turbulent motions and
convert the turbulent energy into plasma heat.  Turbulence
significantly impacts the evolution of astrophysical environments
through this turbulent cascade of energy from large to small
scales. The turbulent energy cascade itself is driven by nonlinear
interactions within the turbulent plasma.

The magnetic field that ubiquitously permeates astrophysical plasmas
significantly influences the nature of the turbulence, endowing it
with substantially different properties from the hydrodynamic
turbulence common in terrestrial environments. In place of the
hierarchy of turbulent eddies of different sizes that make up
hydrodynamic turbulence, turbulent motions in a magnetohydrodynamic
(MHD) plasma are dominated by the physics of \Alfven
waves,\cite{Alfven:1942} traveling disturbances of the plasma and
magnetic field that propagate along the local mean magnetic field.
Early work exploring the nature of plasma turbulence using the
incompressible MHD equations recognized that the nonlinear
interactions underlying the turbulent cascade occur only between
counterpropagating \Alfven wave
packets,\cite{Iroshnikov:1963,Kraichnan:1965} often referred to as
\Alfven wave ``collisions.'' 

This nonlinear interaction, in fact, persists under more general
plasma conditions than required by the MHD approximation, particularly
under the weakly collisional conditions relevant to many space and
astrophysical plasma environments.\cite{Howes:2006,Schekochihin:2009}
The nonlinear term responsible for this counterpropagating wave
interaction in the kinetic equation that governs weakly collisional
plasma dynamics is generically labeled the $\V{E} \times
\V{B}$ nonlinearity, as highlighted, for example, in equation~(35) of
Howes \emph{et al.}\cite{Howes:2006} This name is an appropriate
general term for this nonlinearity because, as shown in
\secref{sec:eb}, the lowest order contribution to the plasma fluid
velocity is given by $\V{u}_\perp
\simeq c\V{E} \times
\V{B}_0/B_0^2$; therefore, the $\V{E} \times
\V{B}$ nonlinearity takes the familiar form of the nonlinearity appearing in the fluid equations,
$\V{u}_\perp \cdot \nabla$. The $\V{E}\times \V{B}$ nonlinearity
requires that the \Alfven wave fronts vary spatially in the plane
perpendicular to the local magnetic field ($k_\perp \ne 0$), as will
be discussed in more detail in \secref{sec:prop}.  A number of other
nonlinearities can occur in a magnetized plasma, including parametric
instabilities driven by gradients parallel to the magnetic field, such
as the decay,\cite{Galeev:1963,Sagdeev:1969,Hasegawa:1976a,
Derby:1978,Goldstein:1978,Spangler:1982,Sakai:1983,Spangler:1986,
Terasawa:1986,Jayanti:1993,Hollweg:1994,Shevchenko:2003,Voitenko:2005}
modulational,\cite{Lashmore-Davies:1976,Mjolhus:1976,Mio:1976b,
Sakai:1983,Wong:1986,Hollweg:1994,Shukla:2007} and
beat\cite{Wong:1986,Hollweg:1994} instabilities, and other possible
nonlinear
interactions.\cite{Lacombe:1980,Brodin:1988,Kucherenko:1993,Yukhimuk:2000,Chandran:2005,
Shukla:2006,Brodin:2007,Mottez:2012} Here we propose the working
hypothesis that \emph{the $\V{E} \times
\V{B}$ nonlinearity is the dominant nonlinear mechanism
underlying the anisotropic cascade of energy in magnetized plasma
turbulence}. Furthermore, we contend that this nonlinearity remains the
dominant driver of turbulent energy transfer even under conditions
beyond the MHD limit, in particular for turbulence in collisionless
plasmas as well as for the turbulent dynamics at small scales below
the characteristic ion kinetic scales, where the plasma waves become
dispersive.

The expected dominance of the $\V{E} \times \V{B}$ nonlinearity over
other potential nonlinear mechanisms in astrophysical plasma
turbulence is easily explained. A nonlinear mechanism will contribute
significantly to the dynamics if the associated nonlinear term in the
equations of evolution approaches or surpasses the order of magnitude
of the linear terms. The relative magnitude of the terms in the
evolution equations can be quantified by a characteristic timescale,
or frequency, associated with each
term.\cite{Goldreich:1995,Howes:2008b,Howes:2011b} The linear
frequency of \Alfven waves in a magnetized plasma is given by
$\omega=k_\parallel v_A$, and thus depends on the gradient parallel to
the equilibrium magnetic field. Nonlinearities that depend on the
parallel gradients in the plasma, such as parametric instabilities,
have an associated nonlinear frequency that scales as $\omega_{NL}
\sim k_\parallel \delta v$. Therefore, parallel nonlinearities
significantly affect the evolution when $\omega_{NL}\sim \omega $,
which requires a fluctuation amplitude of order the \Alfven velocity,
$\delta v/v_A \sim 1$.  In contrast, the $\V{E} \times \V{B}$
nonlinearity depends on the perpendicular gradient in the plasma, with
an associated nonlinear frequency $\omega_{NL} \sim k_\perp \delta v$.
This nonlinearity will likewise significantly influence the evolution
of a turbulent plasma when $\omega_{NL}\sim \omega $, a condition
known in MHD turbulence theory as critical
balance.\cite{Sridhar:1994,Goldreich:1995} In this case, the $\V{E}
\times \V{B}$ nonlinearity can be strong even  for small
fluctuation amplitudes, $\delta v/v_A \ll 1$, as long as the turbulent
fluctuations are significantly anisotropic, $k_\perp / k_\parallel \gg
1$. 

Theoretical arguments have
suggested\cite{Sridhar:1994,Goldreich:1995,Boldyrev:2006} and
numerical
simulations,\cite{Shebalin:1983,Cho:2000,Maron:2001,Cho:2004,Cho:2009,TenBarge:2012a}
laboratory
experiments,\cite{Robinson:1971,Zweben:1979,Montgomery:1981} and solar
wind observations\cite{Belcher:1971,Sahraoui:2010b,Narita:2011} have
demonstrated that magnetized plasma turbulence inherently generates
small-scale fluctuations with precisely this sense of anisotropy,
$k_\perp / k_\parallel \gg 1$. The amplitude of turbulent fluctuations
decreases with scale, so that deep into the inertial range, at length
scales much smaller than the driving scale, the amplitude of
fluctuations is small, $\delta v/v_A \ll 1$. The featureless power-law
appearance of the entire turbulent inertial range in magnetized plasma
turbulence suggests that a single nonlinear mechanism is responsible
for the underlying nonlinear energy transfer to smaller scales.  Since
the $\V{E} \times \V{B}$ nonlinearity must dominate deep within the
inertial range where $\delta v/v_A \ll 1$, this implies that the
$\V{E} \times \V{B}$ nonlinearity indeed dominates throughout the
entire inertial range.  Although a detailed evaluation of the relative
importance of the $\V{E} \times \V{B}$ nonlinearity to other potential
nonlinearities in magnetized plasma turbulence is beyond the scope of
this paper, we believe the evidence supports our working hypothesis
that the $\V{E} \times \V{B}$ nonlinearity is the dominant nonlinear
mechanism underlying the turbulent energy cascade in magnetized
plasmas. This hypothesis motivates the primary aim of this paper, to
solve for the evolution of the nonlinear interaction between
counterpropagating \Alfven waves due to the $\V{E} \times \V{B}$
nonlinearity in an incompressible, ideal MHD plasma. This nonlinear
interaction is the fundamental building block of magnetized plasma
turbulence.

The mathematical properties of the nonlinear interaction between
counterpropagating \Alfven waves have provided indispensable guidance
in the construction and refinement of modern theories for anisotropic
turbulence in a magnetized
plasma.\cite{Iroshnikov:1963,Kraichnan:1965,Montgomery:1982,Shebalin:1983,Higdon:1984a,
Sridhar:1994,Goldreich:1995,Montgomery:1995,Ng:1996,Goldreich:1997,Galtier:2000,
Lithwick:2001,Lithwick:2003,Boldyrev:2006} Therefore, the derivation
of an explicit solution of the evolution of this nonlinear interaction
can provide valuable insight into the fundamental nature of the
turbulence. Here we highlight the role that the consideration of the
mathematical properties of the nonlinear interaction has played in the
development of the modern theory for weak and strong MHD turbulence.

Iroshnikov\cite{Iroshnikov:1963} and Kraichnan\cite{Kraichnan:1965}
independently developed an isotropic theory for incompressible MHD
turbulence, first recognizing that the nonlinear interaction
underlying the turbulence occurs only between counterpropagating
\Alfven wave packets. Shebalin, Matthaeus, and
Montgomery\cite{Shebalin:1983} showed that the particular form of the
nonlinear term leads to three-wave matching conditions that require
one of the waves to have $k_\parallel=0$, leading to an anisotropic
turbulent cascade that transfers energy to smaller scales in the
perpendicular direction, but not the parallel direction.  Based on the
development of the reduced MHD equations by Strauss\cite{Strauss:1976}
and their use in exploring anisotropic MHD turbulence by Montgomery and
Turner,\cite{Montgomery:1981,Montgomery:1982} Higdon\cite{Higdon:1984a}
developed an alternative model of highly anisotropic incompressible
MHD turbulence in the presence of a strong mean magnetic field,
consisting of two-dimensional velocity and magnetic field fluctuations
in the plane perpendicular to the magnetic field, unrelated to the
propagating \Alfven wave fluctuations.

Sridhar and Goldreich\cite{Sridhar:1994} incorporated anisotropy into
the Iroshnikov-Kraichnan (IK) theory, pointing out that the IK theory
is a theory for weak turbulence based on resonant three-wave
interactions. But they claimed that the IK theory was incorrect
because these three-wave interactions involve one mode that has
$\omega=0$, and since linear wave modes with $
\omega=0$ possess no
power, the coupling coefficients for three-wave interactions are
trivially zero.  Sridhar and Goldreich then proceeded to develop a
quantitative model for weak incompressible MHD turbulence based on
four-wave interactions.  This four-wave theory had two important
qualitative predictions: (i) there occurs no parallel cascade of
energy, and (ii) the nonlinear interactions strengthen as the
turbulent cascade proceeds to smaller perpendicular scales, eventually
transitioning to a strong turbulent cascade. Goldreich and
Sridhar\cite{Goldreich:1995} pushed on to describe the nature of
strong incompressible MHD turbulence, suggesting that a broadening of
the condition for the resonant frequency matching leads to the onset
of a parallel cascade such that the nonlinear and linear terms in the
equations of evolution remain in a state of critical balance as the
turbulence transfers energy to smaller scales.  The conjecture of
critical balance has the important implication that the parallel and
perpendicular scales of turbulent fluctuations are correlated, leading
to a scale-dependent anisotropy.

Subsequently, Montgomery and Matthaeus\cite{Montgomery:1995} claimed
that Sridhar and Goldreich\cite{Sridhar:1994} were incorrect in
concluding that three-wave interactions played no role in weak
incompressible MHD turbulence, arguing that three-wave interactions do
not vanish because fluctuations with $k_\parallel=0$ can exist,
although they cannot be treated as linear waves. Ng and
Bhattacharjee\cite{Ng:1996} used perturbation theory to calculate the
interaction between counterpropagating \Alfven wave packets, proving
that the three-wave interactions do not vanish if the wave packets have a
$k_\parallel=0$ component, and that when these interactions are
nonzero, they dominate over four-wave interactions.  Admitting that
their original weak turbulence formulation had effectively neglected
the possibility of magnetic field line wander, which leads to a
$k_\parallel=0$ component in the interaction, Goldreich and
Sridhar\cite{Goldreich:1997} constructed a theory for weak MHD
turbulence based on three-wave interactions.  The two key predictions,
that no parallel cascade occurs and that nonlinear interactions the
strengthen as the turbulence cascades to smaller scale, persisted in
this refined theory.  Furthermore, Goldreich and Sridhar suggested
that the displacement of magnetic field lines would lead to a
correlation relating successive interactions between different
counterpropagating \Alfven wave packets, causing a perturbative
approach to fail in describing the turbulent evolution.  Galtier and
coworkers\cite{Galtier:2000} argued, however, the perturbation theory
remains valid even for weak MHD turbulence dominated by three-wave
interactions, and Lithwick and Goldreich\cite{Lithwick:2003}
demonstrated the validity of this claim using the example of a linear
random oscillator.

Finally, Boldyrev\cite{Boldyrev:2006} pointed out that the vector
nature of the nonlinear term in the incompressible MHD equations leads
to the development of a dynamic alignment between velocity and
magnetic field fluctuations in strong incompressible MHD turbulence.
This important refinement leads to the qualitative prediction that MHD
turbulence generates sheet-like rather than filamentary structures at
small scales, as widely observed in MHD turbulence simulations, and to
a key quantitative change in the predicted magnetic energy spectrum,
bringing the theory into agreement with numerical results.

Much of the progress in the development of the modern anisotropic
theory for weak and strong MHD turbulence outlined above is founded
upon the intuitive concept of nonlinear interactions between
counterpropagating \Alfven waves as the fundamental building block of
the turbulent cascade.  In this paper, we illuminate the general
properties of these nonlinear interactions by using an asymptotic
analysis to explicitly solve for the nonlinear evolution of
counterpropagating \Alfven wave interactions in the weakly nonlinear
limit. As we shall demonstrate, the resulting solution provides
insight into the nature of weak incompressible MHD turbulence,
particularly the relevance of three-wave and four-wave interactions.
The solution also provides a potential explanation for the common
observation of an \Alfven ratio greater than unity at large scales in
plasma turbulence, as observed in the solar wind
\cite{Borovsky:2012} and numerical simulations.\cite{Boldyrev:2011}

In \secref{sec:prop} we highlight the general properties of magnetized
plasma turbulence. The detailed asymptotic analytical solution to the
nonlinear evolution of counterpropagating \Alfven wave interactions in
the weakly nonlinear limit is presented in \secref{sec:solution}. The
general characteristics of the solution are summarized and their
implications are discussed in \secref{sec:discuss}. The verification
of these analytical solutions using nonlinear gyrokinetic simulations
is presented in a companion paper, Nielson, Howes, and
Dorland,\cite{Nielson:2013a} hereafter Paper II.

\section{General Properties of Magnetized Plasma Turbulence}
\label{sec:prop}
\subsection{Incompressible MHD Equations and Characteristic Linear Wave Modes}
To explore the fundamental properties of turbulence in a magnetized
plasma, in particular the nonlinear interaction between
counterpropagating \Alfven waves, a useful point of departure is the
set of incompressible MHD equations, expressed here in the symmetrized
Els\"asser form,\citep{Elsasser:1950}
\begin{equation}
\frac{\partial \V{z}^{\pm}}{\partial t} 
\mp \V{v}_A \cdot \nabla \V{z}^{\pm} 
=-  \V{z}^{\mp}\cdot \nabla \V{z}^{\pm} -\nabla P/\rho_0,
\label{eq:elsasserpm}
\end{equation}
\begin{equation}
\nabla\cdot  \V{z}^{\pm}=0
\label{eq:div0}
\end{equation}
where the magnetic field is decomposed into equilibrium and
fluctuating parts $\V{B}=\V{B}_0+ \delta
\V{B} $, $\V{v}_A =\V{B}_0/\sqrt{4 \pi\rho_0}$ is the \Alfven velocity 
due to the equilibrium field $\V{B}_0=B_0 \zhat$, $P$ is total pressure (thermal
plus magnetic), $\rho_0$ is mass density, and $\V{z}^{\pm}(x,y,z,t) =
\V{u} \pm \delta \V{B}/\sqrt{4 \pi \rho_0}$ are the Els\"asser 
fields given by the sum and difference of the velocity fluctuation
$\V{u}$ and the magnetic field fluctuation $\delta \V{B}$ expressed in
velocity units. At a glance, one might think that an equation of state for $P$
is necessary to close the system of equations given by \eqref{eq:elsasserpm} and
\eqref{eq:div0}, but in fact the divergence free condition
\eqref{eq:div0} closes the system,\cite{Montgomery:1982} as follows. 
Taking the divergence of \eqref{eq:elsasserpm}, the terms on the
left-hand side are zero using \eqref{eq:div0}, leaving the following
expression for the pressure,
\begin{equation}
\nabla^2 P/\rho_0 =  - \nabla \cdot \left(   \V{z}^{\mp}\cdot \nabla \V{z}^{\pm}\right)
=- \frac{\partial }{\partial x_i} z^{-}_j \frac{\partial }{\partial x_j} z^{+}_i ,
\label{eq:press}
\end{equation}
where summation of over repeated indices is implied. This
demonstrates that the pressure can be computed by solving a Poisson
equation, where nonlinear products of the gradients of the Els\"asser
fields act as a source. It is worthwhile noting that the pressure at any
point responds instantaneously to changes in the Els\"asser fields at
any other point; the speed of sound is effectively infinite in the
incompressible limit.

Since \eqref{eq:press} shows that the pressure term in incompressible
MHD arises nonlinearly, both of the terms on the right-hand side of 
\eqref{eq:elsasserpm} are nonlinear. Linearization of the equations yields
two wave modes, both obeying the same linear dispersion relation,
$\omega=\pm k_\parallel v_A$: (i) the \Alfven waves, with
$\V{z}^{\pm}$ polarized in the $\zhat \times \hat{\V{k}}$ direction,
and (ii) the pseudo-Alfven waves, the incompressible limit of the slow
magnetosonic wave from compressible MHD, with $\V{z}^{\pm}$ polarized
in the $\hat{\V{k}} \times (\zhat \times \hat{\V{k}})$
direction.\cite{Maron:2001} Note that the fast magnetosonic wave from
compressible MHD is eliminated from the system by the infinite sound
speed implied by the condition of incompressibility.

The symmetrized Els\"asser form of the incompressible MHD equations
lends itself to a particularly simple physical interpretation, helping
to illuminate the fundamental nature of magnetized plasma turbulence.
The Els\"asser field $\V{z}^{+}$ ($\V{z}^{-}$) represents either the
\Alfven or pseudo-\Alfven wave traveling down (up) the equilibrium magnetic
field.  The second term on the left-hand side of \eqref{eq:elsasserpm}
is the linear term representing the propagation of the Els\"asser
fields along the mean magnetic field at the \Alfven speed, the first
term on the right-hand side is the nonlinear term representing the
interaction between counterpropagating waves, and the second term on
the right-hand side is a nonlinear term that ensures incompressibility
through \eqref{eq:press}.

\subsection{Nonlinear Properties of the Incompressible MHD Equations}
\label{sec:nlprop}
The mathematical form of the nonlinear wave interaction term in
\eqref{eq:elsasserpm} reveals several important properties of incompressible 
MHD turbulence. Consider the nonlinear interaction between two plane
waves with wavevectors $\V{k}_1$ and $\V{k}_2$, where we adopt the
convention that the wave frequency $\omega \ge 0$, so that the sign of
$k_{\parallel}$ determines the direction of propagation of the wave
along the equilibrium magnetic field.  First, the nonlinear
interaction term may be nonzero only if both $\V{z}^{+} \ne 0$ and
$\V{z}^{-} \ne 0$, so the two waves must propagate in opposite
directions along the magnetic field, implying $k_{\parallel 1}$ and
$k_{\parallel 2}$ have opposite
signs.\cite{Iroshnikov:1963,Kraichnan:1965} Physically, the waves
represented by the $\V{z}^{-}$ Els\"asser fields cause nonlinear
distortion of the counterpropagating $\V{z}^{+}$ Els\"asser fields,
and vice-versa.  Therefore, when $\V{z}^{-}=0$, an arbitrary waveform
$\V{z}^{+}(x,y,z+v_At)$ is an exact nonlinear solution of the
equations, representing a finite amplitude \Alfven or pseudo-\Alfven
wavepacket traveling nondispersively in the $-\zhat$ direction, and
vice-versa.\cite{Goldreich:1995}

Next, we consider the nonlinear distortion of a downward propagating
wave $\V{z}^{+}$ with wavevector $\V{k}^+$ by an upward propagating
wave $\V{z}^{-}$ with wavevector $\V{k}^-$, as given by the nonlinear
term $\V{z}^{-}\cdot \nabla \V{z}^{+}$ in \eqref{eq:elsasserpm}. We
will separately consider the distortion due to an \Alfven wave
$\V{z}_A^{-}$ and a pseudo-\Alfven wave $\V{z}_P^{-}$. Note that the
dot product in the nonlinear term selects the gradient of the
$\V{z}^{+}$ wave along the direction of polarization of the
$\V{z}^{-}$ wave, so first let us compute simple expressions for the
directions of polarization for the \Alfven and pseudo-\Alfven wave.
For a wave with wavevector $\V{k}$, let us specify an orthonormal
basis $(\zhat,\kphat, \zhat \times \kphat)$, where the unit vector in
the wavevector direction is given by $\hat{\V{k}}= (k_\parallel \zhat
+ k_\perp \kphat)/k$. The \Alfven wave is polarized in the $\zhat
\times
\hat{\V{k}}$ direction, which simplifies to  the $\zhat \times
\kphat$ direction of the orthonormal basis. The pseudo-\Alfven wave is  polarized in the 
$\hat{\V{k}} \times (\zhat \times \hat{\V{k}})$ direction, which simplifies to a
unit vector direction $(-k_\parallel/k) \kphat + (k_\perp/k) \zhat$,
so this demonstrates that the pseudo-\Alfven wave inhabits the plane
defined by the other two directions of the orthonormal basis.

For the distortion of any $\V{k}^+= \V{k}_{\perp}^+ - k_\parallel^+
\zhat$ wave by a counterpropagating
\Alfven wave $\V{k}_A^-$, it can be easily shown that the nonlinear
term is proportional to $\zhat \cdot (\hat{\V{k}}_{\perp A}^-  \times
\V{k}_\perp^+)$. This implies that  a nonzero  nonlinear interaction due to a
counterpropagating \Alfven wave requires $\hat{\V{k}}_{\perp A}^- 
\times \hat{\V{k}}_{\perp}^+  \ne 0$, that the two waves must have perpendicular
components that are not colinear. For the distortion by a
counterpropagating pseudo-\Alfven wave $\V{k}_P^-$, the nonlinear
term is proportional to
\begin{equation} - k_\perp^+ \frac{ k_{\parallel
P}^-}{k_P^-} (\hat{\V{k}}_{\perp P}^- \cdot \hat{\V{k}}_{\perp}^+) - k_\parallel^+ \frac{ k_{\perp
P}^-}{k_P^-}
\end{equation}
Although this nonlinear term is generally nonzero, it is important to
note that each term involves the parallel component of the wavevector
of one of the waves. As we explain next, this property leads to the
prediction that, at sufficiently small scales within the inertial
range, the contribution of the pseudo-\Alfven waves to the nonlinear
interactions driving the turbulent cascade in an incompressible MHD
plasma is subdominant to the contribution of the \Alfven waves.
 
\subsection{The Anisotropic Limit $k_\perp \gg k_\parallel$  and the Relation to Reduced MHD}
\label{sec:rmhd}
To place the calculation presented in \secref{sec:solution} into the
proper context with respect to related work, it is important to
specify exactly the relation between incompressible MHD and another
widely used system of equations called \emph{reduced
MHD}.\cite{Strauss:1976,Montgomery:1981,Montgomery:1982,Schekochihin:2009}
The reduced MHD equations describe low-frequency, anisotropic
fluctuations in a magnetized plasma, and were originally derived as a
reduced description of the MHD dynamics in tokamak
experiments.\cite{Strauss:1976} Early derivations of the reduced MHD
equations assumed a strong magnetic field, or a $\beta \ll 1$ limit of
plasma,\cite{Strauss:1976,Montgomery:1981,Montgomery:1982} where the
plasma beta is the ratio of the thermal to the magnetic pressure,
$\beta = 8 \pi n T/B_0^2$.  In fact, however, the
derivation\cite{Schekochihin:2009} minimally requires only the
assumption of anisotropic fluctuations $k_\perp \gg k_\parallel$, so
reduced MHD indeed is applicable for arbitrary plasma $\beta$.  This is an
important point, because the limitation to $\beta \ll 1$ plasmas would
severely limit the applicability of the reduced MHD equations for astrophysical
applications, where one often finds $\beta \gtrsim 1$. In this paper,
we refer to the limit $k_\perp \gg k_\parallel$ as the
\emph{anisotropic limit}.

The development of anisotropy is a widely recognized property of
magnetized plasma turbulence, supported by laboratory
experiments,\cite{Robinson:1971,Zweben:1979,Montgomery:1981} numerical
simulations,\cite{Shebalin:1983,Cho:2000,Maron:2001,Cho:2004,Cho:2009,TenBarge:2012a}
and solar wind
observations.\cite{Belcher:1971,Sahraoui:2010b,Narita:2011} Even for
the turbulence driven isotropically with $k_\perp \sim k_\parallel
\sim 1/L$ at the outer scale of the inertial range, the turbulent
fluctuations preferentially transfer energy to smaller scales in the
direction perpendicular to the magnetic field, generating increasingly
anisotropic fluctuations as the turbulent energy cascades to smaller
scales. At perpendicular scales sufficiently smaller than the driving
scale $L$, or $k_\perp L \gg 1$, this scale-dependent anisotropy leads
to small-scale turbulent fluctuations that are highly elongated along
the direction of the magnetic field, described by the anisotropic
limit $k_\perp\gg k_\parallel$.  Therefore, even if the anisotropic
limit is not satisfied at the largest scales of the turbulent cascade,
the anisotropic limit becomes increasingly well satisfied as the
turbulent fluctuations cascade to smaller scales through the inertial
range.

The simplicity of the reduced MHD equations and of the plasma dynamics
in the anisotropic limit have enabled a number of important properties
of anisotropic magnetized plasma turbulence to be identified.  First,
we note that, similar to the case for incompressible MHD, the fast
magnetosonic wave is ordered out of the reduced MHD
system.\cite{Schekochihin:2009} The unimportance of fast-wave fluctuations in the turbulent solar wind is supported by a recent
observational study that finds a statistically negligible contribution
from fast waves to the turbulence.\cite{Howes:2012a} Second, in the
anisotropic limit, reduced MHD provides a rigorous description of the
turbulent dynamics of \Alfvenic fluctuations for both collisional and
collisionless plasmas\cite{Schekochihin:2009} for perpendicular length
scales down to the scale of the ion Larmor radius, $k_\perp \rho_i
\sim 1$. Third, for collisional plasmas in the anisotropic limit, the
general turbulent cascade separates into five decoupled channels: two
for the upward and downward \Alfven waves, two for the upward and
downward slow waves, and one for the non-propagating entropy
fluctuations.\cite{Lithwick:2001,Schekochihin:2009} This decoupling of
turbulent energy cascades is supported by the results of numerical
simulations.\cite{Maron:2001,Cho:2002,Cho:2003} Complementing the
reduced MHD equations which describe the dynamics of the \Alfven
waves, in the collisional limit, the slow wave cascade is described by
equations for the parallel velocity and parallel magnetic field
fluctuations, and the entropy mode cascade is determined by an
equation involving the density and parallel magnetic field
fluctuations.\cite{Schekochihin:2009} In the collisionless limit, the
slow wave and entropy mode cascades are replaced by compressible
fluctuations described by a kinetic equation for the ion distribution
function.  The system described by the usual reduced MHD equations in
addition to this ion kinetic equation leads to the equations of
\emph{kinetic reduced MHD}.\cite{Schekochihin:2009} Fourth, in the
anisotropic limit, the cascade of \Alfven waves is unaffected by the
slow waves and entropy modes, but these cascades of compressible
fluctuations undergo no cascade on their own, their nonlinear transfer
instead being controlled by the turbulent dynamics of the \Alfven
waves.\cite{Lithwick:2001,Maron:2001,Cho:2002,Cho:2003,Schekochihin:2009}
As a corollary, if no slow wave or entropy mode fluctuations are
injected into the turbulence, none will be generated by the nonlinear
interaction between counterpropagating \Alfven waves.

Now we return to consider the nonlinear evolution of any wave due to
its interaction with a counterpropagating pseudo-\Alfven wave compared
to that due to a counterpropagating \Alfven wave in an incompressible
MHD plasma. We explore how this result, in the anisotropic limit,
corresponds to the properties of reduced MHD outlined above. We
specify the anisotropic limit with the ordering $k_\parallel /k_\perp
\sim \epsilon \ll 1$. To simplify matters, we choose the following
wavevectors for each of the interacting waves to obtain the maximum
possible nonlinear interaction in each case: a distorted downward wave
(either an \Alfven or a pseudo-\Alfven wave) with $\V{k}^+=
k_{\perp}\xhat - k_\parallel
\zhat$, a distorting upward \Alfven wave with $\V{k}_A^- =  k_{\perp}\yhat + k_\parallel
\zhat$, and  a distorting upward pseudo-\Alfven wave 
with $\V{k}_P^- = k_{\perp}\xhat + k_\parallel
\zhat$. In this case, the pseudo-\Alfven wave interaction is proportional to
$-2 k_\parallel k_\perp /k$ and the \Alfven wave interaction is
proportional to $-k_\perp$.  The ratio of the pseudo-\Alfven wave
interaction to the \Alfven wave interaction is therefore $2
k_\parallel/k \simeq 2 k_\parallel/k_\perp \sim
\epsilon$. This result proves that, in the anisotropic limit, the pseudo-\Alfven 
wave contribution to the nonlinear interaction is subdominant to the
\Alfven wave contribution for incompressible MHD. This is consistent
with the result from reduced MHD that the slow waves do not affect the
turbulent cascade of \Alfven waves, and that the slow waves are
cascaded by \Alfven waves.  Note that the unit vector in the direction
of polarization for the pseudo-\Alfven wave, given by $
(-k_\parallel/k) \kphat + (k_\perp/k) \zhat$, is dominated by the
$\zhat$ component in the anisotropic limit $k_\perp \gg k_\parallel$,
so the subdominant nonlinearity due to the pseudo-\Alfven wave is
sometimes referred to as the parallel nonlinearity.

As we shall see below, the equations for the evolution of the
\Alfvenic fluctuations in incompressible MHD are identical to the 
equations of reduced MHD. These equations are formally rigorous in the
anisotropic limit $k_\perp\gg k_\parallel$, but are likely to remain a
reasonable description of the lowest order \Alfvenic dynamics even for only a
moderate anisotropy.  The correspondence of the \Alfvenic dynamics
between reduced MHD and incompressible MHD is not particularly
surprising since \Alfven waves are incompressible fluctuations. For
the remainder of this paper, we focus on the dynamics of the \Alfven
waves, which necessarily become dominant and uncoupled from
pseudo-\Alfven wave fluctuations. Even if the pseudo-\Alfven waves
alter the nonlinear evolution of the largest-scale turbulent
fluctuations, they become increasingly negligible as the turbulent
cascade progresses to smaller scales through the inertial range.

\subsection{Other Properties of the Incompressible MHD Equations }
\label{sec:otherprop}

A few other properties of the incompressible MHD equations
\eqref{eq:elsasserpm} are relevant to a discussion of the nonlinear evolution in 
turbulence. First, the upward and downward waves do not exchange
energy.\cite{Maron:2001,Schekochihin:2009} This can be seen by taking
the dot product of \eqref{eq:elsasserpm} with $\V{z}^\pm$ and
integrating over space. Assuming either periodic boundary conditions
or that the Els\"asser fields $\V{z}^\pm$ vanish at infinity, the
nonlinear terms vanish and one obtains the key result
\begin{equation}
\frac{d}{dt} \int d^3\V{r} \  |\V{z}^\pm|^2 = 0.
\end{equation}
This implies that the nonlinear interactions do not lead to any
exchange of energy between upward and downward waves---the energy
fluxes of the waves in each of these directions is conserved.  One may
consider the nonlinear interaction of an upward propagating \Alfven
wave packet with a downward propagating \Alfven wave packet as an
elastic ``collision'' between the wave packets.  The wave packets may
be scattered (generally to higher wavenumber) but do not change
energy. This is the root of the common terminology describing
nonlinear interactions between counterpropagating \Alfven waves as
\Alfven wave collisions, a physical process  unrelated to the
collisionality of plasma species. It is worthwhile noting that, 
in the anisotropic limit, the same property holds for the two slow
wave cascades and the one entropy mode cascade, that no energy is
exchanged among these uncoupled cascades.\cite{Schekochihin:2009}

Second, note that the vector form of \eqref{eq:elsasserpm} readily
demonstrates that turbulence in an incompressible MHD plasma is an
inherently three-dimensional phenomenon. The linear term $\V{v}_A
\cdot \nabla \V{z}^{\pm}$ represents propagation of the \Alfven waves 
along the equilibrium magnetic field and is nonzero only when the
parallel wavenumber $k_\parallel \ne 0$, requiring the inclusion of
the field-parallel dimension. As derived above in \secref{sec:nlprop},
the nonlinearity arising from the interaction of counterpropagating
\Alfven waves is proportional to $\zhat \cdot (\hat{\V{k}}_\perp^-
\times
\V{k}_\perp^+)$. In order for $\hat{\V{k}}_\perp^- \times
\V{k}_\perp^+ \ne 0$, variation in both directions perpendicular to the magnetic field is required. This implies that both perpendicular dimensions must be
included for the nonlinear term to be represented properly. The
manifestly three-dimensional nature of plasma turbulence not only
applies to incompressible MHD plasmas, but persists as a general
characteristic of the turbulence for more complex plasmas, such as
compressible MHD plasmas or kinetic plasmas.\cite{Howes:2011a}

Finally, it is worth explicitly stating that the dominance of the
\Alfven waves in the inherently anisotropic cascade of magnetized
plasma turbulence combined  with the requirement that $\V{k}_\perp^-
\times \V{k}_\perp^+ \ne 0$ implies that \emph{the fundamental building block
of plasma turbulence is the nonlinear interaction between
perpendicularly polarized, counterpropagating
\Alfven waves.}

\subsection{The Els\"asser Potential Equations}

The presence of the pressure term in \eqref{eq:elsasserpm}
significantly complicates the use of asymptotic methods to solve for
the evolution of the nonlinear interaction between counterpropagating
\Alfven waves in the weakly nonlinear limit.  We can focus on the
dynamics of the \Alfven waves and avoid this by difficulty by
expressing the Els\"asser fields $\V{z}^{\pm}$ in terms of Els\"asser
potentials $\zeta^{\pm}$, defined by the relation by $\V{z}^{\pm} =
\zhat \times \nabla_\perp \zeta^{\pm}$.  This procedure projects the 
vector Els\"asser fields onto the polarization direction of the
\Alfven waves, $\zhat \times \hat{\V{k}}$, eliminating the 
pseudo-\Alfven waves from the system.  It also has the added benefit that
that the resulting \Alfvenic Els\"asser fields $\V{z}^{\pm}$
automatically satisfy the incompressibility condition \eqref{eq:div0}.

Substituting the Els\"asser potentials $\zeta^{\pm}$ into
\eqref{eq:elsasserpm} and simplifying yields the \emph{Els\"asser potential
equations}, as given by eq.~(21) of Schekochihin \emph{et
al.},\cite{Schekochihin:2009}
\begin{equation}
\frac{\partial \nabla_\perp^2 \zeta^{\pm}}{\partial t} 
\mp v_A \frac{\partial \nabla_\perp^2 \zeta^{\pm} }{\partial z} =-
\frac{1}{2} \left[ \{\zeta^{+}, \nabla_\perp^2 \zeta^{-}\} +
\{\zeta^{-}, \nabla_\perp^2 \zeta^{+}\} \mp \nabla_\perp^2 \{\zeta^{+},
\zeta^{-}\} \right]
\label{eq:zetapm}
\end{equation}
where the Poisson bracket is defined by
\begin{equation}
 \{f,g\}=\zhat \cdot ( \nabla_\perp f \times \nabla_\perp g)
\label{eq:poisson}
\end{equation} 
In \eqref{eq:zetapm}, the left-hand side of the equation describes the
linear evolution of the Els\"asser potentials $\zeta^{\pm}$, while the
right-hand side contains the nonlinear terms.  The form of the
Els\"asser potential equations lends them to solution by an asymptotic
expansion in the weakly nonlinear limit, where the terms on the right-hand side of 
\eqref{eq:zetapm} are small compared to the terms on the left-hand side.

Note that the Els\"asser potential equations \eqref{eq:zetapm} are
equivalent to the reduced MHD equations, which describe the
low-frequency \Alfvenic dynamics in the anisotropic limit of
compressible MHD.\cite{Schekochihin:2009} Since these equations were
derived by the projection of the incompressible MHD equations onto the
\Alfven wave polarization, they are formally correct to lowest order 
only when the pseudo-\Alfven wave dynamics do not affect the evolution
of the \Alfven waves.  As proven in \secref{sec:rmhd} above, this is
rigorously true only in the anisotropic limit, $k_\perp \gg
k_\parallel$.  However, even for a moderate anisotropy, $k_\perp >
k_\parallel$, the effect of interactions with counterpropagating
\Alfven waves remains the dominant contribution to the nonlinear
evolution of an \Alfven wave, since any anisotropy of this sense
lessens the contribution of interactions with counterpropagating
pseudo-\Alfven waves.

\section{Asymptotic Analytical Solution}
\label{sec:solution}
In this section, we derive an asymptotic solution for the nonlinear
evolution of the interaction between two counterpropagating,
perpendicularly polarized \Alfven waves in the weakly nonlinear limit.

\subsection{Initial Conditions}
\label{sec:ics}
\begin{figure}
\resizebox{5.0in}{!}{\includegraphics{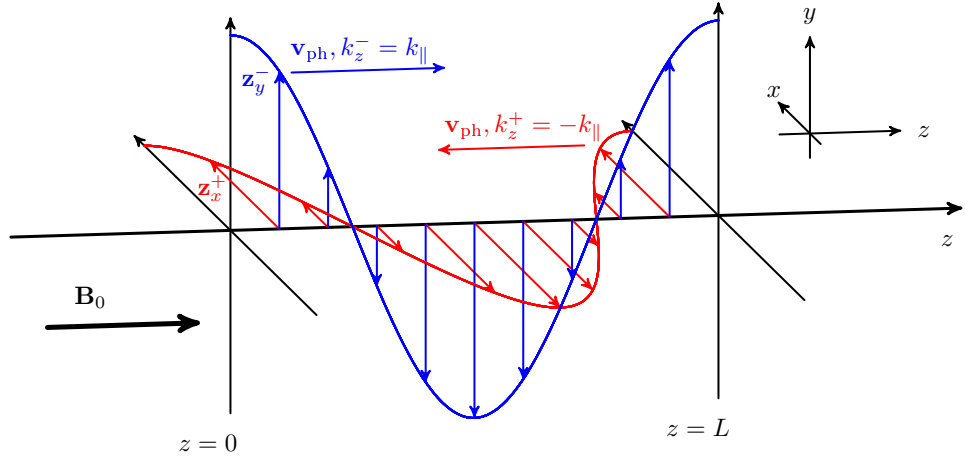}}
\caption{Schematic of the initial conditions specifying two perpendicularly 
polarized, counterpropagating \Alfven waves overlapping within a
periodic domain.
\label{fig:setup}}
\end{figure}

We begin by specifying a magnetized, incompressible ideal MHD plasma
with a uniform equilibrium magnetic field $\V{B}_0 = B_0 \zhat$ in a
triply periodic domain of dimensions $L_x \times L_y \times L_z$.
We choose initial conditions that specify two counterpropagating,
perpendicularly polarized \Alfven waves described by the Els\"asser
variables
\begin{equation}
\V{z}^{+} = z_+ \cos (k_\perp x -k_\parallel z -\omega_0 t)  \yhat 
\label{eq:zinitplus}
\end{equation}
\begin{equation}
\V{z}^{-}  =z_-  \cos (k_\perp y +k_\parallel z -\omega_0 t)\xhat
\label{eq:zinitminus}
\end{equation}
where $z_+$ and $z_-$ are constants that specify the initial
amplitudes for each wave.  The variation perpendicular to the
equilibrium magnetic field of each \Alfven wave occurs at the lowest
nonzero perpendicular wavenumber in the domain, $k_\perp = 2 \pi/L_x=
2 \pi/L_y$. Similarly, the variation parallel to the equilibrium
magnetic field occurs at the lowest nonzero parallel wavenumber
in the domain, $k_\parallel = 2 \pi/L_z$.  We adopt the convention
that the frequency of a plane \Alfven wave is always non-negative, so
the sign of the parallel component of the wavevector indicates the
direction of propagation of the plane wave mode.  Therefore, in this
calculation, $k_\perp>0$, $k_\parallel>0$, and $\omega_0>0$ are always
positive constants.  The direction of propagation of the initial waves
in \eqref{eq:zinitplus} and \eqref{eq:zinitminus} is explicitly
included by the sign of the $k_z$ component of the wavevector, chosen
to be consistent with the meanings of the Els\"asser variables
$\V{z}^{+}$ and $\V{z}^{-}$, specifying that these initial
\Alfven waves are counterpropagating. Note that, for a linear \Alfven wave 
eigenmode, the frequency must satisfy $\omega_0=k_\parallel v_A$, a
relation that will be confirmed by the necessary conditions for the
$\Order(\epsilon)$ solution. These initial conditions are depicted
schematically in \figref{fig:setup}.

The Els\"asser potentials corresponding to \eqref{eq:zinitplus} and \eqref{eq:zinitminus} 
above are given by 
\begin{equation}
\zeta^{+} = \frac{z_+ }{ k_\perp} \sin (k_\perp x -k_\parallel z -\omega_0 t)
\label{eq:zetainitplus}
\end{equation}
\begin{equation}
\zeta^{-}  = \frac{- z_- }{ k_\perp } \sin  (k_\perp y +k_\parallel z -\omega_0 t)
\label{eq:zetainitminus}
\end{equation}
We use these Els\"asser potentials to define the initial conditions of
the calculation that define two counterpropagating, perpendicularly
polarized \Alfven waves.  Note that the usual Dirichlet initial
conditions for the calculation can be obtained by setting $t=0$ in
\eqref{eq:zetainitplus} and \eqref{eq:zetainitminus} and their time derivatives.

Note that since $k_\perp$ and $k_\parallel$ are taken to be positive
constants in this calculation, the description of the various modes
that arise through nonlinear evolution uses the notation
$(k_x/k_\perp, k_y/k_\perp,k_z/k_\parallel)$.  For example, the
secondary modes that arise are $(1,1,0)$ and $(-1,1,-2)$. In addition,
we describe only the modes with $k_y\ge0$, since modes with $k_y<0$
are equivalent to changing the sign of the amplitude of modes with
$k_y>0$.

\subsection{Conversion from $\zeta^{\pm}$ to $\V{E}_\perp$ and $\V{B}_\perp$} 
\label{sec:eb}

The perpendicular velocity fluctuation $\V{u}_\perp$ can be written in
terms of a stream function $\Phi$, and the perpendicular magnetic
field fluctuation $\V{B}_\perp$ can be written in terms of a flux
function \cite{Schekochihin:2009} $\Psi$,
\begin{equation}
\V{u}_\perp = \zhat \times \nabla_\perp \Phi,
\end{equation}
\begin{equation}
\frac{\V{B}_\perp}{\sqrt{4 \pi \rho_0}} = \zhat \times \nabla_\perp \Psi.
\end{equation}
The Els\"asser potentials can be expressed as the sum and difference
of these stream and flux functions, $\zeta^{\pm} = \Phi \pm
\Psi$. Therefore, the Els\"asser
potentials may be converted to the perpendicular velocity and magnetic
field fluctuations using
\begin{equation}
\frac{\V{u}_\perp}{v_A} = \zhat \times \nabla_\perp \frac{1}{2 v_A}(\zeta^+ + \zeta^-)
\end{equation}
\begin{equation}
\frac{\V{B}_\perp}{B_0} = \zhat \times \nabla_\perp \frac{1}{2 v_A}(\zeta^+ - \zeta^-)
\label{eq:conv2b}
\end{equation}

In the laboratory and in spacecraft measurements of plasma turbulence,
the direct measurement of perpendicular velocity fluctuations
$\V{u}_\perp$  often presents a significant challenge.  It is frequently
advantageous to use the measurements of the perpendicular electric
field $\V{E}_\perp$ as a proxy for the perpendicular velocity
$\V{u}_\perp$.  For an ideal MHD plasma, Ohm's Law defines the
relation between the perpendicular electric field and perpendicular
velocity fluctuations, $\V{E} + \V{u}
\times \V{B}/c=0$. For the common case that magnetic field fluctuations are small
compared to the equilibrium magnetic field, $|\delta \V{B}| \ll
|\V{B}_0|$, the dominant, lowest order contribution to the 
perpendicular velocity fluctuation is simply given by the $\V{E} \times
\V{B}$ velocity with respect to the equilibrium magnetic field,
 $\V{u}_\perp \simeq c\V{E} \times
\V{B}_0/B_0^2$,
so we obtain,
\begin{equation}
\frac{ c \V{E}_\perp}{v_AB_0} = - \nabla_\perp \frac{1}{2v_A}(\zeta^+ + \zeta^-)
\label{eq:conv2e}
\end{equation}

Therefore, our initial conditions are given by 
\begin{equation}
\frac{\V{B}_\perp}{B_0} =  \frac{ z_+}{2 v_A} \cos (k_\perp x -k_\parallel z -\omega_0 t)  \yhat  - \frac{ z_-}{2 v_A} \cos (k_\perp y +k_\parallel z -\omega_0 t)  \xhat,
\label{eq:binit}
\end{equation}
\begin{equation}
\frac{ c \V{E}_\perp}{v_A B_0} = -\frac{ z_+}{2 v_A} \cos (k_\perp x -k_\parallel z -\omega_0 t)  \xhat  + \frac{ z_-}{2 v_A} \cos (k_\perp y +k_\parallel z -\omega_0 t)  \yhat,
\label{eq:einit}
\end{equation}
and the Poynting flux $\V{S} = (c/4\pi) \V{E} \times\V{B}$ associated
with the initial conditions is given, in dimensionless form, by
\begin{equation}
\frac{\V{S}}{\rho_0 v_A^3 } 
= \frac{ c \V{E}_\perp}{v_A B_0}  \times \frac{\V{B}_\perp}{B_0}  
= \frac{ z_+^2}{4 v_A^2} \cos^2 (k_\perp x -k_\parallel z -\omega_0 t)  (-\zhat)  
+ \frac{ z_-^2}{4 v_A^2} \cos^2 (k_\perp y +k_\parallel z -\omega_0 t)  \zhat.
\label{eq:init}
\end{equation}
In the Poynting flux, it is clear that the energy flux due to $z_+$ in
the first term propagates in the $-\zhat$ direction, and the energy
flux due to the $z_-$ in the second term propagates in the $\zhat$
direction, as expected.
\subsection{Relation of Nonlinear Solutions to Linear Wave Modes} 
\label{sec:linear}
Interpretation of the physical meaning of the nonlinear solutions
derived here is aided by the identification of some parts of the
nonlinear solution as \emph{linear} \Alfven wave modes of the
system. The linear \Alfven wave modes of \eqref{eq:elsasserpm} satisfy
the following two conditions: (i) a frequency given by the linear
dispersion relation, $\omega=\pm k_\parallel v_A$; and (ii) an
eigenfunction that satisfies $\V{u}_\perp/v_A = \pm \V{B}_\perp /B_0$.
Using the relations to express the lowest order of $\V{u}_\perp$ in
terms of $\V{E}_\perp$ given in \secref{sec:eb}, the eigenfunction
condition may be alternatively expressed by
\begin{equation}
\frac{\V{B}_\perp}{B_0} =   \pm \frac{ c \V{E}_\perp}{v_AB_0}\times \zhat.
\label{eq:alfeigen}
\end{equation}
If the nonlinear solutions for $\V{E}_\perp$ and $\V{B}_\perp$ have
components that satisfy these two conditions, we identify that part of
the nonlinear solution as a particular linear \Alfven wave mode.  The
physical interpretation is that two counterpropagating,
perpendicularly polarized \Alfven waves interact nonlinearly to
transfer energy to other linear \Alfven wave modes with larger values
of $k_\perp$.  This is the fundamental physical mechanism that
underlies the turbulent cascade of energy from large scales to small
scales.

In addition, the nonlinear solution may contain components that do not
satisfy the conditions for a linear wave, and so these parts of the
solution are inherently nonlinear fluctuations. This issue relates to
an important question at the forefront of research on \Alfvenic plasma
turbulence: how much energy is contained in linear wave modes compared
to that in inherently nonlinear fluctuations? The asymptotic solution
presented here represents an important step in the effort to answer
this question in a rigorous, quantitative manner.

\subsection{Transformation to Characteristic Variables}
The Els\"asser potentials are functions of the three spatial
dimensions and time, $\zeta^{\pm}(x,y,z,t)$, and the linear part on
the left-hand side of \eqref{eq:zetapm} is a partial differential
equation in $z$ and $t$. We use the method of characteristics to
convert the linear part of these equations to an ordinary differential
equation in terms of characteristic variables,
\begin{equation}
\phi_\pm = z \pm v_A t.
\label{eq:phipm}
\end{equation}
In terms of these characteristic variables, the set of Els\"asser
potential equations becomes
\begin{equation}
\frac{\partial  \nabla_\perp^2 \zeta^{\pm}}{\partial \phi_\mp}
= \pm \frac{1}{4v_A} \left[ \{\zeta^{\pm},  \nabla_\perp^2 \zeta^{\mp}\}
+  \{\zeta^{\mp},  \nabla_\perp^2 \zeta^{\pm}\} 
\mp \nabla_\perp^2 \{\zeta^{\pm},   \zeta^{\mp}\} \right].
\label{eq:zetaphipm}
\end{equation}
Using the linear dispersion relation $\omega_0 =k_\parallel v_A$, the initial conditions
are expressed in terms of $\phi_\pm$,
\begin{equation}
\zeta^{+} = \frac{ z_+ }{ k_\perp} \sin (k_\perp x -k_\parallel \phi_+)
\label{eq:zeta1plusphi}
\end{equation}
\begin{equation}
\zeta^{-}=  \frac{-z_-}{ k_\perp} \sin  (k_\perp y +k_\parallel \phi_-)
\label{eq:zeta1minusphi}
\end{equation}
Recall that, by convention, we take $\omega_0 >0$ and $k_\parallel >
0$, so the direction of propagation for each mode is specified by the
sign of the $k_\parallel$ term explicitly.

\subsection{Asymptotic Expansion}
We solve for the nonlinear evolution of this system asymptotically in
the weak turbulence limit of small initial wave amplitudes compared to
the \Alfven speed, where we  define the ordering parameter for
our asymptotic expansion by
\begin{equation}
\frac{z_\pm}{v_A} \sim \epsilon \ll 1
\end{equation}
To solve this asymptotically, we will expand the solutions for
$\zeta^{\pm}$ in orders of $\epsilon$,
\begin{equation}
\zeta^{\pm} = \zeta^{\pm}_0 + \epsilon \zeta^{\pm}_1 + \epsilon^2 \zeta^{\pm}_2 + \epsilon^3 \zeta^{\pm}_3 + \cdots
\label{eq:zpm}
\end{equation}
We assume that the $\Order(1)$ solution is zero, $\zeta^{\pm}_0 =0$.

Substituting \eqref{eq:zpm} into \eqref{eq:zetaphipm}, we can find the
equations for each order. At $\Order(\epsilon)$, we obtain
\begin{equation}
\frac{\partial  \nabla_\perp^2 \zeta^{\pm}_1}{\partial \phi_\mp} = 0.
\label{eq:zetaphipm_O1}
\end{equation}
At $\Order(\epsilon^2)$, we obtain
\begin{equation}
\frac{\partial  \nabla_\perp^2 \zeta^{\pm}_2}{\partial \phi_\mp}
= \pm \frac{1}{4v_A} \left[ \{\zeta^{\pm}_1,  \nabla_\perp^2 \zeta^{\mp}_1\}
+  \{\zeta^{\mp}_1,  \nabla_\perp^2 \zeta^{\pm}_1\} 
\mp \nabla_\perp^2 \{\zeta^{\pm}_1,   \zeta^{\mp}_1\} \right].
\label{eq:zetaphipm_O2}
\end{equation}
At $\Order(\epsilon^3)$, we obtain
\begin{equation}
\frac{\partial  \nabla_\perp^2 \zeta^{\pm}_3}{\partial \phi_\mp}
= \pm \frac{1}{4v_A} \left[ \{\zeta^{\pm}_1,  \nabla_\perp^2 \zeta^{\mp}_2\}
+  \{\zeta^{\pm}_2,  \nabla_\perp^2 \zeta^{\mp}_1\}
+  \{\zeta^{\mp}_1,  \nabla_\perp^2 \zeta^{\pm}_2\} 
+  \{\zeta^{\mp}_2,  \nabla_\perp^2 \zeta^{\pm}_1\} 
\mp \nabla_\perp^2 \{\zeta^{\pm}_1,   \zeta^{\mp}_2\} 
\mp \nabla_\perp^2 \{\zeta^{\pm}_2,   \zeta^{\mp}_1\} \right].
\label{eq:zetaphipm_O3}
\end{equation}
We now proceed to solve successively for the evolution of the system
at each order.

\subsection{Primary $\Order(\epsilon)$ Solution}
The $\Order(\epsilon)$ solution describes the linear evolution of the
system.  By inspection, it is clear that the initial conditions
specified by \eqref{eq:zeta1plusphi} and \eqref{eq:zeta1minusphi}
satisfy the $\Order(\epsilon)$ equation \eqref{eq:zetaphipm_O1} so
long as the solvability condition $\omega_0=k_\parallel v_A$ is
satisfied, equivalent to the linear dispersion relation for \Alfven
waves.  Therefore, the $\Order(\epsilon)$ solutions
$\zeta^{\pm}_1(x,y,z,t)$ are simply given by \eqref{eq:zetainitplus}
and \eqref{eq:zetainitminus}, so that the initial conditions for all
higher order solutions $n>1$ are zero, $\zeta^{\pm}_n=0$.  We can use
\eqref{eq:conv2b} and (\ref{eq:conv2e}) to obtain the solutions for
the electromagnetic fields $\V{B}_{\perp 1}$ and $\V{E}_{\perp 1}$
given by \eqref{eq:binit} and (\ref{eq:einit}).

\subsection{Secondary $\Order(\epsilon^2)$ Solution}
\label{sec:O2}
The $\Order(\epsilon^2)$ equations describe the lowest order nonlinear
evolution. At this order, $\zeta^{\pm}_1$ are known functions, so we
may substitute them into the right-hand side of
\eqref{eq:zetaphipm_O2} and solve for $\zeta^{\pm}_2$.

For the particularly symmetric solutions for $\zeta^{\pm}_1$, given by
\eqref{eq:zetainitplus} and \eqref{eq:zetainitminus}, the first and
second terms on the right-hand side of \eqref{eq:zetaphipm_O2} cancel,
so the nonlinear energy transfer from the $\Order(\epsilon)$ solution
to the $\Order(\epsilon^2)$ solution is due entirely to the third
nonlinear term, yielding the equations
\begin{equation}
\frac{\partial  \nabla_\perp^2 \zeta^{\pm}_2}{\partial \phi_\mp}
= \frac{-k_\perp^2 z_+z_-}{4v_A} \left\{ 
\cos\left[k_\perp x+k_\perp y- k_\parallel (\phi_+ - \phi_-)\right] 
+  \cos\left[k_\perp x - k_\perp y - k_\parallel (\phi_+ + \phi_-)\right] 
 \right\}
\label{eq:zpm_a}
\end{equation}

Integration of these equations to obtain solutions for $\nabla_\perp^2 \zeta^{\pm}_2$ 
is fairly straight-forward, but the limits of this
integration must be handled carefully. Physically, we want to
integrate this initial value problem from time $t'=0$ to time $t'=t$
to obtain the solution $\zeta^{\pm}_2(t)$ in terms of
$\left.\zeta^{\pm}_2(t)\right|_{t=0}$.  The initial time $t=0$ corresponds to the case
when the characteristic variables are equal, $\phi_+=\phi_-$.
Therefore, to  perform the $\phi'_-$ integration to obtain
$\zeta^{+}_2$ using \eqref{eq:zpm_a}, the time integration from $t'=0$
to time $t'=t$ corresponds to an integration of the $\phi'_-$ variable
from $\phi'_-=\phi_+$ to $\phi'_-=\phi_-$.  Similarly, to obtain
$\zeta^{-}_2$, we integrate the $\phi'_+$ variable from
$\phi'_+=\phi_-$ to $\phi'_+=\phi_+$.  Therefore, the integral
\begin{equation}
\int_{\phi_+}^{\phi_-} \frac{\partial  \nabla_\perp^2 \zeta^{+}_2 (x,y,\phi_+,\phi_-')}{\partial \phi'_-}
d \phi'_-= \nabla_\perp^2 \zeta^{+}_2(x,y,\phi_+,\phi_-) - \nabla_\perp^2
\zeta^{+}_2(x,y,\phi_+,\phi_+) = \nabla_\perp^2 \zeta^{+}_2(x,y,\phi_+,\phi_-),
\end{equation}
where the last equality holds because of the initial condition that
$\left.\zeta^{+}_2(t)\right|_{t=0}=0$, so that $\zeta^{+}_2(x,y,\phi_+,\phi_+) = 0$.

After integrating \eqref{eq:zpm_a} over $\phi_-'$ with the integration
limits specified above, the $\nabla_\perp^2$ may be trivially
eliminated due to the sinusoidal nature of the solutions in the
perpendicular plane.  Thus, we obtain the $\Order(\epsilon^2)$
solutions
\begin{eqnarray}
\zeta^{+}_2 & = \frac{ z_+ z_-}{8 \omega_0} &\left\{ 
\sin[k_\perp x + k_\perp y - k_\parallel (\phi_+-\phi_-)] - \sin[k_\perp x + k_\perp y ]  \right. 
\nonumber \\
& &-  \left.  \sin[k_\perp x - k_\perp y - k_\parallel (\phi_+ + \phi_-) ] 
+\sin[k_\perp x - k_\perp y - 2 k_\parallel \phi_+ ] \right\}
\label{eq:zeta2plusphi}
\end{eqnarray}
\begin{eqnarray}
\zeta^{-}_2 & = -\frac{ z_+ z_-}{8 \omega_0} &\left\{ 
\sin[k_\perp x + k_\perp y - k_\parallel (\phi_+-\phi_-)] - \sin[k_\perp x + k_\perp y ]  \right. 
\nonumber \\
& &+  \left.  \sin[k_\perp x - k_\perp y - k_\parallel (\phi_+ + \phi_-) ] 
-\sin[k_\perp x - k_\perp y - 2 k_\parallel \phi_- ] \right\}
\label{eq:zeta2minusphi}
\end{eqnarray}
Converting this solution from characteristic variables $\phi_+$ and
$\phi_-$ back to $z$ and $t$ gives
\begin{eqnarray}
\zeta^{+}_2 & = \frac{ z_+ z_-}{8 \omega_0} &\left\{ 
\sin[k_\perp x + k_\perp y - 2 \omega_0 t ] - 
\sin[k_\perp x + k_\perp y ]  \right. \nonumber \\
& &-  \left.  \sin[k_\perp x - k_\perp y - 2 k_\parallel z ] 
+\sin[k_\perp x - k_\perp y - 2 k_\parallel z - 2 \omega_0 t ] \right\}
\label{eq:zeta2plus}
\end{eqnarray}
\begin{eqnarray}
\zeta^{-}_2 & = -\frac{ z_+ z_-}{8 \omega_0} &\left\{ 
\sin[k_\perp x + k_\perp y - 2 \omega_0 t ] - \sin[k_\perp x + k_\perp y ]  \right. \nonumber \\
& &+  \left.  \sin[k_\perp x - k_\perp y - 2 k_\parallel z] 
-\sin[k_\perp x - k_\perp y - 2 k_\parallel z + 2 \omega_0 t  ] \right\}
\label{eq:zeta2minus}
\end{eqnarray}
It is easily verified that these $\Order(\epsilon^2)$ solutions indeed
satisfy the initial conditions $\left.\zeta^{\pm}_2(t)\right|_{t=0}=0$.

We may then convert from the Els\"asser potentials $\zeta^\pm_2$ to
the electromagnetic fields $\V{B}_{\perp 2}$ and $\V{E}_{\perp 2}$ using 
\eqref{eq:conv2b} and (\ref{eq:conv2e}) to obtain
\begin{eqnarray}
\frac{\V{B}_{\perp 2}}{B_0}&  = \frac{  z_+ z_-}{16 v_A^2} 
\frac{  k_\perp }{k_\parallel}
&\left\{ 
\left[ 2 \cos(k_\perp x + k_\perp y - 2 \omega_0 t ) - 
2 \cos(k_\perp x + k_\perp y )  \right] (-\xhat+ \yhat) \right. \nonumber \\
& &+  \left. \left[ \cos(-k_\perp x + k_\perp y +2 k_\parallel z + 2 \omega_0 t ) 
-\cos(- k_\perp x + k_\perp y + 2 k_\parallel z - 2 \omega_0 t ) 
\right](\xhat+ \yhat) \right\} \label{eq:O2B}
\end{eqnarray}
\begin{eqnarray}
\frac{ c \V{E}_{\perp 2}}{v_A B_0} & = -\frac{  z_+ z_-}{16 v_A^2} 
\frac{  k_\perp }{k_\parallel}
&\left\{ \left[ 2 \cos(-k_\perp x + k_\perp y +2 k_\parallel z ) -
\cos(-k_\perp x + k_\perp y +2 k_\parallel z + 2 \omega_0 t ) 
 \right. \right. \nonumber \\
& &  \left. \left. - \cos(- k_\perp x + k_\perp y + 2 k_\parallel z - 2 \omega_0 t ) 
\right](-\xhat+ \yhat)  \right\}
\end{eqnarray}
where we have used $\omega_0=k_\parallel v_A$ to simplify the coefficient.

The $\Order(\epsilon^2)$ nonlinear solutions for $\V{B}_{\perp 2}$ and
$\V{E}_{\perp 2}$ enable the identification of several important
characteristics of the fluctuations at this order. First, there is no
secular transfer of energy to modes at this order. Second, the
$\Order(\epsilon^2)$ nonlinear solution is comprised of two spatial
Fourier components $(k_x/k_\perp,
k_y/k_\perp,k_z/k_\parallel)=(1,1,0)$ and $(-1,1,2)$. Note that the
Fourier components arising at this order are given by $\V{k}^- \pm
\V{k}^+$, where the wavevectors of the initial \Alfven waves are
$\V{k}^+= k_\perp \xhat - k_\parallel \zhat$ and $\V{k}^-= k_\perp
\yhat + k_\parallel \zhat$, or $(1,0,-1)$ and $(0,1,1)$.  The
$(-1,1,2)$ modes satisfy the two conditions outlined in
\secref{sec:linear} to be identified as two counterpropagating linear
\Alfven waves: (i) the modes satisfy the linear dispersion relation
$\omega=\pm k_z v_A$, given the frequency $\omega=\pm 2\omega_0$ and
the component of the wavevector along the equilibrium magnetic field
$k_z=2 k_\parallel$; and (ii) the modes satisfy the eigenfunction
relation for linear \Alfven waves given by \eqref{eq:alfeigen}. One of
these modes propagates as a linear \Alfven wave down the equilibrium
magnetic field, and the other propagates up the equilibrium magnetic
field. For the symmetric initial conditions specified in this problem,
the linear superposition of these two counterpropagating linear
\Alfven waves generates a standing wave in the magnetic field
polarized in the $\xhat+ \yhat$ direction and a standing wave in the
electric field polarized in the $-\xhat+ \yhat$ direction. Third, the
$(1,1,0)$ mode is a purely magnetic fluctuation polarized in the
$-\xhat+ \yhat$ direction.  It is important to note that this mode has
no spatial variation along the equilibrium magnetic field, $k_z=0$,
but it oscillates at frequency $\omega=2
\omega_0$, therefore this mode does \emph{not} satisfy the linear
dispersion relation for \Alfven waves, $\omega=\pm k_z v_A$.  Hence,
this mode represents an inherently nonlinear fluctuation in the
magnetic field with no associated fluctuation in the electric field.
Finally, note that the terms in the $\Order(\epsilon^2)$ solution with
no time dependence arise to satisfy the zero initial conditions at
this order, $\left.\zeta^{\pm}_2(x,y,z,t)\right|_{t=0}=0$.

\subsection{Tertiary $\Order(\epsilon^3)$ Solution}
\label{sec:O3}
The $\Order(\epsilon^3)$ equations govern the next higher order of the
nonlinear evolution.  As we shall see, it is necessary to carry out
the asymptotic expansion to this order to capture the secular transfer
of energy from the primary $\Order(\epsilon)$ linear \Alfven waves to linear
\Alfven waves with higher perpendicular wavenumber. At this order, the
solutions for $\zeta^{\pm}_2$, given by
\eqref{eq:zeta2plus} and \eqref{eq:zeta2minus}, are known, so we may 
substitute them into the right-hand side of \eqref{eq:zetaphipm_O3}
and solve for $\zeta^{\pm}_3$.  The procedure followed to obtain the
$\Order(\epsilon^3)$ solution is exactly the same as outlined in
detail for the $\Order(\epsilon^2)$ solution in \secref{sec:O2}
above. We omit the many of the mathematical steps of this derivation,
focusing only on the important differences that arise that lead to the
secular transfer of energy to $\Order(\epsilon^3)$ linear \Alfven wave
modes.

Evaluating the six terms on the right hand side of
\eqref{eq:zetaphipm_O3} using the solutions for $\zeta^{\pm}_1$, given
by \eqref{eq:zetainitplus} and \eqref{eq:zetainitminus}, and the
solutions for $\zeta^{\pm}_2$, given by
\eqref{eq:zeta2plus} and \eqref{eq:zeta2minus}, we obtain the two evolution equations, 
in  terms of the characteristic variables,
\begin{eqnarray}
\frac{\partial \nabla_\perp^2 \zeta^{+}_3}{\partial \phi_-} & = \frac{ z_+^2 z_- k_\perp^3}{64 \omega_0 v_A} &\left\{ 
4 \cos[2 k_\perp x + k_\perp y - k_\parallel \phi_+] 
- 4 \cos[2 k_\perp x + k_\perp y - 2 k_\parallel \phi_+ + k_\parallel \phi_-]  \right. \label{eq:eq3plus}
\\
& & \left. + 4 \cos[-2 k_\perp x + k_\perp y + 2 k_\parallel \phi_+ +  k_\parallel \phi_-] -4 \cos[-2 k_\perp x + k_\perp y + k_\parallel \phi_+ + 2 k_\parallel \phi_-] \right\}\nonumber \\
& + \frac{ z_+ z_-^2 k_\perp^3}{64 \omega_0 v_A} &\left\{ 
6 \cos[ k_\perp x + 2 k_\perp y  + k_\parallel \phi_-] 
- 6 \cos[ k_\perp x + 2 k_\perp y - k_\parallel \phi_+ + 2 k_\parallel \phi_-]  \right. \nonumber\\
& &  +6 \cos[ -k_\perp x + 2 k_\perp y + k_\parallel \phi_+ + 2 k_\parallel \phi_-] 
 -6 \cos[ -k_\perp x + 2 k_\perp y +2 k_\parallel \phi_+ + k_\parallel \phi_-] 
 \nonumber\\
&&  \left. + 2  \cos[ k_\perp x - k_\parallel \phi_-] 
- 2 \cos[ k_\perp x -2  k_\parallel \phi_+ + k_\parallel \phi_-] 
 \right\}\nonumber 
\end{eqnarray}
\begin{eqnarray}
\frac{\partial \nabla_\perp^2 \zeta^{-}_3}{\partial \phi_+} & = \frac{ z_+^2 z_- k_\perp^3}{64 \omega_0 v_A} &\left\{ 
6 \cos[ 2 k_\perp x +  k_\perp y  - k_\parallel \phi_+] 
- 6 \cos[2 k_\perp x + k_\perp y - 2k_\parallel \phi_+ +  k_\parallel \phi_-]  \right.\label{eq:eq3minus}
 \\
& &  +6 \cos[ -2 k_\perp x +  k_\perp y + 2 k_\parallel \phi_+ +  k_\parallel \phi_-] 
 -6 \cos[ -2 k_\perp x +  k_\perp y + k_\parallel \phi_+ +2  k_\parallel \phi_-] 
 \nonumber\\
&&  \left. + 2  \cos[ k_\perp y + k_\parallel \phi_+] 
- 2 \cos[ k_\perp y - k_\parallel \phi_+ +2  k_\parallel \phi_-] 
 \right\} \nonumber  \\
& + \frac{ z_+ z_-^2 k_\perp^3}{64 \omega_0 v_A}  &\left\{ 
 4 \cos[ k_\perp x + 2 k_\perp y -  k_\parallel \phi_+ + 2 k_\parallel \phi_-] 
- 4 \cos[ k_\perp x + 2 k_\perp y + k_\parallel \phi_-] 
 \right. \nonumber\\
& & \left. + 4 \cos[- k_\perp x + 2 k_\perp y + 2 k_\parallel \phi_+ +  k_\parallel \phi_-] 
-4 \cos[- k_\perp x + 2 k_\perp y + k_\parallel \phi_+ + 2 k_\parallel \phi_-] \right\}\nonumber 
\end{eqnarray}
Inspection of \eqref{eq:eq3plus} and \eqref{eq:eq3minus} shows the two
terms that will be responsible for secular energy transfer to modes at
$\Order(\epsilon^3)$. The first term on the right-hand side of
\eqref{eq:eq3plus} does not depend on $\phi_-$, so integrating the
equation over $\phi_-'$ from $\phi'_-=\phi_+$ to $\phi'_-=\phi_-$
leads to a factor of $(\phi_- -\phi_+ )= - 2 v_A t$ in the coefficient
multiplying the cosine function. Therefore, the amplitude of the
resulting mode will increase secularly with $t$.  Similarly, the
eighth term on the right-hand side of \eqref{eq:eq3minus} does not
depend on $\phi_+$, so integrating the equation over $\phi_+'$ from
$\phi'_+=\phi_-$ to $\phi'_+=\phi_+$ leads to a factor of $(\phi_+
-\phi_- )= 2 v_A t$ in the coefficient, causing a secular increase of
the mode amplitude with time. This fundamental mathematical difference
in the $\Order(\epsilon^3)$ solution is the underlying cause for the
nonlinear transfer of energy from the primary counterpropagating
\Alfven waves to the two linear \Alfven wave modes of the $\Order(\epsilon^3)$ 
solution.

After integrating \eqref{eq:eq3plus} and \eqref{eq:eq3minus} and
eliminating the $\nabla_\perp^2$ following the steps outlined in
\secref{sec:O2}, we obtain the $\Order(\epsilon^3)$ solutions for
$\zeta^{\pm}_3$ in terms of the characteristic variables,
\begin{eqnarray}
\zeta^{+}_3 & = \frac{ z_+^2 z_- k_\perp}{320 \omega_0^2} &\left\{ 
8 \omega_0 t \cos[2 k_\perp x + k_\perp y - k_\parallel \phi_+]  \right.\nonumber \\
& &  + 4 \sin[2 k_\perp x + k_\perp y - 2 k_\parallel \phi_+ + k_\parallel \phi_-] 
-4 \sin[2 k_\perp x + k_\perp y - k_\parallel \phi_+] \nonumber \\
& & \left. + 2 \sin[-2 k_\perp x + k_\perp y + k_\parallel \phi_+ + 2 k_\parallel \phi_-] 
+ 2  \sin[-2 k_\perp x + k_\perp y + 3k_\parallel \phi_+] 
-4 \sin[-2 k_\perp x + k_\perp y + 2 k_\parallel \phi_+ +  k_\parallel \phi_-] \right\}\nonumber \\
& + \frac{ z_+ z_-^2 k_\perp}{320 \omega_0^2} &\left\{ 
 3 \sin[ k_\perp x + 2 k_\perp y - k_\parallel \phi_+ + 2 k_\parallel \phi_-] 
+ 3 \sin[ k_\perp x + 2 k_\perp y + k_\parallel \phi_+] 
- 6 \sin[ k_\perp x + 2 k_\perp y  + k_\parallel \phi_-] \right.\nonumber \\
& &  +6 \sin[ -k_\perp x + 2 k_\perp y +2 k_\parallel \phi_+ + k_\parallel \phi_-] 
- 3 \sin[ -k_\perp x + 2 k_\perp y + 3 k_\parallel \phi_+] 
-3 \sin[ -k_\perp x + 2 k_\perp y + k_\parallel \phi_+ + 2k_\parallel \phi_-] \nonumber \\
& & \left. +10 \sin[ k_\perp x -2  k_\parallel \phi_+ + k_\parallel \phi_-] 
+ 10  \sin[ k_\perp x - k_\parallel \phi_-] 
-20 \sin[ k_\perp x - k_\parallel \phi_+] \right\}\nonumber
\label{eq:zeta3plusphi}
\end{eqnarray}
\begin{eqnarray}
\zeta^{-}_3 &  = \frac{ z_+^2 z_- k_\perp}{320 \omega_0^2} &\left\{ 
6 \sin[ 2 k_\perp x +  k_\perp y - k_\parallel \phi_+] 
- 3 \sin[ 2 k_\perp x +  k_\perp y - 2 k_\parallel \phi_+ + k_\parallel \phi_-] 
- 3 \sin[ 2 k_\perp x +  k_\perp y -  k_\parallel \phi_-] \right.\nonumber \\
& & + 6 \sin[ -2 k_\perp x +  k_\perp y +k_\parallel \phi_+ +2 k_\parallel \phi_-] 
- 3 \sin[ -2 k_\perp x +  k_\perp y + 2 k_\parallel \phi_+ + k_\parallel \phi_-] 
- 3 \sin[ -2 k_\perp x +  k_\perp y +3  k_\parallel \phi_-] \nonumber \\
& & \left. +20 \sin[ k_\perp y + k_\parallel \phi_-]
- 10 \sin[ k_\perp y -  k_\parallel \phi_+ + 2 k_\parallel \phi_-] 
- 10  \sin[ k_\perp y + k_\parallel \phi_+]  \right\} \nonumber \\
& + \frac{ z_+ z_-^2 k_\perp}{320 \omega_0^2} &\left\{ 
8 \omega_0 t \cos[ k_\perp x + 2 k_\perp y + k_\parallel \phi_-]  \right.\nonumber \\
& &  +4 \sin[ k_\perp x + 2 k_\perp y -  k_\parallel \phi_+ + 2 k_\parallel \phi_-] 
-4 \sin[ k_\perp x + 2 k_\perp y + k_\parallel \phi_-] \nonumber \\
& & \left.
+ 2 \sin[- k_\perp x + 2 k_\perp y + 3k_\parallel \phi_-] 
+ 2 \sin[- k_\perp x + 2 k_\perp y + 2 k_\parallel \phi_+ +  k_\parallel \phi_-]
 - 4 \sin[- k_\perp x + 2 k_\perp y + k_\parallel \phi_+ + 2 k_\parallel \phi_-] 
 \right\}\nonumber
\label{eq:zeta3minusphi}
\end{eqnarray}

Converting this solution from characteristic variables $\phi_+$ and
$\phi_-$ back to $z$ and $t$ gives
\begin{eqnarray}
\zeta^{+}_3 & = \frac{ z_+^2 z_- k_\perp}{320 \omega_0^2} &\left\{ 
8 \omega_0 t \cos[2 k_\perp x + k_\perp y- k_\parallel z -\omega_0 t ]  \right.\nonumber \\
& &  + 4 \sin[2 k_\perp x + k_\perp y - k_\parallel z -3\omega_0 t] 
-4 \sin[2 k_\perp x + k_\perp y  - k_\parallel z -\omega_0 t ] \nonumber \\
& & \left. + 2 \sin[-2 k_\perp x + k_\perp y + 3 k_\parallel z -\omega_0 t ] 
+ 2  \sin[-2 k_\perp x + k_\perp y  +3 k_\parallel z +3\omega_0 t] 
-4 \sin[-2 k_\perp x + k_\perp y  +3 k_\parallel z +\omega_0 t] \right\}\nonumber \\
& + \frac{ z_+ z_-^2 k_\perp}{320 \omega_0^2} &\left\{ 
3 \sin[ k_\perp x + 2 k_\perp y + k_\parallel z -3\omega_0 t ] 
+ 3 \sin[ k_\perp x + 2 k_\perp y  + k_\parallel z +\omega_0 t ] 
- 6 \sin[ k_\perp x + 2 k_\perp y   + k_\parallel z -\omega_0 t ] \right.\nonumber \\
& &  +6 \sin[ -k_\perp x + 2 k_\perp y  +3 k_\parallel z +\omega_0 t ] 
- 3 \sin[ -k_\perp x + 2 k_\perp y  +3 k_\parallel z +3\omega_0 t ] 
-3 \sin[ -k_\perp x + 2 k_\perp y  +3 k_\parallel z -\omega_0 t ] \nonumber \\
& & \left. +10 \sin[ k_\perp x - k_\parallel z -3 \omega_0 t ] 
+ 10  \sin[ k_\perp x - k_\parallel z +\omega_0 t ] 
-20 \sin[ k_\perp x - k_\parallel z -\omega_0 t ] \right\}\nonumber
\label{eq:zeta3plus}
\end{eqnarray}
\begin{eqnarray}
\zeta^{-}_3 &  = \frac{ z_+^2 z_- k_\perp}{320 \omega_0^2} &\left\{ 
 6 \sin[ 2 k_\perp x +  k_\perp y - k_\parallel z -\omega_0 t] 
- 3 \sin[ 2 k_\perp x +  k_\perp y - k_\parallel z -3\omega_0 t ] 
- 3 \sin[ 2 k_\perp x +  k_\perp y  - k_\parallel z +\omega_0 t] \right.\nonumber \\
& & + 6 \sin[ -2 k_\perp x +  k_\perp y +3 k_\parallel z -\omega_0 t ] 
- 3 \sin[ -2 k_\perp x +  k_\perp y  +3 k_\parallel z +\omega_0 t] 
- 3 \sin[ -2 k_\perp x +  k_\perp y  +3 k_\parallel z -3\omega_0 t] \nonumber \\
& & \left. +20 \sin[ k_\perp y + k_\parallel z -\omega_0 t]
- 10 \sin[ k_\perp y  + k_\parallel z -3 \omega_0 t] 
- 10  \sin[ k_\perp y + k_\parallel z +\omega_0 t ]  \right\} \nonumber \\
& + \frac{ z_+ z_-^2 k_\perp}{320 \omega_0^2} &\left\{ 
8 \omega_0 t \cos[ k_\perp x + 2 k_\perp y + k_\parallel z -\omega_0 t]  \right.\nonumber \\
& &  +4 \sin[ k_\perp x + 2 k_\perp y + k_\parallel z -3\omega_0 t] 
-4 \sin[ k_\perp x + 2 k_\perp y + k_\parallel z -\omega_0 t] \nonumber \\
& & \left.
+ 2 \sin[- k_\perp x + 2 k_\perp y +3 k_\parallel z +\omega_0 t] 
+ 2 \sin[- k_\perp x + 2 k_\perp y +3 k_\parallel z -3\omega_0 t]
 - 4 \sin[- k_\perp x + 2 k_\perp y  +3 k_\parallel z -\omega_0 t ] 
 \right\}\nonumber
\label{eq:zeta3minus}
\end{eqnarray}

We may then convert from the Els\"asser potentials $\zeta^\pm_3$ to
the electromagnetic fields $\V{B}_{\perp 3}$ and $\V{E}_{\perp 3}$
using \eqref{eq:conv2b} and (\ref{eq:conv2e}) to obtain
\begin{eqnarray}
\frac{ \V{B}_{\perp 3}}{B_0}&  =  \frac{ z_+^2 z_- }{640 v_A^3} 
\frac{k_\perp^2}{k_\parallel^2}
&\left\{  \left[ 
-8 \omega_0 t\sin(2k_\perp x + k_\perp y - k_\parallel z -\omega_0 t ) 
+ 3\cos(2k_\perp x + k_\perp y - k_\parallel z +\omega_0 t ) \right. \right. \label{eq:bperp3}
 \\
&& \left. \left. -10\cos(2k_\perp x + k_\perp y - k_\parallel z -\omega_0 t )
+7 \cos(2k_\perp x + k_\perp y - k_\parallel z -3\omega_0 t ) \right](-\xhat+ 2\yhat) \right.  \nonumber\\
&& \left. 
+ \left[ -2\cos(-2k_\perp x + k_\perp y +3 k_\parallel z +3\omega_0 t ) 
+ \cos(-2k_\perp x + k_\perp y +3 k_\parallel z +\omega_0 t ) \right. \right. \nonumber\\
&& \left. \left. +4\cos(-2k_\perp x + k_\perp y +3 k_\parallel z -\omega_0 t )
-3 \cos(-2k_\perp x + k_\perp y +3 k_\parallel z -3\omega_0 t ) \right](\xhat+ 2\yhat) \right.  \nonumber\\
&& \left. 
+ \left[-10\cos(k_\perp y + k_\parallel z +\omega_0 t ) 
+20 \cos(k_\perp y + k_\parallel z -\omega_0 t ) 
- 10 \cos(k_\perp y + k_\parallel z -3 \omega_0 t )\right]\xhat \right\}  \nonumber\\
&+ \frac{ z_+ z_-^2 }{640 v_A^3} 
\frac{k_\perp^2}{k_\parallel^2} &\left\{ \left[ 
8 \omega_0 t\sin(k_\perp x + 2 k_\perp y + k_\parallel z -\omega_0 t ) 
+ 3\cos(k_\perp x + 2 k_\perp y + k_\parallel z +\omega_0 t ) \right. \right. \nonumber\\
&& \left. \left. -2\cos(k_\perp x + 2 k_\perp y + k_\parallel z -\omega_0 t )
- \cos(k_\perp x + 2 k_\perp y + k_\parallel z -3\omega_0 t ) \right](-2\xhat+ \yhat) \right.  \nonumber\\
&& \left. 
+ \left[ 3\cos(-k_\perp x + 2 k_\perp y +3 k_\parallel z +3\omega_0 t ) 
-4 \cos(-k_\perp x + 2k_\perp y +3 k_\parallel z +\omega_0 t ) \right. \right. \nonumber\\
&& \left. \left. -\cos(-k_\perp x +2 k_\perp y +3 k_\parallel z -\omega_0 t )
+2 \cos(-k_\perp x + 2k_\perp y +3 k_\parallel z -3\omega_0 t ) \right](2\xhat+ \yhat) \right.  \nonumber\\
&& \left. 
+ \left[ 10\cos(k_\perp x - k_\parallel z +\omega_0 t ) 
-20 \cos(k_\perp x - k_\parallel z -\omega_0 t ) 
+ 10 \cos(k_\perp x - k_\parallel z -3 \omega_0 t )\right]\yhat \right\} \nonumber
\end{eqnarray}
\begin{eqnarray}
\frac{ c \V{E}_{\perp 3}}{v_A B_0} & =   \frac{ z_+^2 z_- }{640 v_A^3} 
\frac{k_\perp^2}{k_\parallel^2}
&\left\{ \left[ 
8 \omega_0 t\sin(2k_\perp x + k_\perp y - k_\parallel z -\omega_0 t ) 
+ 3\cos(2k_\perp x + k_\perp y - k_\parallel z +\omega_0 t ) \right. \right.\label{eq:eperp3}
 \\
&& \left. \left. -2\cos(2k_\perp x + k_\perp y - k_\parallel z -\omega_0 t )
-\cos(2k_\perp x + k_\perp y - k_\parallel z -3\omega_0 t ) \right](2\xhat+ \yhat)\right.  \nonumber\\
&& \left. 
+ \left[ -2\cos(-2k_\perp x + k_\perp y +3 k_\parallel z +3\omega_0 t ) 
+ 7\cos(-2k_\perp x + k_\perp y +3 k_\parallel z +\omega_0 t ) \right. \right. \nonumber\\
&& \left. \left. -8\cos(-2k_\perp x + k_\perp y +3 k_\parallel z -\omega_0 t )
+3 \cos(-2k_\perp x + k_\perp y +3 k_\parallel z -3\omega_0 t ) \right](-2\xhat+ \yhat) \right.  \nonumber\\
&& \left. 
+ \left[10\cos(k_\perp y + k_\parallel z +\omega_0 t ) 
-20 \cos(k_\perp y + k_\parallel z -\omega_0 t ) 
+ 10 \cos(k_\perp y + k_\parallel z -3 \omega_0 t )\right]\yhat \right\}  \nonumber\\
&  + \frac{ z_+ z_-^2 }{640 v_A^3} 
\frac{k_\perp^2}{k_\parallel^2}  & \left\{ \left[ 
8 \omega_0 t\sin(k_\perp x + 2k_\perp y + k_\parallel z -\omega_0 t ) 
- 3\cos(k_\perp x + 2k_\perp y + k_\parallel z +\omega_0 t ) \right. \right. \nonumber\\
&& \left. \left. +10\cos(k_\perp x + 2k_\perp y + k_\parallel z -\omega_0 t )
-7 \cos(k_\perp x + 2k_\perp y + k_\parallel z -3\omega_0 t ) \right](\xhat+2\yhat)\right.  \nonumber\\
&& \left. 
+ \left[ 3\cos(-k_\perp x + 2k_\perp y +3 k_\parallel z +3\omega_0 t ) 
-8\cos(-k_\perp x + 2k_\perp y +3 k_\parallel z +\omega_0 t ) \right. \right. \nonumber\\
&& \left. \left. +7\cos(-k_\perp x + 2k_\perp y +3 k_\parallel z -\omega_0 t )
-2 \cos(-k_\perp x + 2k_\perp y +3 k_\parallel z -3\omega_0 t ) \right](-\xhat+ 2\yhat) \right.  \nonumber\\
&& \left. 
+ \left[ -10\cos(k_\perp x - k_\parallel z +\omega_0 t ) 
+20 \cos(k_\perp x - k_\parallel z -\omega_0 t ) 
 -10 \cos(k_\perp x - k_\parallel z -3 \omega_0 t )\right]\xhat  \right\}\nonumber
\end{eqnarray}

The $\Order(\epsilon^3)$ nonlinear solution demonstrates several
important physical characteristics. The first and most important
point is that the nonlinear evolution leads to a secular increase of
amplitude for two of the resulting $\Order(\epsilon^3)$ spatial
Fourier modes, $(k_x/k_\perp, k_y/k_\perp,k_z/k_\parallel)=(2,1,-1)$
and $(1,2,1)$, as indicated by the terms with the factor of $t$ in the
coefficient in \eqref{eq:bperp3} and \eqref{eq:eperp3}.  Both of these
secularly increasing modes can be identified as linear
\Alfven waves: (i) the  modes satisfy the
linear dispersion relation for \Alfven waves $\omega= \pm k_z v_A$,
since $\omega=\omega_0$ and $k_z=\mp k_\parallel$ for these two modes;
(ii) the four terms with the secularly increasing coefficients in
\eqref{eq:bperp3} and \eqref{eq:eperp3} satisfy the eigenfunction
relation for linear \Alfven waves given by \eqref{eq:alfeigen}.  The
$(2,1,-1)$ \Alfven wave propagates down the magnetic field (in the
$-\zhat$ direction) at the \Alfven speed $v_A$, whereas the $(1,2,1)$
\Alfven wave propagates up the magnetic field. The amplitude of these 
two \Alfven waves increases linearly with time, and the waves are
phase-shifted by $-\pi/2$ relative to the primary, $\Order(\epsilon)$
\Alfven waves (since this mode appears as a sine function, while all
other modes appear as cosine functions). The secular increase of
energy in these nonlinearly generated Alfven waves is a consequence of
the nonlinear transfer of energy from the primary counterpropagating
\Alfven waves with perpendicular wavenumber $k_\perp$ and parallel wavenumbers 
$\pm k_\parallel$ to \Alfven waves with a larger perpendicular
wavenumber $\sqrt{5} k_\perp$ but the same parallel wavenumbers $\pm
k_\parallel$.  For the symmetric problem at hand, this is the
fundamental nonlinear interaction underlying the turbulent cascade of
energy to higher perpendicular wavenumbers.

Second, the remainder of the $\Order(\epsilon^3)$ modes, which do not
gain energy secularly, consist of a mixture of upward and downward
propagating linear \Alfven waves and inherently nonlinear
electromagnetic fluctuations. Note that, unlike the strictly magnetic
$(1,1,0)$ mode at $\Order(\epsilon^2)$, all of the $\Order(\epsilon^3)$
modes have both electric and magnetic field fluctuations.  All of the
modes appearing at this order fall into a set of six wavevectors.  The
four wavevectors $(2,1,-1)$, $(1,2,1)$, $(-1,2,3)$, and $(-2,1,3)$ are
newly generated modes with a higher perpendicular wavenumber of
magnitude $\sqrt{5} k_\perp$. The remaining two wavevectors are the
same as the primary $\Order(\epsilon)$ waves, $(1,0,-1)$ and
$(0,1,1)$, and these fluctuations at $\Order(\epsilon^3)$ serve to
diminish the amplitudes of the fluctuations in these wavevectors.
Finally, note that the terms in the $\Order(\epsilon^3)$ solution
clearly cancel at $t=0$ to satisfy the zero initial conditions at this
order, $\left.\zeta^{\pm}_3(x,y,z,t)\right|_{t=0}=0$.

\section{Discussion}
\label{sec:discuss}

Here we aim to make a connection between this asymptotic analytical
solution for the nonlinear interaction between counterpropagating
\Alfven waves and the present state of understanding of weak MHD
turbulence in an incompressible MHD plasma.

\subsection{Qualitative Picture of Counterpropagating \Alfven Wave Collisions}
\label{sec:qual}

\begin{figure}
\resizebox{5.0in}{!}{\includegraphics{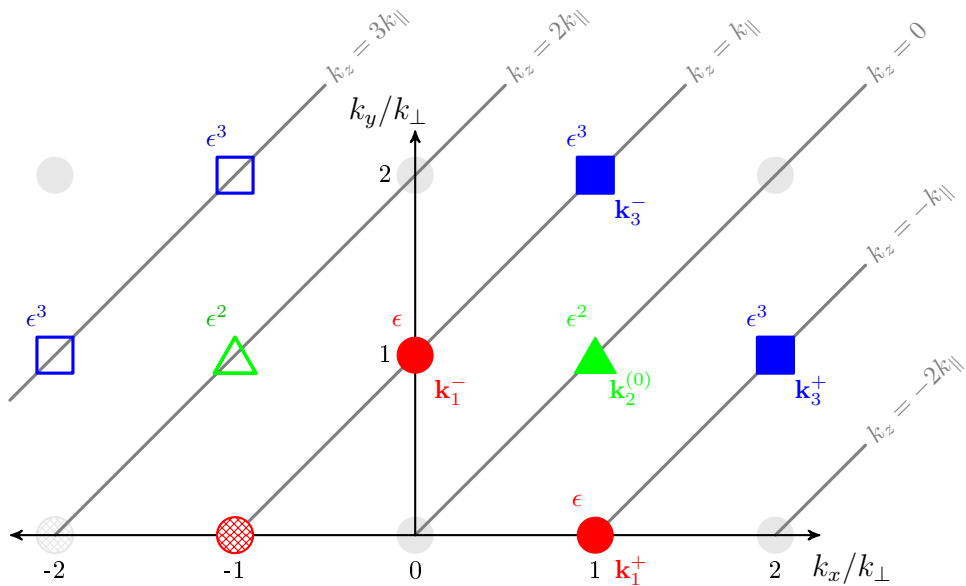}}
\caption{Schematic diagram of the Fourier modes in the
 $(k_x,k_y)$ perpendicular plane arising in the asymptotic
 solution. The Fourier modes depicted are the primary
 $\Order(\epsilon)$ modes (circles), secondary $\Order(\epsilon^2)$
 modes (triangles), and tertiary $\Order(\epsilon^3)$ modes
 (squares). Filled symboles denote the key Fourier modes that play a
 role in the secular transfer of energy to small scales in the
\Alfven wave collision. The parallel wavenumber $k_z$
for each of the modes is indicated by the diagonal grey lines, a
consequence of the resonance conditions for the wavevector.
\label{fig:modes}}
\end{figure}

To establish an intuitive foundation upon which to develop a refined
understanding of the energy flow to small scales in plasma turbulence,
we present here the qualitative picture of the nonlinear evolution of
the interaction between counterpropagating \Alfven waves,
corresponding to the asymptotic analytical solution derived in
\secref{sec:solution}. We focus first on the Fourier modes that
play a key role in the secular transfer of energy to smaller scales,
or equivalently to larger wavenumber. These key modes are represented
by the filled symbols in \figref{fig:modes}, a plot of wavevector
space $(k_x,k_y)$ in the plane perpendicular to the equilibrium
magnetic field.

In \figref{fig:modes}, the primary modes (filled circles) at
$\Order(\epsilon)$ correspond to the initial conditions consisting of
two perpendicularly polarized, counterpropagating \Alfven waves with
wavevectors $\V{k}_1^+ = (k_x/k_\perp, k_y/k_\perp,
k_z/k_\parallel)=(1,0,-1)$ and $\V{k}_1^- =(0,1,1)$, as depicted in
\figref{fig:setup}. The frequency of both of these initial linear
\Alfven wave modes is $\omega_0=k_\parallel v_A$.  These two primary
\Alfven waves interact nonlinearly, due to the terms on the right-hand
side of \eqref{eq:elsasserpm}, generating a secondary mode (filled
triangle) at $\Order(\epsilon^2)$ that is a strictly magnetic
fluctuation with wavevector $\V{k}_2^{(0)}=(1,1,0)$. This secondary
mode has no variation along the equilibrium magnetic field, $k_z=0$,
but it oscillates with frequency $2 \omega_0$; since this mode does
not satisfy either the linear \Alfven wave dispersion relation or the
\Alfvenic eigenfunction
\eqref{eq:alfeigen}, it is an inherently nonlinear fluctuation. 

When this secondary mode $\V{k}_2^{(0)}$ has nonzero amplitude, the
primary modes $\V{k}_1^+$ and $\V{k}_1^-$ nonlinearly interact with it
to transfer energy to the tertiary modes (filled squares) at
$\Order(\epsilon^3)$, with wavevectors $\V{k}_3^+ = (2,1,-1)$ and
$\V{k}_3^- =(1,2,1)$, respectively. As demonstrated in
\secref{sec:O3}, each of these tertiary modes is a linear \Alfven wave
with frequency $\omega_0$ that propagates in the same direction along
the magnetic field as the primary mode that generated it, consistent
with the constraint that the upward and downward waves conserve energy
separately.  The amplitudes of these tertiary \Alfven waves increase
linearly with time.  Thus, this sequence of interactions---the
interaction $\V{k}_1^+ +\V{k}_1^-=\V{k}_2^{(0)}$ followed by
$
\V{k}_1^\pm + \V{k}_2^{(0)}=\V{k}_3^\pm $---leads to the secular transfer of
energy from the primary waves at low wavenumber with $|\V{k}_1^\pm |
=\sqrt{k_\perp^2 + k_\parallel^2}$ to the tertiary waves at higher
wavenumber with $|\V{k}_3^\pm | =\sqrt{5k_\perp^2 + k_\parallel^2}$.
Therefore, the asymptotic solution presented in this paper provides a
detailed picture of the fundamental building block of plasma
turbulence---the nonlinear interaction between perpendicularly
polarized, counterpropagating \Alfven waves---responsible for the
turbulent cascade of energy from large to small scales.  This forward
cascade of energy to small scales is often the most significant impact
that turbulence has on astrophysical environments.  In agreement with
the heuristic model of weak
turbulence,\citep{Sridhar:1994,Montgomery:1995,Goldreich:1997,Galtier:2000,Lithwick:2003}
the nonlinear energy transfer is strictly perpendicular to higher
$k_\perp$, with no parallel cascade of energy to higher $k_\parallel$.

The vital component of this physical picture of the nonlinear energy
transfer in \Alfvenic turbulence is the generation of the secondary
$\V{k}_2^{(0)}$ mode, a purely magnetic fluctuation that is
self-consistently generated by the nonlinear interaction between
counterpropagating \Alfven waves. A physical interpretation of this
mode is illuminating. This $\V{k}_2^{(0)}=(1,1,0)$ magnetic mode
corresponds to a shear in the direction of the total magnetic field
(including the equilibrium field $\V{B}_0=B_0 \zhat$ and the
fluctuation due to the $\V{k}_2^{(0)}$ mode) across perpendicular
plane. Because it arises as a consequence of the interaction between
the primary counterpropagating \Alfven waves, the sense of this shear
reverses with the frequency of the mode, $2\omega_0$. The propagation of
the primary \Alfven waves along this sheared magnetic field is the
essential mechanism by which these waves are distorted
nonlinearly. Mathematically, this nonlinear distortion is represented
by the transfer of energy to the tertiary \Alfven waves
$\V{k}_3^\pm$. The identification of the secondary mode as a magnetic
shear establishes a crucial connection to the discussion of the
relative importance of three-wave vs.~four-wave interactions in weak
incompressible MHD turbulence, as discussed below in
\secref{sec:34wave}.

Note that although this physical picture of the nonlinear energy
transfer in \Alfvenic turbulence is formally valid only in the limit
of weak turbulence, many of the general qualitative properties of this
process will persist even in the more important limit of strong
turbulence. Future study will explore in detail the applicability of
this physical mechanism as the turbulence becomes strong.

Above we have focused on the five modes, represented by the filled
symbols in \figref{fig:modes}, that mediate the fundamental transfer
of energy to small scales in plasma turbulence. Other Fourier modes
arise in the asymptotic solution outlined in \secref{sec:solution},
depicted by the open symbols in \figref{fig:modes}. The secondary mode
(open triangle) at $(k_x/k_\perp, k_y/k_\perp,k_z/k_\parallel)=(-1,1,2)$,
as explained in \secref{sec:O2}, consists of two counterpropagating 
\Alfven waves with parallel wavenumber $k_z=2 k_\parallel$ and frequency
$\omega=\pm 2\omega_0$, but these waves do not secularly gain energy.
The tertiary modes (open squares) at $(-2,1,3)$ and $(-1,2,3)$ consist
of a mixture of upward and downward propagating linear \Alfven waves
and inherently nonlinear electromagnetic fluctuations.  Because these
modes do not gain energy secularly, their role in the nonlinear
transfer of energy is negligible compared to the tertiary modes at
$\V{k}_3^+ = (2,1,-1)$ and $\V{k}_3^- =(1,2,1)$ (filled squares).
Finally, note that we show the primary mode at $(-1,0,1)$
(cross-hatched circle) in \figref{fig:modes} because a complex
representation of the perpendicular Fourier modes requires only the
upper half-plane $k_y
\ge 0$ when the reality condition $B(k_x,k_y)=B^*(-k_x,-k_y)$ is
imposed. This reality condition implies that modes with $k_x <0 $ and
$k_y=0$ are the complex conjugates of the modes with $k_x >0 $ and
$k_y=0$, so this complex coefficient of the $(-1,0,1)$ mode is nonzero.
In summary, although there is some nonlinear energy transfer to the
Fourier modes with $k_x<0$ (open symbols), these modes generally
receive a negligible amount of energy compared to the secular transfer
of energy to the tertiary modes $\V{k}_3^+ = (2,1,-1)$ and $\V{k}_3^-
=(1,2,1)$.

The diagonal gray lines in \figref{fig:modes} illustrate another
important property of the nonlinear evolution of the interaction
between counterpropagating Alfven waves that was solved in
\secref{sec:solution}. In general, the three-wave 
nonlinear interaction between two modes $\V{k}_a$ and $\V{k}_b$, that
results in the transfer of energy to a third mode $\V{k}_c$, must
satisfy the condition $\V{k}_c = \V{k}_a + \V{k}_b$.  This 
condition is equivalent to the conservation of momentum. For the
particular initial conditions specified in \secref{sec:ics}, this
requirement determines the structure of the parallel wavenumber $k_z$
across the $(k_x,k_y)$ plane, as shown in \figref{fig:modes}.  The
result is that lines of constant $k_z$ follow along diagonals of unity
slope.  The lack of a parallel cascade of energy to higher $k_z$ means
that the nonlinear transfer of energy to higher values of
perpendicular wavenumber $\sqrt{k_x^2+k_y^2}$ follows primarily along
the lines of constant $k_z$ given by $k_z=k_\parallel$, $k_z=0$, and
$k_z=-k_\parallel$. This property will be demonstrated numerically in
a future extension of this work.

\subsection{Three-wave vs.~Four-wave Interactions in Weak MHD Turbulence}
\label{sec:34wave}
In the development of the theory for weak incompressible MHD
turbulence, significant controversy arose regarding the importance of
three-wave interactions in the
turbulence,\cite{Sridhar:1994,Montgomery:1995,Ng:1996,
Goldreich:1997,Galtier:2000,Lithwick:2003} as outlined above in the
introduction.  Here we place the explicit solution derived in
\secref{sec:solution} into the context of this discussion of 
three-wave and four-wave interactions.

Three-wave interactions must satisfy the resonance
conditions\cite{Shebalin:1983,Sridhar:1994,Montgomery:1995,Ng:1996,Galtier:2000} on
wavevector and  frequency 
\begin{equation}
\V{k}_a + \V{k}_b = \V{k}_c, \ \ \ \omega_a + \omega_b = \omega_c.
\label{eq:3wave_res}
\end{equation}
Here we adopt the shorthand $\omega_i=\omega(\V{k}_i)$, where the
linear dispersion relation determines the frequency $\omega_i$ of a
plane wave with wavevector $\V{k}_i$. For \Alfven waves in a plasma
with an equilibrium magnetic field $\V{B}_0 = B_0 \zhat$, the linear
dispersion relation is $\omega = |k_\parallel| v_A$, where we have
maintained here the convention from \secref{sec:nlprop} that the wave
frequency $\omega \ge 0$, so that the sign of $k_{\parallel}$
determines the direction of propagation of the wave. As discussed in
\secref{sec:nlprop}, since only counterpropagating
\Alfven waves interact
nonlinearly,\cite{Iroshnikov:1963,Kraichnan:1965,Shebalin:1983,Sridhar:1994}
 $k_{\parallel a}$ and
$k_{\parallel b}$ must have opposite signs, so we choose $k_{\parallel
a} \ge 0$ and $k_{\parallel b} \le0$ without loss of generality.  In
this case, taking the $\zhat$ component of the wavevector condition 
and substituting the linear dispersion relation into the frequency
condition in \eqref{eq:3wave_res}, we obtain
\begin{equation}
{k}_{\parallel a} + {k}_{\parallel b} = {k}_{\parallel c}, \ \ \ 
{k}_{\parallel a} - {k}_{\parallel b} = {k}_{\parallel c}
\label{eq:3wave_simp}
\end{equation}
These equations have a solution only if ${k}_{\parallel b}=0$.  This
property suggests that three-wave interactions require one of the
interacting waves to have $k_\parallel=0$ to yield resonant nonlinear energy
transfer.\cite{Shebalin:1983,Sridhar:1994,Montgomery:1995,Ng:1996,Galtier:2000}
In addition, since $k_{\parallel a} = k_{\parallel c}$, there is no parallel
cascade of energy.

It is important to note here that the three-wave resonance condition
for frequency \eqref{eq:3wave_res} arises when the solution is
obtained using a Fourier transform from the time to the frequency
domain. When considering the solution of the nonlinear interaction on
a periodic spatial domain over a finite time interval, the necessary
periodicity in time requires that the time interval is an integral
number of wave periods. Alternatively, this frequency condition also
applies in the investigation of the interaction of localized
wave packets over an infinite spatial domain over a sufficiently long
time interval for the wave packets to cease interacting.\cite{Ng:1996}
\emph{For intermediate times that do not satisfy one of these criteria, 
one may obtain a solution that demonstrates nonresonant energy
transfer that need not satisfy the resonant condition on  frequency in
\eqref{eq:3wave_res}.}

Analogous to three-wave interactions, the four-wave interactions must
satisfy the resonance conditions\cite{Sridhar:1994,Ng:1996} on
wavevector and frequency
\begin{equation}
\V{k}_a + \V{k}_b = \V{k}_c+ \V{k}_d, \ \ \ \omega_a + \omega_b = \omega_c + \omega_d.
\label{eq:4wave_res}
\end{equation}
In this case, a similar procedure can be used, assuming $k_{\parallel
a} > 0$ and $k_{\parallel b} < 0$, to obtain simplified conditions\cite{Sridhar:1994}
\begin{equation}
{k}_{\parallel a} + {k}_{\parallel b} = {k}_{\parallel c} + {k}_{\parallel d}  , \ \ \ 
{k}_{\parallel a} - {k}_{\parallel b} = {k}_{\parallel c} - {k}_{\parallel d}
\label{eq:4wave_simp}
\end{equation}
In this case, the solution requires $k_{\parallel a} = k_{\parallel
c}$ and $k_{\parallel b} = k_{\parallel d}$, and once again there is
no parallel cascade of energy for the four-wave interaction.\cite{Sridhar:1994}

Let us now discuss the explicit solution derived in \secref{sec:solution} in
terms of three- and four-wave interactions. First, we note that the
initial conditions of the problem contain no $k_\parallel=0$
component, so the expectation arising from a consideration of the
resonance conditions \eqref{eq:3wave_res} is that there will be no
energy transfer from three-wave interactions. The secondary solution
at $\Order(\epsilon^2)$ arises from the three-wave interaction
$\V{k}_1^+ +\V{k}_1^-=\V{k}_2^{(0)}$, where the amplitude of the
$\V{k}_2^{(0)}$ mode scales $\propto z_+z_-$, as expected for
three-wave interactions.\cite{Sridhar:1994,Ng:1996} From inspection of
the first line of \eqref{eq:O2B}, it is clear that there is indeed a
nonlinear transfer of energy to mode $\V{k}_2^{(0)}=(1,1,0)$. There
is, however, no secular change in the amplitude; in fact, at times
$t_n= n \pi /\omega_0$ for $n=0,1,2, \ldots$, the net transfer of energy
to this mode is zero. The nonlinear transfer of energy to this mode is
\emph{nonresonant},\cite{Goldreich:1997} yielding no net transfer of
energy over an integral number of periods.  This result agrees with
the analytical findings of Ng and Bhattacharjee,\cite{Ng:1996} that
the three-wave resonant interactions vanish if there is no
$k_\parallel=0$ component to the interacting \Alfven waves.

The tertiary solution at $\Order(\epsilon^3)$ arises from the
interaction $\V{k}_1^\pm + \V{k}_2^{(0)}=\V{k}_3^\pm $. Since
$\V{k}_2^{(0)}=\V{k}_1^+ +\V{k}_1^-$, it is apparent that this
tertiary solution arises through a four-wave interaction, although the
detailed mechanism mediating the energy transfer to the modes
$\V{k}_3^+ = (2,1,-1)$ and $\V{k}_3^- =(1,2,1)$ looks somewhat
different from the four-wave interactions discussed in the
literature\cite{Sridhar:1994} and summarized above by the resonance conditions
\eqref{eq:4wave_res}. Indeed, the interaction $\V{k}_1^\pm + \V{k}_2^{(0)}=\V{k}_3^\pm $
satisfies the requirements for resonant three-wave interactions
required by the resonance conditions \eqref{eq:3wave_res}, namely that
the interacting mode $\V{k}_2^{(0)}$ has $k_\parallel =0$. But a
crucial insight arising from the solution derived here is that this
$k_\parallel =0$ mode arises self-consistently from the interaction
between the primary counterpropagating \Alfven waves $\V{k}_1^+$ and
$\V{k}_1^-$.  An important property of this self-consistently
generated $\V{k}_2^{(0)}$ mode is that the energy in the mode rises
and falls with frequency $2 \omega_0$. As mentioned above in
\secref{sec:qual}, this mode can be interpreted physically as
an oscillating shear in the magnetic field direction. The primary
modes $\V{k}_1^+$ and $\V{k}_1^-$ propagate along this sheared field
and become distorted, thereby transferring energy to the tertiary modes
$\V{k}_3^+$ and $\V{k}_3^-$.  The net result of this detailed process
is described by the four-wave interactions $2\V{k}_1^+ +\V{k}_1^-
=\V{k}_3^+$ and $\V{k}_1^+ +2
\V{k}_1^- =\V{k}_3^-$. The amplitude of the $\V{k}_3^+$ and $\V{k}_3^-$
tertiary modes increases linearly with time, resulting in a secular
transfer of energy from $\V{k}_1^+$ to $\V{k}_3^+$ and from
$\V{k}_1^-$ to $\V{k}_3^-$ that is due to \emph{resonant} four-wave
interactions. In addition, according to \eqref{eq:bperp3}
and \eqref{eq:eperp3}, the $\V{k}_3^+$ \Alfven wave has an amplitude that scales
$\propto z_+^2z_-$ and the $\V{k}_3^-$ \Alfven wave has an amplitude that scales
$\propto z_+z_-^2$, as expected for four-wave
interactions.\cite{Sridhar:1994,Ng:1996}

Note that, although the nonlinear energy transfer in our solution is
dominated by resonant four-wave interactions, the role played by the
inherently nonlinear $\V{k}_2^{(0)}$ magnetic mode establishes a
direct connection to the reasoning used to argue for the dominance of
three-wave interactions in weak incompressible MHD
turbulence.\cite{Montgomery:1995,Ng:1996,
Goldreich:1997,Galtier:2000,Lithwick:2003}.  Qualitatively, the
three-wave interactions are nonzero if the magnetic field lines wander
away from each other in a weakly turbulent plasma.  Such field line
wander requires a shear in the direction of the magnetic field across
the plane perpendicular to the field, which is mathematically
represented by a Fourier component of the background magnetic field
with  $k_\perp \ne 0 $ but $k_\parallel=0$. The
$\V{k}_2^{(0)}=(1,1,0)$ mode arising in our solution has precisely
this property, and thus mediates nonlinear energy transfer in the same
way that magnetic field line wander does. The difference between these
two pictures is that the $\V{k}_2^{(0)}$ mode arises at
$\Order(\epsilon^2)$, whereas magnetic field line wander would
correspond to a $k_\parallel=0$ Fourier component at
$\Order(\epsilon)$; therefore, the $\V{k}_2^{(0)}$ mode transfers
energy through resonant four-wave interactions, whereas magnetic field
line wander transfers energy through resonant three-wave interactions.

It is worthwhile here to compare our explicit asymptotic solution to
the closely related work by Ng and Bhattacharjee,\cite{Ng:1996} which
presents an analytical and numerical treatment of the nonlinear
interaction between two counterpropagating \Alfven wave packets of
finite extent. Using perturbation theory, they calculate analytically
in closed form the three- and four-wave interactions between colliding
\Alfven wave packets of arbitrary form to determine which of these
mechanisms dominates in the weak turbulence limit.  They proved that,
if the wave packets have a nonzero $k_\parallel=0$ component, then
three-wave interactions dominate over the four-wave interactions. The
nature of our explicit analytical solution is completely consistent
with the properties of the \Alfven wave interactions that are
elucidated in their work, although there is one subtle issue worthy of
comment.  By assuming that the functions describing the primary
interacting wave packets are separable, $z^\pm(x,y,z,t) = f_\perp^\pm
(x,y) f_\parallel^\pm(z,t)$, they obtain the surprising result that
the four-wave interactions also vanish if the initial waves do not
contain a $k_\parallel=0$ component. This finding appears to be
inconsistent with the secular energy transfer mediated by resonant
four-wave interactions in the solution derived in \secref{sec:solution}. The plane wave
form of the primary \Alfven waves given by
\eqref{eq:zinitplus} and \eqref{eq:zinitminus}, however, is not 
separable in the manner assumed by Ng and Bhattacharjee, and therefore
our findings remain consistent. Since the plane wave decomposition
used here is a valuable basis that can be used to describe the
interaction of an arbitrary \Alfvenic fluctuation, we believe that the
development of a detailed understanding of the dynamics of the
nonlinear interaction between \Alfven plane waves provides an
important foundation for developing a detailed intuitive picture of
this fundamental building block of astrophysical plasma turbulence.

\subsection{Implications for Turbulence in Astrophysical Plasmas}
\label{sec:implications}

In this paper, we explore in detail the process of nonlinear energy
transfer in astrophysical plasma turbulence due to the $\V{E}\times
\V{B}$ nonlinearity.  As long as there is a nonzero \Alfven wave energy flux
traveling in both directions along the magnetic field, this mechanism
is very effective at transferring energy nonlinearly to higher
$k_\perp$, and, in the anisotropic limit $k_\perp \gg k_\parallel$
that naturally develops in plasma turbulence, this mechanism does not
even require a large fluctuation amplitude for the nonlinearity to be
strong, as explained in the introduction. Unless particularly special
circumstances exist to eliminate the $\V{E}\times \V{B}$
nonlinearity---such as all wave energy traveling in one direction
along the magnetic field, all
\Alfvenic fluctuations polarized in the same plane, or all
fluctuations with $k_\perp=0$---we expect that the $\V{E}\times \V{B}$
nonlinearity will dominate over other mechanisms for nonlinear energy
transfer, such as parametric instabilities or nonlinear Landau
damping.

The detailed nature of the asymptotic analytical solution in the weak
turbulence limit lends valuable insight into the properties of plasma
turbulence.  The eigenfunction of linear \Alfven waves (see
\secref{sec:linear}) implies that if plasma turbulence is dominated by 
\Alfvenic fluctuations, one should expect that the turbulent fluctuations 
have equal kinetic and magnetic energies, where this equipartition of
energies applies scale by scale.  Thus, the turbulent kinetic and
magnetic energies should have the same wavenumber spectra. This has
been quantified in the literature by a study of the \Alfven ratio
$R_A(k) \equiv E_V(k)/E_B(k)$ or the residual energy $E_R (k)\equiv
E_V(k)-E_B(k)$, where a normalized version of the residual energy is
defined by $\sigma_R \equiv (E_V-E_B)/(E_V+E_B) = (R_A -1
)/(R_A+1)$. Equipartition therefore corresponds to $R_A=1$ or
$\sigma_R=0$.  But, within the inertial range of solar wind
turbulence, corresponding to the approximate spacecraft-frame
frequency range from $10^{-4}$~Hz to $0.4$~Hz (where the Taylor
hypothesis\cite{Taylor:1938} relates the frequency spectrum to the
wavenumber spectrum),
observations\cite{Matthaeus:1982b,Roberts:1987a,Bruno:1985,Goldstein:1995,Tu:1995,Bavassano:1998,Bavassano:2000a,Bruno:2005,Podesta:2007,Salem:2009,Boldyrev:2011}
show that the magnetic energy exceeds the kinetic energy, especially
at large scales, leading to an \Alfven ratio $R_A<1$ or a negative
residual energy $\sigma_R<0$. This excess of magnetic energy has also
been observed in MHD simulations of weak\cite{Boldyrev:2009b} and
strong\cite{Muller:2005,Boldyrev:2011} turbulence. Based on the
results of MHD turbulence simulations, Boldyrev and
Perez\cite{Boldyrev:2009b} suggested that the negative residual energy
was a consequence of the breakdown of mirror symmetry in imbalanced
turbulence (turbulence with a greater \Alfven wave energy flux in one
direction along the magnetic field than along the opposite
direction). Subsequently, Wang \emph{et al.}\cite{Wang:2011}
demonstrated analytically that the development of a negative residual
energy arises naturally through the nonlinear interaction of \Alfven
waves, leading to a ``condensate'' of energy in $k_\parallel \approx
0$ modes, and further suggested that this condensate plays an
essential role in the turbulent dynamics. These findings are supported
by the analytical solution presented in \secref{sec:solution}.  In
fact, the qualitative picture of the nonlinear energy transfer due to
the interaction between counterpropagating \Alfven waves, described in
\secref{sec:qual}, illuminates the physical mechanisms that generate 
the negative residual energy.  In particular, the inherently nonlinear
$\V{k}_2^{(0)}$ mode in our solution is a purely magnetic mode,
thereby contributing to the magnetic energy but not to the kinetic
energy, leading to $R_A<1$ or $\sigma_R<0$, consistent with solar wind
observations and MHD simulations. Although our solution rigorously
holds only in the weak turbulence limit, it is likely that many of the
qualitative properties of our solution persist into the strongly
nonlinear regime.  Future work will explore the applicability on this
picture of the nonlinear evolution in the strongly nonlinear limit.

The analytical solution in \secref{sec:solution} applies for the
interaction between counterpropagating plane \Alfven waves that remain
correlated over all time.  The nonlinear energy transfer from the
primary $\V{k}_1^\pm$ linear \Alfven waves to the tertiary
$\V{k}_3^\pm$ linear \Alfven waves results in tertiary wave amplitudes
that increase linearly with time, as demonstrated in \eqref{eq:bperp3}
and \eqref{eq:eperp3}, or equivalently, an energy of the tertiary mode
that increases as $E \propto t^2$.  At first, this may seem
inconsistent with the expectation from turbulence theories that the
energy increase due to nonlinear energy transfer from wavenumber $k$
to wavenumber $2k$ scales as $E \propto t$.  The difference arises
because the energy transfer remains coherent in the idealized model of
two interacting plane \Alfven waves solved here, whereas in a
turbulent system, the typical model involves a localized \Alfven
wave packet interacting with many uncorrelated
\Alfven wave packets that are propagating in the opposite
direction. The connection between these models can be described as
follows. The fundamental building block of astrophysical plasma
turbulence is the interaction between two counterpropagating \Alfven
wave packets. The nature of each of these individual interactions is
informed by the analytical solution provided here, where the energy
transferred within a single collision between wave packets scales as $E
\propto t^2$.  But each of these collisions effectively provides a
step in the energy of a wave packet, $\Delta E$. After interaction with
many uncorrelated counterpropagating wave packets, yielding a random
walk in energy with step size $\Delta E$, the statistical result is a
transferred energy that scales as $E
\propto t$, in agreement with theoretical expectations. It is therefore necessary 
to account for the incoherent nature of successive wave packet
collisions when exploiting the intuition gained from the solution
derived here in any attempt to refine the theory of astrophysical plasma
turbulence. In particular, this intuition may be useful in the attempt
to incorporate the role of the $k_\parallel=0$ modes into a refined
theory.\cite{Wang:2011}

An important property of plasma turbulence that is highlighted by the
solution obtained here is the inherently three-dimensional nature of
turbulence in a magnetized
plasma\citep{Howes:2011a,Tronko:2012}, as discussed in
\secref{sec:otherprop}. Motivated by the power arising 
at $k_\parallel \approx 0$ in MHD turbulence
simulations,\cite{Dmitruk:2009,Boldyrev:2009} to achieve higher
spatial resolution, many recent studies of plasma turbulence have
employed two-dimensional simulations in the plane perpendicular to the
mean magnetic
field.\cite{Parashar:2009,Servidio:2009,Parashar:2010,Servidio:2010,Servidio:2011,Markovskii:2011,Parashar:2011,Servidio:2012,Vasquez:2012} However,
these 2D simulations cannot describe the dynamics of \Alfven waves,
which are generally considered to be fundamental to the
turbulence. The solution here demonstrates two important points about
the nature of incompressible MHD turbulence: (i) the interaction
between counterpropagating \Alfven waves naturally generates modes
with $k_\parallel =0$, and (ii) these $k_\parallel =0$ modes play an
essential role in mediating the energy transfer from the initial
interacting \Alfven waves to generate \Alfven waves with larger
perpendicular wavenumber. We consider that this mechanism, described
in \secref{sec:qual}, is the fundamental building block of plasma
turbulence, and it cannot be described without a full three-dimensional
treatment.  Therefore, although 2D turbulence simulations are
possible, their relevance to turbulence in astrophysical plasmas
remains to be established.

The nonlinear solution obtained here also exposes an important
misconception that has arisen relating to the importance of linear
wave modes in plasma turbulence. A number of papers have suggested
that, if linear wave modes play an important role in plasma
turbulence, then one should be able to ``see'' the signature of the
wave's linear dispersion relation in a plot of turbulent power on the
$\omega$-$k$ plane.\cite{Dmitruk:2009,Svidzinski:2009,Parashar:2011,Narita:2011,Hunana:2011}
This is not correct.  We have clearly demonstrated in \secref{sec:O3} that
the secularly increasing tertiary modes in this solution are linear
\Alfven waves. However, because the amplitudes of these linear waves
increase as a function of time, the Fourier transform of a finite time
interval to the frequency domain will not generate a clear signal at
the linear wave frequency; instead, the power will be spread over a
range of frequencies.  In addition, if linear \Alfven wave power is
confined spatially to localized wave packets, the frequency spectrum
resulting from a probe sampling the wave packet as it passes by will be
similarly broadened. For turbulence consisting of many nonlinearly
interacting linear wave modes with a spectrum of different
wavevectors, each of which has a broadened frequency content, the
combination of these modes is not likely to display any distinct
features along the $\omega$--$k$ diagram, even if it is composed of
nothing but linear eigenmodes. For this reason, the frequency is not a
good property to use for identifying linear wave modes.  The
eigenfunction---which, for a particular linear wave mode, defines the
amplitude and phase relations between the fluctuations in the density,
pressure, and the components of the magnetic field and the velocity
field---is a much more valuable property for evaluating the importance
linear wave properties in plasma turbulence. Approaches that exploit
the characteristics of the linear eigenfunctions have yielded valuable
results that constrain the nature of the turbulent fluctuations in the
solar wind.\cite{Salem:2012,Howes:2012a} In fact, a novel method for
the analysis and interpretation of spacecraft measurements of
turbulence utilizes the properties of the linear eigenfunctions to
construct synthetic spacecraft data.\cite{Klein:2012}

Although the problem solved here---the interaction of overlapping plane
\Alfven waves on a periodic domain---is significantly idealized, it
provides an important intuitive foundation upon which to refine our
understanding of plasma turbulence.  Specifically, this problem models
interactions that are local in wavenumber space, since the initial
upward and downward
\Alfven waves have $k_\perp^+= k_\perp^-$ and $k_\parallel^+=
-k_\parallel^-$. In addition, the plane waves are initially
overlapping in space before the nonlinear interaction is allowed to
begin.  Nevertheless, the asymptotic analytical solution derived here
provides an intuitive picture of the nature of interactions between
counterpropagating \Alfven waves, which consitute the fundamental
building blocks of astrophysical plasma turbulence.  Future studies
will relax the simplifications used in this paper, allowing for the
investigation of nonlocal interactions ($k_\perp^+\ne k_\perp^-$
and/or $k_\parallel^+\ne -k_\parallel^-$), interactions between wave
packets that do not initially overlap (similar to a previous
investigation\cite{Ng:1996}), and the modification of the turbulence
as the \Alfvenic fluctuations reach the small scales, $k_\perp \rho_i
\gtrsim 1$, where the wave modes become the dispersive kinetic and
inertial \Alfven waves.

\section{Conclusions}

Turbulence significantly impacts the evolution of a wide range of
astrophysical plasma environments, from galaxy clusters and accretion
disks to interstellar medium of the Galaxy to the solar corona and
solar wind.  The primary effect of this plasma turbulence is to
mediate a cascade of energy from the large scales where the turbulent
motions are driven down to sufficiently small scales where dissipative
mechanisms bring about the ultimate conversion of the turbulent energy
to plasma heat. Nonlinearity in the equations governing the plasma
dynamics is the underlying physical mechanism responsible for this
turbulent cascade of energy to small scales. Although a number of
different nonlinear mechanisms are possible in a magnetized plasma,
the dominant nonlinearity in plasma turbulence is the $\V{E}\times
\V{B}$ nonlinearity, responsible for the nonlinear interaction between
perpendicularly polarized, counterpropagating \Alfven waves.  This
nonlinear interaction, commonly called an \Alfven wave ``collision,''
constitutes the \emph{fundamental building block of astrophysical
plasma turbulence}.

In this paper, we present an asymptotic analytical solution for the
nonlinear evolution of the interaction between counterpropagating
\Alfven waves in an incompressible MHD plasma in limit of weak
nonlinearity. This solution provides a firm intuitive foundation upon
which to develop a more detailed understanding of the physical
mechanisms driving the cascade of energy in plasma turbulence.
Rigorously, this calculation requires sufficient anisotropy $k_\perp
\gg k_\parallel$ so that the pseudo-\Alfven waves have no effect on 
the \Alfvenic turbulent dynamics. A similar simplification occurs 
for compressible MHD plasmas in the anisotropic limit, $k_\perp
\gg k_\parallel$, since it has been shown that the compressible 
fluctuations associated with the slow waves and entropy modes decouple
from the reduced MHD equations that govern the \Alfven wave
cascade.\cite{Schekochihin:2009} Note that the fast magnetosonic wave
is ordered out of the dynamics by assuming either incompressibility or
anisotropy.

The primary result of this paper is a qualitative picture of the
nonlinear interactions driving the turbulent cascade of energy to
small scales, involving the key Fourier modes depicted by the filled
symbols in \figref{fig:modes}. It describes the interaction between
two perpendicularly polarized, counterpropagating plane \Alfven waves,
with frequency $\omega_0$ and wavevectors $\V{k}_1^+= k_\perp \xhat -
k_\parallel \zhat$ and $\V{k}_1^-= k_\perp
\yhat + k_\parallel \zhat$, schematically shown in \figref{fig:setup}. 
First, these primary \Alfven waves interact nonlinearly to generate a
secondary mode with wavevector $\V{k}_2^{(0)}= k_\perp \xhat +k_\perp
\yhat $.  This mode satisfies neither the frequency nor the
eigenfunction conditions for a linear \Alfven wave, and therefore it
is an inherently nonlinear fluctuation. It is purely magnetic and has
no parallel variation, and therefore corresponds to a shear in the
magnetic field direction that reverses with a frequency $2
\omega_0$. The energy transfer to this secondary mode is due to a 
nonresonant three-wave interaction. Next, the two primary \Alfven
waves $\V{k}_1^\pm$ each interact with this secondary mode
$\V{k}_2^{(0)}$ to transfer energy to two tertiary \Alfven waves with
frequency $\omega_0$ and wavevectors $\V{k}_3^+ = 2 k_\perp \xhat
+k_\perp \yhat - k_\parallel \zhat$ and $\V{k}_3^- = k_\perp \xhat
+2k_\perp \yhat + k_\parallel \zhat$. This resonant four-wave
interaction leads to a secular increase in the amplitude of these
tertiary modes, which satisfy both the frequency and eigenfunction
conditions for linear \Alfven waves. The tertiary \Alfven waves have the
same parallel wavevectors as the primary waves, so there is no parallel
cascade. In addition, the upward primary wave transfers energy to the
upward tertiary wave, while the downward primary wave transfers energy
to the downward tertiary wave, so it is clear that these interactions
obey the constraint that the upward and downward wave energy fluxes
are conserved by the nonlinear equations. In summary, it is this
process that forms the fundamental mechanism supporting the nonlinear
cascade of energy in a turbulent, magnetized plasma. 

The detailed properties of this interaction between counterpropagating
\Alfven waves in the limit of weak nonlinearity provide crucial insight into
the nature of both weak and strong plasma turbulence. First, this
solution provides a detailed analytical form against which the results
of nonlinear simulations can be compared.  Second, this solution
illustrates clearly how $k_\parallel =0$ modes play an important role
in both three-wave and four-wave resonant interactions, and informs
the debate about the which of these mechanisms is dominant in weak MHD
turbulence. An extension of this work will consider the interaction
between two counterpropagating plane \Alfven waves with arbitrary
wavevectors.  Third, the solution immediately demonstrates the
development of purely magnetic fluctuations as a result of the
collision between \Alfven waves, and this is the likely explanation
for the observations of a negative residual energy in solar wind
turbulence.\cite{Wang:2011} Fourth, the energy of the tertiary mode in
this solution increases with the square of time, $E \propto t^2$, due
to the coherent nature of the interaction.  For turbulence consisting
of collisions with many uncorrelated \Alfven wave packets, the
statistically averaged energy increase is expected to scale as $E
\propto t$. Fifth, this solution highlights the inherently
three-dimensional nature of turbulence in a magnetized plasma, clearly
demonstrating that the interaction between counterpropagating \Alfven
waves naturally generates modes with $k_\parallel =0$, and that these
modes play an essential role in mediating the nonlinear energy
transfer. Finally, the nonlinearly generated modes in this weak
turbulence problem are linear \Alfven waves, but the secular increase
in the mode amplitudes will obscure a clear linear dispersion relation
signature in an $\omega$--$k$ diagram.

Ultimately, this asymptotic analytical solution was derived to provide
the necessary theoretical basis for the design of an experimental
program to measure, for the first time, the nonlinear interaction
between counterpropagating \Alfven waves.  The successful completion
of this experimental effort has confirmed this fundamental nonlinear
interaction in a laboratory plasma, even under the weakly collisional
conditions relevant to many turbulent space and astrophysical plasma
systems, and has demonstrated that theoretical models developed under
simplified plasma descriptions, such as incompressible MHD, remain
applicable under more general plasma conditions.\cite{Howes:2012b} As
a complement to the analytical derivation presented here, quantitative
verification of this analytical solution using gyrokinetic numerical
simulations is presented in Paper II.\cite{Nielson:2013a} A detailed
description of the theoretical considerations involved in the
experimental design and supporting numerical simulations is contained
in Paper III,\cite{Howes:2013b} and the specifics of the experimental
procedure and data analysis appear in Paper IV.\cite{Drake:2013}
Future extensions of this work will explore the transition to the
limit of strong turbulence and the nature of the nonlinear energy
transfer in the small-scale, dispersive regime of kinetic and inertial
\Alfven waves.

\begin{acknowledgments}
This work was supported by NSF PHY-10033446, NSF CAREER AGS-1054061,
and NASA NNX10AC91G. 
\end{acknowledgments}

%



\begin{thebibliography}{10}%
\makeatletter
\providecommand \@ifxundefined [1]{%
 \ifx #1\undefined \expandafter \@firstoftwo
 \else \expandafter \@secondoftwo
\fi
}%
\providecommand \@ifnum [1]{%
 \ifnum #1\expandafter \@firstoftwo
 \else \expandafter \@secondoftwo
\fi
}%
\providecommand \enquote [1]{``#1''}%
\providecommand \bibnamefont  [1]{#1}%
\providecommand \bibfnamefont [1]{#1}%
\providecommand \citenamefont [1]{#1}%
\providecommand\href[0]{\@sanitize\@href}%
\providecommand\@href[1]{\endgroup\@@startlink{#1}\endgroup\@@href}%
\providecommand\@@href[1]{#1\@@endlink}%
\providecommand \@sanitize [0]{\begingroup\catcode`\&12\catcode`\#12\relax}%
\@ifxundefined \pdfoutput {\@firstoftwo}{%
 \@ifnum{\z@=\pdfoutput}{\@firstoftwo}{\@secondoftwo}%
}{%
 \providecommand\@@startlink[1]{\leavevmode}%
 \providecommand\@@endlink[0]{}%
}{%
 \providecommand\@@startlink[1]{%
  \leavevmode
  \pdfstartlink
   attr{/Border[0 0 1 ]/H/I/C[0 1 1]}%
   user{/Subtype/Link/A<</Type/Action/S/URI/URI(#1)>>}%
  \relax
 }%
 \providecommand\@@endlink[0]{\pdfendlink}%
}%
\providecommand \url  [0]{\begingroup\@sanitize \@url }%
\providecommand \@url [1]{\endgroup\@href {#1}{\urlprefix}}%
\providecommand \urlprefix [0]{URL }%
\providecommand \Eprint[0]{\href }%
\@ifxundefined \urlstyle {%
  \providecommand \doi [1]{doi:\discretionary{}{}{}#1}%
}{%
  \providecommand \doi [0]{doi:\discretionary{}{}{}\begingroup
  \urlstyle{rm}\Url }%
}%
\providecommand \doibase [0]{http://dx.doi.org/}%
\providecommand \Doi[1]{\href{\doibase#1}}%
\providecommand \selectlanguage [0]{\@gobble}%
\providecommand \bibinfo [0]{\@secondoftwo}%
\providecommand \bibfield [0]{\@secondoftwo}%
\providecommand \translation [1]{[#1]}%
\providecommand \BibitemOpen[0]{}%
\providecommand \bibitemStop [0]{}%
\providecommand \bibitemNoStop [0]{.\EOS\space}%
\providecommand \EOS [0]{\spacefactor3000\relax}%
\providecommand \BibitemShut [1]{\csname bibitem#1\endcsname}%
\bibitem{Alfven:1942}%
  \BibitemOpen
  \bibfield{author}{%
  \bibinfo {author} {\bibfnamefont{H.}~\bibnamefont{{Alfv{\'e}n}}},\ }%
  \bibfield{title}{%
  \enquote{\bibinfo {title} {{Existence of Electromagnetic-Hydrodynamic
  Waves}},}\ }%
  \bibfield{journal}{%
  \Doi{10.1038/150405d0}{\bibinfo {journal} {Nature}}\ }%
  \textbf{\bibinfo {volume} {150}},\ \bibinfo {pages} {405--406} (\bibinfo
  {month} {Oct.}\ \bibinfo {year} {1942})\BibitemShut{NoStop}%
\bibitem{Iroshnikov:1963}%
  \BibitemOpen
  \bibfield{author}{%
  \bibinfo {author} {\bibfnamefont{R.~S.}\ \bibnamefont{Iroshnikov}},\ }%
  \bibfield{title}{%
  \enquote{\bibinfo {title} {The turbulence of a conducting fluid in a strong
  magnetic field},}\ }%
  \bibfield{journal}{%
  \bibinfo {journal} {Astron. Zh.}\ }%
  \textbf{\bibinfo {volume} {40}},\ \bibinfo {pages} {742} (\bibinfo {year}
  {1963}),\ \bibinfo {note} {{English} Translation: Sov. Astron., 7 566
  (1964)}\BibitemShut{NoStop}%
\bibitem{Kraichnan:1965}%
  \BibitemOpen
  \bibfield{author}{%
  \bibinfo {author} {\bibfnamefont{R.~H.}\ \bibnamefont{Kraichnan}},\ }%
  \bibfield{title}{%
  \enquote{\bibinfo {title} {Inertial range spectrum of hyromagnetic
  turbulence},}\ }%
  \bibfield{journal}{%
  \bibinfo {journal} {Phys.~Fluids}\ }%
  \textbf{\bibinfo {volume} {8}},\ \bibinfo {pages} {1385--1387} (\bibinfo
  {year} {1965})\BibitemShut{NoStop}%
\bibitem{Howes:2006}%
  \BibitemOpen
  \bibfield{author}{%
  \bibinfo {author} {\bibfnamefont{G.~G.}\ \bibnamefont{{Howes}}}, \bibinfo
  {author} {\bibfnamefont{S.~C.}\ \bibnamefont{{Cowley}}}, \bibinfo {author}
  {\bibfnamefont{W.}~\bibnamefont{{Dorland}}}, \bibinfo {author}
  {\bibfnamefont{G.~W.}\ \bibnamefont{{Hammett}}}, \bibinfo {author}
  {\bibfnamefont{E.}~\bibnamefont{{Quataert}}},\ and\ \bibinfo {author}
  {\bibfnamefont{A.~A.}\ \bibnamefont{{Schekochihin}}},\ }%
  \bibfield{title}{%
  \enquote{\bibinfo {title} {{Astrophysical Gyrokinetics: Basic Equations and
  Linear Theory}},}\ }%
  \bibfield{journal}{%
  \Doi{10.1086/506172}{\bibinfo {journal} {Astrophys.~J.}}\ }%
  \textbf{\bibinfo {volume} {651}},\ \bibinfo {pages} {590--614} (\bibinfo
  {month} {Nov.}\ \bibinfo {year} {2006}),\
  \Eprint{http://arxiv.org/abs/astro-ph/0511812}{astro-ph/0511812}\BibitemShut%
{NoStop}%
\bibitem{Schekochihin:2009}%
  \BibitemOpen
  \bibfield{author}{%
  \bibinfo {author} {\bibfnamefont{A.~A.}\ \bibnamefont{{Schekochihin}}},
  \bibinfo {author} {\bibfnamefont{S.~C.}\ \bibnamefont{{Cowley}}}, \bibinfo
  {author} {\bibfnamefont{W.}~\bibnamefont{{Dorland}}}, \bibinfo {author}
  {\bibfnamefont{G.~W.}\ \bibnamefont{{Hammett}}}, \bibinfo {author}
  {\bibfnamefont{G.~G.}\ \bibnamefont{{Howes}}}, \bibinfo {author}
  {\bibfnamefont{E.}~\bibnamefont{{Quataert}}},\ and\ \bibinfo {author}
  {\bibfnamefont{T.}~\bibnamefont{{Tatsuno}}},\ }%
  \bibfield{title}{%
  \enquote{\bibinfo {title} {{Astrophysical Gyrokinetics: Kinetic and Fluid
  Turbulent Cascades in Magnetized Weakly Collisional Plasmas}},}\ }%
  \bibfield{journal}{%
  \Doi{10.1088/0067-0049/182/1/310}{\bibinfo {journal} {Astrophys.~J.~Supp.}}\
  }%
  \textbf{\bibinfo {volume} {182}},\ \bibinfo {pages} {310--377} (\bibinfo
  {month} {May}\ \bibinfo {year} {2009})\BibitemShut{NoStop}%
\bibitem{Galeev:1963}%
  \BibitemOpen
  \bibfield{author}{%
  \bibinfo {author} {\bibfnamefont{A.~A.}\ \bibnamefont{{Galeev}}}\ and\
  \bibinfo {author} {\bibfnamefont{V.~N.}\ \bibnamefont{{Oraevskii}}},\ }%
  \bibfield{title}{%
  \enquote{\bibinfo {title} {{The Stability of Alfv{\'e}n Waves}},}\ }%
  \bibfield{journal}{%
  \bibinfo {journal} {Soviet Physics Doklady}\ }%
  \textbf{\bibinfo {volume} {7}},\ \bibinfo {pages} {988} (\bibinfo {month}
  {May}\ \bibinfo {year} {1963})\BibitemShut{NoStop}%
\bibitem{Sagdeev:1969}%
  \BibitemOpen
  \bibfield{author}{%
  \bibinfo {author} {\bibfnamefont{R.~Z.}\ \bibnamefont{{Sagdeev}}}\ and\
  \bibinfo {author} {\bibfnamefont{A.~A.}\ \bibnamefont{{Galeev}}},\ }%
  \emph{\bibinfo {title} {Nonlinear Plasma Theory, New York: Benjamin, 1969}}\
  (\bibinfo {year} {1969})\BibitemShut{NoStop}%
\bibitem{Hasegawa:1976a}%
  \BibitemOpen
  \bibfield{author}{%
  \bibinfo {author} {\bibfnamefont{A.}~\bibnamefont{{Hasegawa}}},\ }%
  \bibfield{title}{%
  \enquote{\bibinfo {title} {{Kinetic theory of MHD instabilities in a
  nonuniform plasma}},}\ }%
  \bibfield{journal}{%
  \Doi{10.1007/BF00152271}{\bibinfo {journal} {Sol.~Phys.}}\ }%
  \textbf{\bibinfo {volume} {47}},\ \bibinfo {pages} {325--330} (\bibinfo
  {month} {Mar.}\ \bibinfo {year} {1976})\BibitemShut{NoStop}%
\bibitem{Derby:1978}%
  \BibitemOpen
  \bibfield{author}{%
  \bibinfo {author} {\bibfnamefont{N.~F.}\ \bibnamefont{{Derby}},
  \bibfnamefont{Jr.}},\ }%
  \bibfield{title}{%
  \enquote{\bibinfo {title} {{Modulational instability of finite-amplitude,
  circularly polarized Alfven waves}},}\ }%
  \bibfield{journal}{%
  \Doi{10.1086/156451}{\bibinfo {journal} {Astrophys.~J.}}\ }%
  \textbf{\bibinfo {volume} {224}},\ \bibinfo {pages} {1013--1016} (\bibinfo
  {month} {Sep.}\ \bibinfo {year} {1978})\BibitemShut{NoStop}%
\bibitem{Goldstein:1978}%
  \BibitemOpen
  \bibfield{author}{%
  \bibinfo {author} {\bibfnamefont{M.~L.}\ \bibnamefont{{Goldstein}}},\ }%
  \bibfield{title}{%
  \enquote{\bibinfo {title} {{An instability of finite amplitude circularly
  polarized Alfven waves}},}\ }%
  \bibfield{journal}{%
  \Doi{10.1086/155829}{\bibinfo {journal} {Astrophys.~J.}}\ }%
  \textbf{\bibinfo {volume} {219}},\ \bibinfo {pages} {700--704} (\bibinfo
  {month} {Jan.}\ \bibinfo {year} {1978})\BibitemShut{NoStop}%
\bibitem{Spangler:1982}%
  \BibitemOpen
  \bibfield{author}{%
  \bibinfo {author} {\bibfnamefont{S.~R.}\ \bibnamefont{{Spangler}}}\ and\
  \bibinfo {author} {\bibfnamefont{J.~P.}\ \bibnamefont{{Sheerin}}},\ }%
  \bibfield{title}{%
  \enquote{\bibinfo {title} {{Properties of Alfven solitons in a finite-beta
  plasma}},}\ }%
  \bibfield{journal}{%
  \Doi{10.1017/S0022377800026519}{\bibinfo {journal} {J.~Plasma Phys.}}\ }%
  \textbf{\bibinfo {volume} {27}},\ \bibinfo {pages} {193--198} (\bibinfo
  {month} {Apr.}\ \bibinfo {year} {1982})\BibitemShut{NoStop}%
\bibitem{Sakai:1983}%
  \BibitemOpen
  \bibfield{author}{%
  \bibinfo {author} {\bibfnamefont{J.-I.}\ \bibnamefont{{Sakai}}}\ and\
  \bibinfo {author} {\bibfnamefont{U.~O.}\ \bibnamefont{{Sonnerup}}},\ }%
  \bibfield{title}{%
  \enquote{\bibinfo {title} {{Modulational instability of finite amplitude
  dispersive Alfven waves}},}\ }%
  \bibfield{journal}{%
  \Doi{10.1029/JA088iA11p09069}{\bibinfo {journal} {J.~Geophys.~Res.}}\ }%
  \textbf{\bibinfo {volume} {88}},\ \bibinfo {pages} {9069--9079} (\bibinfo
  {month} {Nov.}\ \bibinfo {year} {1983})\BibitemShut{NoStop}%
\bibitem{Spangler:1986}%
  \BibitemOpen
  \bibfield{author}{%
  \bibinfo {author} {\bibfnamefont{S.~R.}\ \bibnamefont{{Spangler}}},\ }%
  \bibfield{title}{%
  \enquote{\bibinfo {title} {{The evolution of nonlinear Alfven waves subject
  to growth and damping}},}\ }%
  \bibfield{journal}{%
  \Doi{10.1063/1.865545}{\bibinfo {journal} {Phys.~Fluids}}\ }%
  \textbf{\bibinfo {volume} {29}},\ \bibinfo {pages} {2535--2547} (\bibinfo
  {month} {Aug.}\ \bibinfo {year} {1986})\BibitemShut{NoStop}%
\bibitem{Terasawa:1986}%
  \BibitemOpen
  \bibfield{author}{%
  \bibinfo {author} {\bibfnamefont{T.}~\bibnamefont{{Terasawa}}}, \bibinfo
  {author} {\bibfnamefont{M.}~\bibnamefont{{Hoshino}}}, \bibinfo {author}
  {\bibfnamefont{J.-I.}\ \bibnamefont{{Sakai}}},\ and\ \bibinfo {author}
  {\bibfnamefont{T.}~\bibnamefont{{Hada}}},\ }%
  \bibfield{title}{%
  \enquote{\bibinfo {title} {{Decay instability of finite-amplitude circularly
  polarized Alfven waves - A numerical simulation of stimulated Brillouin
  scattering}},}\ }%
  \bibfield{journal}{%
  \Doi{10.1029/JA091iA04p04171}{\bibinfo {journal} {J.~Geophys.~Res.}}\ }%
  \textbf{\bibinfo {volume} {91}},\ \bibinfo {pages} {4171--4187} (\bibinfo
  {month} {Apr.}\ \bibinfo {year} {1986})\BibitemShut{NoStop}%
\bibitem{Jayanti:1993}%
  \BibitemOpen
  \bibfield{author}{%
  \bibinfo {author} {\bibfnamefont{V.}~\bibnamefont{{Jayanti}}}\ and\ \bibinfo
  {author} {\bibfnamefont{J.~V.}\ \bibnamefont{{Hollweg}}},\ }%
  \bibfield{title}{%
  \enquote{\bibinfo {title} {{On the dispersion relations for parametric
  instabilities of parallel-progagating Alfv{\'e}n waves}},}\ }%
  \bibfield{journal}{%
  \Doi{10.1029/93JA00920}{\bibinfo {journal} {J.~Geophys.~Res.}}\ }%
  \textbf{\bibinfo {volume} {98}},\ \bibinfo {pages} {13247--13252} (\bibinfo
  {month} {Aug.}\ \bibinfo {year} {1993})\BibitemShut{NoStop}%
\bibitem{Hollweg:1994}%
  \BibitemOpen
  \bibfield{author}{%
  \bibinfo {author} {\bibfnamefont{J.~V.}\ \bibnamefont{{Hollweg}}},\ }%
  \bibfield{title}{%
  \enquote{\bibinfo {title} {{Beat, modulational, and decay instabilities of a
  circularly polarized Alfven wave}},}\ }%
  \bibfield{journal}{%
  \Doi{10.1029/94JA02185}{\bibinfo {journal} {J.~Geophys.~Res.}}\ }%
  \textbf{\bibinfo {volume} {99}},\ \bibinfo {pages} {23431--+} (\bibinfo
  {month} {Dec.}\ \bibinfo {year} {1994})\BibitemShut{NoStop}%
\bibitem{Shevchenko:2003}%
  \BibitemOpen
  \bibfield{author}{%
  \bibinfo {author} {\bibfnamefont{V.~I.}\ \bibnamefont{{Shevchenko}}},
  \bibinfo {author} {\bibfnamefont{R.~Z.}\ \bibnamefont{{Sagdeev}}}, \bibinfo
  {author} {\bibfnamefont{V.~L.}\ \bibnamefont{{Galinsky}}},\ and\ \bibinfo
  {author} {\bibfnamefont{M.~V.}\ \bibnamefont{{Medvedev}}},\ }%
  \bibfield{title}{%
  \enquote{\bibinfo {title} {{The DNLS equation and parametric decay
  instability}},}\ }%
  \bibfield{journal}{%
  \Doi{10.1134/1.1592552}{\bibinfo {journal} {Plasma Phys. Rep.}}\ }%
  \textbf{\bibinfo {volume} {29}},\ \bibinfo {pages} {545--549} (\bibinfo
  {month} {Jul.}\ \bibinfo {year} {2003})\BibitemShut{NoStop}%
\bibitem{Voitenko:2005}%
  \BibitemOpen
  \bibfield{author}{%
  \bibinfo {author} {\bibfnamefont{Y.~M.}\ \bibnamefont{{Voitenko}}}\ and\
  \bibinfo {author} {\bibfnamefont{M.}~\bibnamefont{{Goossens}}},\ }%
  \bibfield{title}{%
  \enquote{\bibinfo {title} {{Nonlinear coupling of Alfv{\'e}n waves with
  widely different cross-field wavelengths in space plasmas}},}\ }%
  \bibfield{journal}{%
  \Doi{10.1029/2004JA010874}{\bibinfo {journal} {J.~Geophys.~Res.}}\ }%
  \textbf{\bibinfo {volume} {110}},\ \bibinfo {eid} {A10S01} (\bibinfo {month}
  {Jul.}\ \bibinfo {year} {2005})\BibitemShut{NoStop}%
\bibitem{Lashmore-Davies:1976}%
  \BibitemOpen
  \bibfield{author}{%
  \bibinfo {author} {\bibfnamefont{C.~N.}\ \bibnamefont{{Lashmore-Davies}}},\
  }%
  \bibfield{title}{%
  \enquote{\bibinfo {title} {{Modulational instability of a finite amplitude
  Alfven wave}},}\ }%
  \bibfield{journal}{%
  \Doi{10.1063/1.861493}{\bibinfo {journal} {Phys.~Fluids}}\ }%
  \textbf{\bibinfo {volume} {19}},\ \bibinfo {pages} {587--589} (\bibinfo
  {month} {Apr.}\ \bibinfo {year} {1976})\BibitemShut{NoStop}%
\bibitem{Mjolhus:1976}%
  \BibitemOpen
  \bibfield{author}{%
  \bibinfo {author} {\bibfnamefont{E.}~\bibnamefont{{Mjolhus}}},\ }%
  \bibfield{title}{%
  \enquote{\bibinfo {title} {{On the modulational instability of hydromagnetic
  waves parallel to the magnetic field}},}\ }%
  \bibfield{journal}{%
  \Doi{10.1017/S0022377800020249}{\bibinfo {journal} {J.~Plasma Phys.}}\ }%
  \textbf{\bibinfo {volume} {16}},\ \bibinfo {pages} {321--334} (\bibinfo
  {month} {Dec.}\ \bibinfo {year} {1976})\BibitemShut{NoStop}%
\bibitem{Mio:1976b}%
  \BibitemOpen
  \bibfield{author}{%
  \bibinfo {author} {\bibfnamefont{K.}~\bibnamefont{{Mio}}}, \bibinfo {author}
  {\bibfnamefont{T.}~\bibnamefont{{Ogino}}}, \bibinfo {author}
  {\bibfnamefont{S.}~\bibnamefont{{Takeda}}},\ and\ \bibinfo {author}
  {\bibfnamefont{K.}~\bibnamefont{{Minami}}},\ }%
  \bibfield{title}{%
  \enquote{\bibinfo {title} {{Modulational instability and envelope-solitons
  for nonlinear Alfven waves propagating along the magnetic field in
  plasmas}},}\ }%
  \bibfield{journal}{%
  \bibinfo {journal} {Journal of the Physical Society of Japan}\ }%
  \textbf{\bibinfo {volume} {41}},\ \bibinfo {pages} {667--673} (\bibinfo
  {month} {Aug.}\ \bibinfo {year} {1976})\BibitemShut{NoStop}%
\bibitem{Wong:1986}%
  \BibitemOpen
  \bibfield{author}{%
  \bibinfo {author} {\bibfnamefont{H.~K.}\ \bibnamefont{{Wong}}}\ and\ \bibinfo
  {author} {\bibfnamefont{M.~L.}\ \bibnamefont{{Goldstein}}},\ }%
  \bibfield{title}{%
  \enquote{\bibinfo {title} {{Parametric instabilities of the circularly
  polarized Alfven waves including dispersion}},}\ }%
  \bibfield{journal}{%
  \Doi{10.1029/JA091iA05p05617}{\bibinfo {journal} {J.~Geophys.~Res.}}\ }%
  \textbf{\bibinfo {volume} {91}},\ \bibinfo {pages} {5617--5628} (\bibinfo
  {month} {May}\ \bibinfo {year} {1986})\BibitemShut{NoStop}%
\bibitem{Shukla:2007}%
  \BibitemOpen
  \bibfield{author}{%
  \bibinfo {author} {\bibfnamefont{P.~K.}\ \bibnamefont{{Shukla}}}, \bibinfo
  {author} {\bibfnamefont{N.}~\bibnamefont{{Shukla}}},\ and\ \bibinfo {author}
  {\bibfnamefont{L.}~\bibnamefont{{Stenflo}}},\ }%
  \bibfield{title}{%
  \enquote{\bibinfo {title} {{Kinetic modulational instability of broadband
  dispersive Alfv{\'e}n waves in plasmas}},}\ }%
  \bibfield{journal}{%
  \Doi{10.1017/S0022377806006271}{\bibinfo {journal} {J.~Plasma Phys.}}\ }%
  \textbf{\bibinfo {volume} {73}},\ \bibinfo {pages} {153--157} (\bibinfo
  {month} {Apr.}\ \bibinfo {year} {2007})\BibitemShut{NoStop}%
\bibitem{Lacombe:1980}%
  \BibitemOpen
  \bibfield{author}{%
  \bibinfo {author} {\bibfnamefont{C.}~\bibnamefont{{Lacombe}}}\ and\ \bibinfo
  {author} {\bibfnamefont{A.}~\bibnamefont{{Mangeney}}},\ }%
  \bibfield{title}{%
  \enquote{\bibinfo {title} {{Non-Linear Interaction of Alfven Waves with
  Compressive Fast Magnetosonic Waves}},}\ \ }%
  \textbf{\bibinfo {volume} {88}},\ \bibinfo {pages} {277} (\bibinfo {month}
  {Aug.}\ \bibinfo {year} {1980})\BibitemShut{NoStop}%
\bibitem{Brodin:1988}%
  \BibitemOpen
  \bibfield{author}{%
  \bibinfo {author} {\bibfnamefont{G.}~\bibnamefont{{Brodin}}}\ and\ \bibinfo
  {author} {\bibfnamefont{L.}~\bibnamefont{{Stenflo}}},\ }%
  \bibfield{title}{%
  \enquote{\bibinfo {title} {{Three-wave coupling coefficients for MHD
  plasmas}},}\ }%
  \bibfield{journal}{%
  \Doi{10.1017/S0022377800013027}{\bibinfo {journal} {J.~Plasma Phys.}}\ }%
  \textbf{\bibinfo {volume} {39}},\ \bibinfo {pages} {277--284} (\bibinfo
  {month} {Apr.}\ \bibinfo {year} {1988})\BibitemShut{NoStop}%
\bibitem{Kucherenko:1993}%
  \BibitemOpen
  \bibfield{author}{%
  \bibinfo {author} {\bibfnamefont{V.~P.}\ \bibnamefont{{Kucherenko}}}\ and\
  \bibinfo {author} {\bibfnamefont{A.~K.}\ \bibnamefont{{Yukhimuk}}},\ }%
  \bibfield{title}{%
  \enquote{\bibinfo {title} {{Nonlinear interaction of kinetic Alfven
  waves}},}\ }%
  \bibfield{journal}{%
  \bibinfo {journal} {Kinematika i Fizika Nebesnykh Tel}\ }%
  \textbf{\bibinfo {volume} {9}},\ \bibinfo {pages} {41--46} (\bibinfo {month}
  {Jun.}\ \bibinfo {year} {1993})\BibitemShut{NoStop}%
\bibitem{Yukhimuk:2000}%
  \BibitemOpen
  \bibfield{author}{%
  \bibinfo {author} {\bibfnamefont{A.~K.}\ \bibnamefont{{Yukhimuk}}}, \bibinfo
  {author} {\bibfnamefont{V.~M.}\ \bibnamefont{{Fedun}}}, \bibinfo {author}
  {\bibfnamefont{E.~K.}\ \bibnamefont{{Sirenko}}}, \bibinfo {author}
  {\bibfnamefont{Y.~M.}\ \bibnamefont{{Voitenko}}},\ and\ \bibinfo {author}
  {\bibfnamefont{A.~D.}\ \bibnamefont{{Voytsekhovskaya}}},\ }%
  \bibfield{title}{%
  \enquote{\bibinfo {title} {{Nonlinear interaction of MHD waves and solar
  corona heating}},}\ }%
  \bibfield{journal}{%
  \bibinfo {journal} {Kinematika i Fizika Nebesnykh Tel Supplement}\ }%
  \textbf{\bibinfo {volume} {3}},\ \bibinfo {pages} {477--480} (\bibinfo
  {month} {Sep.}\ \bibinfo {year} {2000})\BibitemShut{NoStop}%
\bibitem{Chandran:2005}%
  \BibitemOpen
  \bibfield{author}{%
  \bibinfo {author} {\bibfnamefont{B.~D.~G.}\ \bibnamefont{Chandran}},\ }%
  \bibfield{title}{%
  \enquote{\bibinfo {title} {Weak compressible magnetohydrodynamic turbulence
  in the solar corona},}\ }%
  \bibfield{journal}{%
  \bibinfo {journal} {Phys.~Rev.~Lett.}\ }%
  \textbf{\bibinfo {volume} {95}},\ \bibinfo {pages} {265004/1--4} (\bibinfo
  {year} {2005})\BibitemShut{NoStop}%
\bibitem{Shukla:2006}%
  \BibitemOpen
  \bibfield{author}{%
  \bibinfo {author} {\bibfnamefont{P.~K.}\ \bibnamefont{{Shukla}}}, \bibinfo
  {author} {\bibfnamefont{G.}~\bibnamefont{{Brodin}}},\ and\ \bibinfo {author}
  {\bibfnamefont{L.}~\bibnamefont{{Stenflo}}},\ }%
  \bibfield{title}{%
  \enquote{\bibinfo {title} {{Nonlinear interaction between three kinetic
  Alfv{\'e}n waves}},}\ }%
  \bibfield{journal}{%
  \Doi{10.1016/j.physleta.2005.11.079}{\bibinfo {journal} {Physics Letters A}}\
  }%
  \textbf{\bibinfo {volume} {353}},\ \bibinfo {pages} {73--75} (\bibinfo
  {month} {Apr.}\ \bibinfo {year} {2006})\BibitemShut{NoStop}%
\bibitem{Brodin:2007}%
  \BibitemOpen
  \bibfield{author}{%
  \bibinfo {author} {\bibfnamefont{G.}~\bibnamefont{{Brodin}}}, \bibinfo
  {author} {\bibfnamefont{L.}~\bibnamefont{{Stenflo}}},\ and\ \bibinfo {author}
  {\bibfnamefont{P.~K.}\ \bibnamefont{{Shukla}}},\ }%
  \bibfield{title}{%
  \enquote{\bibinfo {title} {{Nonlinear interactions between three inertial
  Alfv{\'e}n waves}},}\ }%
  \bibfield{journal}{%
  \Doi{10.1017/S0022377806004788}{\bibinfo {journal} {J.~Plasma Phys.}}\ }%
  \textbf{\bibinfo {volume} {73}},\ \bibinfo {pages} {9--13} (\bibinfo {month}
  {Jan.}\ \bibinfo {year} {2007})\BibitemShut{NoStop}%
\bibitem{Mottez:2012}%
  \BibitemOpen
  \bibfield{author}{%
  \bibinfo {author} {\bibfnamefont{F.}~\bibnamefont{{Mottez}}},\ }%
  \bibfield{title}{%
  \enquote{\bibinfo {title} {{Non-propagating electric and density structures
  formed through non-linear interaction of Alfv{\'e}n waves}},}\ }%
  \bibfield{journal}{%
  \Doi{10.5194/angeo-30-81-2012}{\bibinfo {journal} {Annales Geophysicae}}\ }%
  \textbf{\bibinfo {volume} {30}},\ \bibinfo {pages} {81--95} (\bibinfo {month}
  {Jan.}\ \bibinfo {year} {2012})\BibitemShut{NoStop}%
\bibitem{Goldreich:1995}%
  \BibitemOpen
  \bibfield{author}{%
  \bibinfo {author} {\bibfnamefont{P.}~\bibnamefont{Goldreich}}\ and\ \bibinfo
  {author} {\bibfnamefont{S.}~\bibnamefont{Sridhar}},\ }%
  \bibfield{title}{%
  \enquote{\bibinfo {title} {{Toward a Theery of Interstellar Turbulence II.
  Strong Alfv\'enic Turbulence}},}\ }%
  \bibfield{journal}{%
  \bibinfo {journal} {Astrophys.~J.}\ }%
  \textbf{\bibinfo {volume} {438}},\ \bibinfo {pages} {763--775} (\bibinfo
  {year} {1995})\BibitemShut{NoStop}%
\bibitem{Howes:2008b}%
  \BibitemOpen
  \bibfield{author}{%
  \bibinfo {author} {\bibfnamefont{G.~G.}\ \bibnamefont{{Howes}}}, \bibinfo
  {author} {\bibfnamefont{S.~C.}\ \bibnamefont{{Cowley}}}, \bibinfo {author}
  {\bibfnamefont{W.}~\bibnamefont{{Dorland}}}, \bibinfo {author}
  {\bibfnamefont{G.~W.}\ \bibnamefont{{Hammett}}}, \bibinfo {author}
  {\bibfnamefont{E.}~\bibnamefont{{Quataert}}},\ and\ \bibinfo {author}
  {\bibfnamefont{A.~A.}\ \bibnamefont{{Schekochihin}}},\ }%
  \bibfield{title}{%
  \enquote{\bibinfo {title} {{A model of turbulence in magnetized plasmas:
  Implications for the dissipation range in the solar wind}},}\ }%
  \bibfield{journal}{%
  \Doi{10.1029/2007JA012665}{\bibinfo {journal} {J.~Geophys.~Res.}}\ }%
  \textbf{\bibinfo {volume} {113}},\ \bibinfo {pages} {A05103} (\bibinfo
  {month} {May}\ \bibinfo {year} {2008}),\
  \Eprint{http://arxiv.org/abs/arXiv:0707.3147}{arXiv:0707.3147}\BibitemShut{N%
oStop}%
\bibitem{Howes:2011b}%
  \BibitemOpen
  \bibfield{author}{%
  \bibinfo {author} {\bibfnamefont{G.~G.}\ \bibnamefont{{Howes}}}, \bibinfo
  {author} {\bibfnamefont{J.~M.}\ \bibnamefont{{Tenbarge}}},\ and\ \bibinfo
  {author} {\bibfnamefont{W.}~\bibnamefont{{Dorland}}},\ }%
  \bibfield{title}{%
  \enquote{\bibinfo {title} {{A weakened cascade model for turbulence in
  astrophysical plasmas}},}\ }%
  \bibfield{journal}{%
  \Doi{10.1063/1.3646400}{\bibinfo {journal} {Phys.~Plasmas}}\ }%
  \textbf{\bibinfo {volume} {18}},\ \bibinfo {pages} {102305} (\bibinfo {year}
  {2011}),\ \Eprint{http://arxiv.org/abs/1109.4158}{arXiv:1109.4158
  [astro-ph.SR]}\BibitemShut{NoStop}%
\bibitem{Sridhar:1994}%
  \BibitemOpen
  \bibfield{author}{%
  \bibinfo {author} {\bibfnamefont{S.}~\bibnamefont{{Sridhar}}}\ and\ \bibinfo
  {author} {\bibfnamefont{P.}~\bibnamefont{{Goldreich}}},\ }%
  \bibfield{title}{%
  \enquote{\bibinfo {title} {{Toward a theory of interstellar turbulence. 1:
  Weak Alfvenic turbulence}},}\ }%
  \bibfield{journal}{%
  \Doi{10.1086/174600}{\bibinfo {journal} {Astrophys.~J.}}\ }%
  \textbf{\bibinfo {volume} {432}},\ \bibinfo {pages} {612--621} (\bibinfo
  {month} {Sep.}\ \bibinfo {year} {1994})\BibitemShut{NoStop}%
\bibitem{Boldyrev:2006}%
  \BibitemOpen
  \bibfield{author}{%
  \bibinfo {author} {\bibfnamefont{S.}~\bibnamefont{{Boldyrev}}},\ }%
  \bibfield{title}{%
  \enquote{\bibinfo {title} {{Spectrum of Magnetohydrodynamic Turbulence}},}\
  }%
  \bibfield{journal}{%
  \Doi{10.1103/PhysRevLett.96.115002}{\bibinfo {journal} {Phys.~Rev.~Lett.}}\
  }%
  \textbf{\bibinfo {volume} {96}},\ \bibinfo {pages} {115002--+} (\bibinfo
  {month} {Mar.}\ \bibinfo {year} {2006}),\
  \Eprint{http://arxiv.org/abs/arXiv:astro-ph/0511290}{arXiv:astro-ph/0511290}%
\BibitemShut{NoStop}%
\bibitem{Shebalin:1983}%
  \BibitemOpen
  \bibfield{author}{%
  \bibinfo {author} {\bibfnamefont{J.~V.}\ \bibnamefont{Shebalin}}, \bibinfo
  {author} {\bibfnamefont{W.~H.}\ \bibnamefont{Matthaeus}},\ and\ \bibinfo
  {author} {\bibfnamefont{D.}~\bibnamefont{Montgomery}},\ }%
  \bibfield{title}{%
  \enquote{\bibinfo {title} {Anisotropy in mhd turbulence due to a mean
  magnetic field},}\ }%
  \bibfield{journal}{%
  \bibinfo {journal} {J.~Plasma Phys.}\ }%
  \textbf{\bibinfo {volume} {29}},\ \bibinfo {pages} {525--547} (\bibinfo
  {year} {1983})\BibitemShut{NoStop}%
\bibitem{Cho:2000}%
  \BibitemOpen
  \bibfield{author}{%
  \bibinfo {author} {\bibfnamefont{J.}~\bibnamefont{Cho}}\ and\ \bibinfo
  {author} {\bibfnamefont{E.~T.}\ \bibnamefont{Vishniac}},\ }%
  \bibfield{title}{%
  \enquote{\bibinfo {title} {{The Anisotropy of Magnetohydrodynamic Alfv\'enic
  Turbulence}},}\ }%
  \bibfield{journal}{%
  \bibinfo {journal} {Astrophys.~J.}\ }%
  \textbf{\bibinfo {volume} {539}},\ \bibinfo {pages} {273--282} (\bibinfo
  {year} {2000})\BibitemShut{NoStop}%
\bibitem{Maron:2001}%
  \BibitemOpen
  \bibfield{author}{%
  \bibinfo {author} {\bibfnamefont{J.}~\bibnamefont{Maron}}\ and\ \bibinfo
  {author} {\bibfnamefont{P.}~\bibnamefont{Goldreich}},\ }%
  \bibfield{title}{%
  \enquote{\bibinfo {title} {Simulations of incompressible magnetohydrodynamic
  turbulence},}\ }%
  \bibfield{journal}{%
  \bibinfo {journal} {Astrophys.~J.}\ }%
  \textbf{\bibinfo {volume} {554}},\ \bibinfo {pages} {1175--1196} (\bibinfo
  {year} {2001})\BibitemShut{NoStop}%
\bibitem{Cho:2004}%
  \BibitemOpen
  \bibfield{author}{%
  \bibinfo {author} {\bibfnamefont{J.}~\bibnamefont{{Cho}}}\ and\ \bibinfo
  {author} {\bibfnamefont{A.}~\bibnamefont{{Lazarian}}},\ }%
  \bibfield{title}{%
  \enquote{\bibinfo {title} {{The Anisotropy of Electron Magnetohydrodynamic
  Turbulence}},}\ }%
  \bibfield{journal}{%
  \Doi{10.1086/425215}{\bibinfo {journal} {Astrophys.~J.~Lett.}}\ }%
  \textbf{\bibinfo {volume} {615}},\ \bibinfo {pages} {L41--L44} (\bibinfo
  {month} {Nov.}\ \bibinfo {year} {2004}),\
  \Eprint{http://arxiv.org/abs/astro-ph/0406595}{astro-ph/0406595}\BibitemShut%
{NoStop}%
\bibitem{Cho:2009}%
  \BibitemOpen
  \bibfield{author}{%
  \bibinfo {author} {\bibfnamefont{J.}~\bibnamefont{{Cho}}}\ and\ \bibinfo
  {author} {\bibfnamefont{A.}~\bibnamefont{{Lazarian}}},\ }%
  \bibfield{title}{%
  \enquote{\bibinfo {title} {{Simulations of Electron Magnetohydrodynamic
  Turbulence}},}\ }%
  \bibfield{journal}{%
  \Doi{10.1088/0004-637X/701/1/236}{\bibinfo {journal} {Astrophys.~J.}}\ }%
  \textbf{\bibinfo {volume} {701}},\ \bibinfo {pages} {236--252} (\bibinfo
  {month} {Aug.}\ \bibinfo {year} {2009}),\
  \Eprint{http://arxiv.org/abs/0904.0661}{arXiv:0904.0661
  [astro-ph.EP]}\BibitemShut{NoStop}%
\bibitem{TenBarge:2012a}%
  \BibitemOpen
  \bibfield{author}{%
  \bibinfo {author} {\bibfnamefont{J.~M.}\ \bibnamefont{{TenBarge}}}\ and\
  \bibinfo {author} {\bibfnamefont{G.~G.}\ \bibnamefont{{Howes}}},\ }%
  \bibfield{title}{%
  \enquote{\bibinfo {title} {{Evidence of critical balance in kinetic
  Alfv{\'e}n wave turbulence simulations}},}\ }%
  \bibfield{journal}{%
  \Doi{10.1063/1.3693974}{\bibinfo {journal} {Phys.~Plasmas}}\ }%
  \textbf{\bibinfo {volume} {19}},\ \bibinfo {pages} {055901} (\bibinfo {month}
  {May}\ \bibinfo {year} {2012})\BibitemShut{NoStop}%
\bibitem{Robinson:1971}%
  \BibitemOpen
  \bibfield{author}{%
  \bibinfo {author} {\bibfnamefont{D.~C.}\ \bibnamefont{Robinson}}\ and\
  \bibinfo {author} {\bibfnamefont{M.~G.}\ \bibnamefont{Rusbridge}},\ }%
  \bibfield{title}{%
  \enquote{\bibinfo {title} {Structure of turbulence in the zeta plasma},}\ }%
  \bibfield{journal}{%
  \bibinfo {journal} {Phys.~Fluids}\ }%
  \textbf{\bibinfo {volume} {14}},\ \bibinfo {pages} {2499--2511} (\bibinfo
  {year} {1971})\BibitemShut{NoStop}%
\bibitem{Zweben:1979}%
  \BibitemOpen
  \bibfield{author}{%
  \bibinfo {author} {\bibfnamefont{S.~J.}\ \bibnamefont{Zweben}}, \bibinfo
  {author} {\bibfnamefont{C.~R.}\ \bibnamefont{Menyuk}},\ and\ \bibinfo
  {author} {\bibfnamefont{R.~J.}\ \bibnamefont{Taylor}},\ }%
  \bibfield{title}{%
  \enquote{\bibinfo {title} {Small-scale magnetic fluctuations inside the
  macrotor tokamak},}\ }%
  \bibfield{journal}{%
  \bibinfo {journal} {Phys.~Rev.~Lett.}\ }%
  \textbf{\bibinfo {volume} {42}},\ \bibinfo {pages} {1270--1274} (\bibinfo
  {year} {1979})\BibitemShut{NoStop}%
\bibitem{Montgomery:1981}%
  \BibitemOpen
  \bibfield{author}{%
  \bibinfo {author} {\bibfnamefont{D.}~\bibnamefont{Montgomery}}\ and\ \bibinfo
  {author} {\bibfnamefont{L.}~\bibnamefont{Turner}},\ }%
  \bibfield{title}{%
  \enquote{\bibinfo {title} {Anisotropic magnetohydrodynamic turbulence in a
  strong external magnetic field},}\ }%
  \bibfield{journal}{%
  \bibinfo {journal} {Phys.~Fluids}\ }%
  \textbf{\bibinfo {volume} {24}},\ \bibinfo {pages} {825--831} (\bibinfo
  {year} {1981})\BibitemShut{NoStop}%
\bibitem{Belcher:1971}%
  \BibitemOpen
  \bibfield{author}{%
  \bibinfo {author} {\bibfnamefont{J.~W.}\ \bibnamefont{{Belcher}}}\ and\
  \bibinfo {author} {\bibfnamefont{L.}~\bibnamefont{{Davis}}},\ }%
  \bibfield{title}{%
  \enquote{\bibinfo {title} {{Large-Amplitude Alfv{\'e}n Waves in the
  Interplanetary Medium, 2}},}\ }%
  \bibfield{journal}{%
  \bibinfo {journal} {J.~Geophys.~Res.}\ }%
  \textbf{\bibinfo {volume} {76}},\ \bibinfo {pages} {3534--3563} (\bibinfo
  {year} {1971})\BibitemShut{NoStop}%
\bibitem{Sahraoui:2010b}%
  \BibitemOpen
  \bibfield{author}{%
  \bibinfo {author} {\bibfnamefont{F.}~\bibnamefont{{Sahraoui}}}, \bibinfo
  {author} {\bibfnamefont{M.~L.}\ \bibnamefont{{Goldstein}}}, \bibinfo {author}
  {\bibfnamefont{G.}~\bibnamefont{{Belmont}}}, \bibinfo {author}
  {\bibfnamefont{P.}~\bibnamefont{{Canu}}},\ and\ \bibinfo {author}
  {\bibfnamefont{L.}~\bibnamefont{{Rezeau}}},\ }%
  \bibfield{title}{%
  \enquote{\bibinfo {title} {{Three Dimensional Anisotropic k Spectra of
  Turbulence at Subproton Scales in the Solar Wind}},}\ }%
  \bibfield{journal}{%
  \Doi{10.1103/PhysRevLett.105.131101}{\bibinfo {journal} {Phys.~Rev.~Lett.}}\
  }%
  \textbf{\bibinfo {volume} {105}},\ \bibinfo {pages} {131101--+} (\bibinfo
  {month} {Sep.}\ \bibinfo {year} {2010})\BibitemShut{NoStop}%
\bibitem{Narita:2011}%
  \BibitemOpen
  \bibfield{author}{%
  \bibinfo {author} {\bibfnamefont{Y.}~\bibnamefont{{Narita}}}, \bibinfo
  {author} {\bibfnamefont{S.~P.}\ \bibnamefont{{Gary}}}, \bibinfo {author}
  {\bibfnamefont{S.}~\bibnamefont{{Saito}}}, \bibinfo {author}
  {\bibfnamefont{K.-H.}\ \bibnamefont{{Glassmeier}}},\ and\ \bibinfo {author}
  {\bibfnamefont{U.}~\bibnamefont{{Motschmann}}},\ }%
  \bibfield{title}{%
  \enquote{\bibinfo {title} {{Dispersion relation analysis of solar wind
  turbulence}},}\ }%
  \bibfield{journal}{%
  \Doi{10.1029/2010GL046588}{\bibinfo {journal} {Geophys.~Res.~Lett.}}\ }%
  \textbf{\bibinfo {volume} {38}},\ \bibinfo {pages} {L05101} (\bibinfo {month}
  {Mar.}\ \bibinfo {year} {2011})\BibitemShut{NoStop}%
\bibitem{Montgomery:1982}%
  \BibitemOpen
  \bibfield{author}{%
  \bibinfo {author} {\bibfnamefont{D.}~\bibnamefont{{Montgomery}}},\ }%
  \bibfield{title}{%
  \enquote{\bibinfo {title} {{Major disruptions, inverse cascades, and the
  Strauss equations}},}\ }%
  \bibfield{journal}{%
  \bibinfo {journal} {Physica Scripta}\ }%
  \textbf{\bibinfo {volume} {T2A}},\ \bibinfo {pages} {83} (\bibinfo {year}
  {1982})\BibitemShut{NoStop}%
\bibitem{Higdon:1984a}%
  \BibitemOpen
  \bibfield{author}{%
  \bibinfo {author} {\bibfnamefont{J.~C.}\ \bibnamefont{Higdon}},\ }%
  \bibfield{title}{%
  \enquote{\bibinfo {title} {Density fluctuations in the interstellar medium:
  Evidence for anisotropic magnetogasdynamic turbulence i. model and
  astrophysical sites},}\ }%
  \bibfield{journal}{%
  \bibinfo {journal} {Astrophys.~J.}\ }%
  \textbf{\bibinfo {volume} {285}},\ \bibinfo {pages} {109--123} (\bibinfo
  {year} {1984})\BibitemShut{NoStop}%
\bibitem{Montgomery:1995}%
  \BibitemOpen
  \bibfield{author}{%
  \bibinfo {author} {\bibfnamefont{D.}~\bibnamefont{{Montgomery}}}\ and\
  \bibinfo {author} {\bibfnamefont{W.~H.}\ \bibnamefont{{Matthaeus}}},\ }%
  \bibfield{title}{%
  \enquote{\bibinfo {title} {{Anisotropic Modal Energy Transfer in Interstellar
  Turbulence}},}\ }%
  \bibfield{journal}{%
  \Doi{10.1086/175910}{\bibinfo {journal} {Astrophys.~J.}}\ }%
  \textbf{\bibinfo {volume} {447}},\ \bibinfo {pages} {706--+} (\bibinfo
  {month} {Jul.}\ \bibinfo {year} {1995})\BibitemShut{NoStop}%
\bibitem{Ng:1996}%
  \BibitemOpen
  \bibfield{author}{%
  \bibinfo {author} {\bibfnamefont{C.~S.}\ \bibnamefont{{Ng}}}\ and\ \bibinfo
  {author} {\bibfnamefont{A.}~\bibnamefont{{Bhattacharjee}}},\ }%
  \bibfield{title}{%
  \enquote{\bibinfo {title} {{Interaction of Shear-Alfven Wave Packets:
  Implication for Weak Magnetohydrodynamic Turbulence in Astrophysical
  Plasmas}},}\ }%
  \bibfield{journal}{%
  \Doi{10.1086/177468}{\bibinfo {journal} {Astrophys.~J.}}\ }%
  \textbf{\bibinfo {volume} {465}},\ \bibinfo {pages} {845--+} (\bibinfo
  {month} {Jul.}\ \bibinfo {year} {1996})\BibitemShut{NoStop}%
\bibitem{Goldreich:1997}%
  \BibitemOpen
  \bibfield{author}{%
  \bibinfo {author} {\bibfnamefont{P.}~\bibnamefont{Goldreich}}\ and\ \bibinfo
  {author} {\bibfnamefont{S.}~\bibnamefont{Sridhar}},\ }%
  \bibfield{title}{%
  \enquote{\bibinfo {title} {Magnetohydrodynamic turbulence revisited},}\ }%
  \bibfield{journal}{%
  \bibinfo {journal} {Astrophys.~J.}\ }%
  \textbf{\bibinfo {volume} {485}},\ \bibinfo {pages} {680--688} (\bibinfo
  {year} {1997})\BibitemShut{NoStop}%
\bibitem{Galtier:2000}%
  \BibitemOpen
  \bibfield{author}{%
  \bibinfo {author} {\bibfnamefont{S.}~\bibnamefont{{Galtier}}}, \bibinfo
  {author} {\bibfnamefont{S.~V.}\ \bibnamefont{{Nazarenko}}}, \bibinfo {author}
  {\bibfnamefont{A.~C.}\ \bibnamefont{{Newell}}},\ and\ \bibinfo {author}
  {\bibfnamefont{A.}~\bibnamefont{{Pouquet}}},\ }%
  \bibfield{title}{%
  \enquote{\bibinfo {title} {{A weak turbulence theory for incompressible
  magnetohydrodynamics}},}\ }%
  \bibfield{journal}{%
  \bibinfo {journal} {J.~Plasma Phys.}\ }%
  \textbf{\bibinfo {volume} {63}},\ \bibinfo {pages} {447--488} (\bibinfo
  {month} {Jun.}\ \bibinfo {year} {2000}),\
  \Eprint{http://arxiv.org/abs/astro-ph/0008148}{astro-ph/0008148}\BibitemShut%
{NoStop}%
\bibitem{Lithwick:2001}%
  \BibitemOpen
  \bibfield{author}{%
  \bibinfo {author} {\bibfnamefont{Y.}~\bibnamefont{Lithwick}}\ and\ \bibinfo
  {author} {\bibfnamefont{P.}~\bibnamefont{Goldreich}},\ }%
  \bibfield{title}{%
  \enquote{\bibinfo {title} {Compressible magnetohydrodynamic turbulence in
  interstellar plasmas},}\ }%
  \bibfield{journal}{%
  \bibinfo {journal} {Astrophys.~J.}\ }%
  \textbf{\bibinfo {volume} {562}},\ \bibinfo {pages} {279--296} (\bibinfo
  {year} {2001})\BibitemShut{NoStop}%
\bibitem{Lithwick:2003}%
  \BibitemOpen
  \bibfield{author}{%
  \bibinfo {author} {\bibfnamefont{Y.}~\bibnamefont{Lithwick}}\ and\ \bibinfo
  {author} {\bibfnamefont{P.}~\bibnamefont{Goldreich}},\ }%
  \bibfield{title}{%
  \enquote{\bibinfo {title} {Imbalanced weak magnetohydrodynamic turbulence},}\
  }%
  \bibfield{journal}{%
  \bibinfo {journal} {Astrophys.~J.}\ }%
  \textbf{\bibinfo {volume} {582}},\ \bibinfo {pages} {1220--1240} (\bibinfo
  {year} {2003})\BibitemShut{NoStop}%
\bibitem{Strauss:1976}%
  \BibitemOpen
  \bibfield{author}{%
  \bibinfo {author} {\bibfnamefont{H.~R.}\ \bibnamefont{{Strauss}}},\ }%
  \bibfield{title}{%
  \enquote{\bibinfo {title} {{Nonlinear, three-dimensional magnetohydrodynamics
  of noncircular tokamaks}},}\ }%
  \bibfield{journal}{%
  \bibinfo {journal} {Phys.~Fluids}\ }%
  \textbf{\bibinfo {volume} {19}},\ \bibinfo {pages} {134--140} (\bibinfo
  {month} {Jan.}\ \bibinfo {year} {1976})\BibitemShut{NoStop}%
\bibitem{Borovsky:2012}%
  \BibitemOpen
  \bibfield{author}{%
  \bibinfo {author} {\bibfnamefont{J.~E.}\ \bibnamefont{{Borovsky}}},\ }%
  \bibfield{title}{%
  \enquote{\bibinfo {title} {{The velocity and magnetic field fluctuations of
  the solar wind at 1 AU: Statistical analysis of Fourier spectra and
  correlations with plasma properties}},}\ }%
  \bibfield{journal}{%
  \Doi{10.1029/2011JA017499}{\bibinfo {journal} {Journal of Geophysical
  Research (Space Physics)}}\ }%
  \textbf{\bibinfo {volume} {117}},\ \bibinfo {eid} {A05104} (\bibinfo {month}
  {May}\ \bibinfo {year} {2012})\BibitemShut{NoStop}%
\bibitem{Boldyrev:2011}%
  \BibitemOpen
  \bibfield{author}{%
  \bibinfo {author} {\bibfnamefont{S.}~\bibnamefont{{Boldyrev}}}, \bibinfo
  {author} {\bibfnamefont{J.~C.}\ \bibnamefont{{Perez}}}, \bibinfo {author}
  {\bibfnamefont{J.~E.}\ \bibnamefont{{Borovsky}}},\ and\ \bibinfo {author}
  {\bibfnamefont{J.~J.}\ \bibnamefont{{Podesta}}},\ }%
  \bibfield{title}{%
  \enquote{\bibinfo {title} {{Spectral Scaling Laws in Magnetohydrodynamic
  Turbulence Simulations and in the Solar Wind}},}\ }%
  \bibfield{journal}{%
  \Doi{10.1088/2041-8205/741/1/L19}{\bibinfo {journal} {Astrophys.~J.~Lett.}}\
  }%
  \textbf{\bibinfo {volume} {741}},\ \bibinfo {eid} {L19} (\bibinfo {month}
  {Nov.}\ \bibinfo {year} {2011}),\
  \Eprint{http://arxiv.org/abs/1106.0700}{arXiv:1106.0700
  [astro-ph.GA]}\BibitemShut{NoStop}%
\bibitem{Nielson:2013a}%
  \BibitemOpen
  \bibfield{author}{%
  \bibinfo {author} {\bibfnamefont{K.~D.}\ \bibnamefont{{Nielson}}}, \bibinfo
  {author} {\bibfnamefont{G.~G.}\ \bibnamefont{{Howes}}},\ and\ \bibinfo
  {author} {\bibfnamefont{W.}~\bibnamefont{{Dorland}}},\ }%
  \bibfield{title}{%
  \enquote{\bibinfo {title} {{Alfven Wave Collisions, The Fundamental Building
  Block of Plasma Turbulence II: Numerical Solution}},}\ }%
  \bibfield{journal}{%
  \bibinfo {journal} {Phys.~Plasmas}}%
   (\bibinfo {year} {2013}),\ \bibinfo {note} {submitted}\BibitemShut{NoStop}%
\bibitem{Elsasser:1950}%
  \BibitemOpen
  \bibfield{author}{%
  \bibinfo {author} {\bibfnamefont{W.~M.}\ \bibnamefont{{Elsasser}}},\ }%
  \bibfield{title}{%
  \enquote{\bibinfo {title} {{The Hydromagnetic Equations}},}\ }%
  \bibfield{journal}{%
  \Doi{10.1103/PhysRev.79.183}{\bibinfo {journal} {Physical Review}}\ }%
  \textbf{\bibinfo {volume} {79}},\ \bibinfo {pages} {183} (\bibinfo {month}
  {Jul.}\ \bibinfo {year} {1950})\BibitemShut{NoStop}%
\bibitem{Howes:2012a}%
  \BibitemOpen
  \bibfield{author}{%
  \bibinfo {author} {\bibfnamefont{G.~G.}\ \bibnamefont{{Howes}}}, \bibinfo
  {author} {\bibfnamefont{S.~D.}\ \bibnamefont{{Bale}}}, \bibinfo {author}
  {\bibfnamefont{K.~G.}\ \bibnamefont{{Klein}}}, \bibinfo {author}
  {\bibfnamefont{C.~H.~K.}\ \bibnamefont{{Chen}}}, \bibinfo {author}
  {\bibfnamefont{C.~S.}\ \bibnamefont{{Salem}}},\ and\ \bibinfo {author}
  {\bibfnamefont{J.~M.}\ \bibnamefont{{TenBarge}}},\ }%
  \bibfield{title}{%
  \enquote{\bibinfo {title} {{The Slow-mode Nature of Compressible Wave Power
  in Solar Wind Turbulence}},}\ }%
  \bibfield{journal}{%
  \Doi{10.1088/2041-8205/753/1/L19}{\bibinfo {journal} {Astrophys.~J.~Lett.}}\
  }%
  \textbf{\bibinfo {volume} {753}},\ \bibinfo {eid} {L19} (\bibinfo {year}
  {2012}),\ \Eprint{http://arxiv.org/abs/1106.4327}{arXiv:1106.4327
  [astro-ph.SR]}\BibitemShut{NoStop}%
\bibitem{Cho:2002}%
  \BibitemOpen
  \bibfield{author}{%
  \bibinfo {author} {\bibfnamefont{J.}~\bibnamefont{{Cho}}}, \bibinfo {author}
  {\bibfnamefont{A.}~\bibnamefont{{Lazarian}}},\ and\ \bibinfo {author}
  {\bibfnamefont{E.~T.}\ \bibnamefont{{Vishniac}}},\ }%
  \bibfield{title}{%
  \enquote{\bibinfo {title} {{Simulations of Magnetohydrodynamic Turbulence in
  a Strongly Magnetized Medium}},}\ }%
  \bibfield{journal}{%
  \Doi{10.1086/324186}{\bibinfo {journal} {Astrophys.~J.}}\ }%
  \textbf{\bibinfo {volume} {564}},\ \bibinfo {pages} {291--301} (\bibinfo
  {month} {Jan.}\ \bibinfo {year} {2002}),\
  \Eprint{http://arxiv.org/abs/astro-ph/0105235}{astro-ph/0105235}\BibitemShut%
{NoStop}%
\bibitem{Cho:2003}%
  \BibitemOpen
  \bibfield{author}{%
  \bibinfo {author} {\bibfnamefont{J.}~\bibnamefont{{Cho}}}\ and\ \bibinfo
  {author} {\bibfnamefont{A.}~\bibnamefont{{Lazarian}}},\ }%
  \bibfield{title}{%
  \enquote{\bibinfo {title} {{Compressible magnetohydrodynamic turbulence: mode
  coupling, scaling relations, anisotropy, viscosity-damped regime and
  astrophysical implications}},}\ }%
  \bibfield{journal}{%
  \Doi{10.1046/j.1365-8711.2003.06941.x}{\bibinfo {journal}
  {Mon.~Not.~Roy.~Astron.~Soc.}}\ }%
  \textbf{\bibinfo {volume} {345}},\ \bibinfo {pages} {325--339} (\bibinfo
  {month} {Oct.}\ \bibinfo {year} {2003}),\
  \Eprint{http://arxiv.org/abs/astro-ph/0301062}{astro-ph/0301062}\BibitemShut%
{NoStop}%
\bibitem{Howes:2011a}%
  \BibitemOpen
  \bibfield{author}{%
  \bibinfo {author} {\bibfnamefont{G.~G.}\ \bibnamefont{Howes}}, \bibinfo
  {author} {\bibfnamefont{J.~M.}\ \bibnamefont{TenBarge}}, \bibinfo {author}
  {\bibfnamefont{W.}~\bibnamefont{Dorland}}, \bibinfo {author}
  {\bibfnamefont{E.}~\bibnamefont{Quataert}}, \bibinfo {author}
  {\bibfnamefont{A.~A.}\ \bibnamefont{Schekochihin}}, \bibinfo {author}
  {\bibfnamefont{R.}~\bibnamefont{Numata}},\ and\ \bibinfo {author}
  {\bibfnamefont{T.}~\bibnamefont{Tatsuno}},\ }%
  \bibfield{title}{%
  \enquote{\bibinfo {title} {Gyrokinetic simulations of solar wind turbulence
  from ion to electron scales},}\ }%
  \bibfield{journal}{%
  \Doi{10.1103/PhysRevLett.107.035004}{\bibinfo {journal} {Phys.~Rev.~Lett.}}\
  }%
  \textbf{\bibinfo {volume} {107}},\ \bibinfo {pages} {035004} (\bibinfo {year}
  {2011})\BibitemShut{NoStop}%
\bibitem{Taylor:1938}%
  \BibitemOpen
  \bibfield{author}{%
  \bibinfo {author} {\bibfnamefont{G.~I.}\ \bibnamefont{{Taylor}}},\ }%
  \bibfield{title}{%
  \enquote{\bibinfo {title} {{The Spectrum of Turbulence}},}\ }%
  \bibfield{journal}{%
  \bibinfo {journal} {{Proc. Roy. Soc. A}}\ }%
  \textbf{\bibinfo {volume} {164}},\ \bibinfo {pages} {476--490} (\bibinfo
  {year} {1938})\BibitemShut{NoStop}%
\bibitem{Matthaeus:1982b}%
  \BibitemOpen
  \bibfield{author}{%
  \bibinfo {author} {\bibfnamefont{W.~H.}\ \bibnamefont{{Matthaeus}}}\ and\
  \bibinfo {author} {\bibfnamefont{M.~L.}\ \bibnamefont{{Goldstein}}},\ }%
  \bibfield{title}{%
  \enquote{\bibinfo {title} {{Measurement of the rugged invariants of
  magnetohydrodynamic turbulence in the solar wind}},}\ }%
  \bibfield{journal}{%
  \bibinfo {journal} {J.~Geophys.~Res.}\ }%
  \textbf{\bibinfo {volume} {87}},\ \bibinfo {pages} {6011--6028} (\bibinfo
  {month} {Aug.}\ \bibinfo {year} {1982})\BibitemShut{NoStop}%
\bibitem{Roberts:1987a}%
  \BibitemOpen
  \bibfield{author}{%
  \bibinfo {author} {\bibfnamefont{D.~A.}\ \bibnamefont{{Roberts}}}, \bibinfo
  {author} {\bibfnamefont{L.~W.}\ \bibnamefont{{Klein}}}, \bibinfo {author}
  {\bibfnamefont{M.~L.}\ \bibnamefont{{Goldstein}}},\ and\ \bibinfo {author}
  {\bibfnamefont{W.~H.}\ \bibnamefont{{Matthaeus}}},\ }%
  \bibfield{title}{%
  \enquote{\bibinfo {title} {{The nature and evolution of magnetohydrodynamic
  fluctuations in the solar wind - Voyager observations}},}\ }%
  \bibfield{journal}{%
  \bibinfo {journal} {J.~Geophys.~Res.}\ }%
  \textbf{\bibinfo {volume} {92}},\ \bibinfo {pages} {11021--11040} (\bibinfo
  {month} {Oct.}\ \bibinfo {year} {1987})\BibitemShut{NoStop}%
\bibitem{Bruno:1985}%
  \BibitemOpen
  \bibfield{author}{%
  \bibinfo {author} {\bibfnamefont{R.}~\bibnamefont{{Bruno}}}, \bibinfo
  {author} {\bibfnamefont{B.}~\bibnamefont{{Bavassano}}},\ and\ \bibinfo
  {author} {\bibfnamefont{U.}~\bibnamefont{{Villante}}},\ }%
  \bibfield{title}{%
  \enquote{\bibinfo {title} {{Evidence for long period Alfven waves in the
  inner solar system}},}\ }%
  \bibfield{journal}{%
  \Doi{10.1029/JA090iA05p04373}{\bibinfo {journal} {J.~Geophys.~Res.}}\ }%
  \textbf{\bibinfo {volume} {90}},\ \bibinfo {pages} {4373--4377} (\bibinfo
  {month} {May}\ \bibinfo {year} {1985})\BibitemShut{NoStop}%
\bibitem{Goldstein:1995}%
  \BibitemOpen
  \bibfield{author}{%
  \bibinfo {author} {\bibfnamefont{M.~L.}\ \bibnamefont{{Goldstein}}}, \bibinfo
  {author} {\bibfnamefont{D.~A.}\ \bibnamefont{{Roberts}}},\ and\ \bibinfo
  {author} {\bibfnamefont{W.~H.}\ \bibnamefont{{Matthaeus}}},\ }%
  \bibfield{title}{%
  \enquote{\bibinfo {title} {{Magnetohydrodynamic Turbulence In The Solar
  Wind}},}\ }%
  \bibfield{journal}{%
  \Doi{10.1146/annurev.aa.33.090195.001435}{\bibinfo {journal}
  {Ann.~Rev.~Astron.~Astrophys.}}\ }%
  \textbf{\bibinfo {volume} {33}},\ \bibinfo {pages} {283--326} (\bibinfo
  {year} {1995})\BibitemShut{NoStop}%
\bibitem{Tu:1995}%
  \BibitemOpen
  \bibfield{author}{%
  \bibinfo {author} {\bibfnamefont{C.-Y.}\ \bibnamefont{{Tu}}}\ and\ \bibinfo
  {author} {\bibfnamefont{E.}~\bibnamefont{{Marsch}}},\ }%
  \bibfield{title}{%
  \enquote{\bibinfo {title} {{MHD structures, waves and turbulence in the solar
  wind: Observations and theories}},}\ }%
  \bibfield{journal}{%
  \bibinfo {journal} {Space Sci.~Rev.}\ }%
  \textbf{\bibinfo {volume} {73}},\ \bibinfo {pages} {1--2} (\bibinfo {month}
  {Jul.}\ \bibinfo {year} {1995})\BibitemShut{NoStop}%
\bibitem{Bavassano:1998}%
  \BibitemOpen
  \bibfield{author}{%
  \bibinfo {author} {\bibfnamefont{B.}~\bibnamefont{{Bavassano}}}, \bibinfo
  {author} {\bibfnamefont{E.}~\bibnamefont{{Pietropaolo}}},\ and\ \bibinfo
  {author} {\bibfnamefont{R.}~\bibnamefont{{Bruno}}},\ }%
  \bibfield{title}{%
  \enquote{\bibinfo {title} {{Cross-helicity and residual energy in solar wind
  turbulence - Radial evolution and latitudinal dependence in the region from 1
  to 5 AU}},}\ }%
  \bibfield{journal}{%
  \Doi{10.1029/97JA03029}{\bibinfo {journal} {J.~Geophys.~Res.}}\ }%
  \textbf{\bibinfo {volume} {103}},\ \bibinfo {pages} {6521} (\bibinfo {month}
  {Apr.}\ \bibinfo {year} {1998})\BibitemShut{NoStop}%
\bibitem{Bavassano:2000a}%
  \BibitemOpen
  \bibfield{author}{%
  \bibinfo {author} {\bibfnamefont{B.}~\bibnamefont{{Bavassano}}}, \bibinfo
  {author} {\bibfnamefont{E.}~\bibnamefont{{Pietropaolo}}},\ and\ \bibinfo
  {author} {\bibfnamefont{R.}~\bibnamefont{{Bruno}}},\ }%
  \bibfield{title}{%
  \enquote{\bibinfo {title} {{Alfv{\'e}nic turbulence in the polar wind: A
  statistical study on cross helicity and residual energy variations}},}\ }%
  \bibfield{journal}{%
  \Doi{10.1029/2000JA900004}{\bibinfo {journal} {J.~Geophys.~Res.}}\ }%
  \textbf{\bibinfo {volume} {105}},\ \bibinfo {pages} {12697--12704} (\bibinfo
  {month} {Jun.}\ \bibinfo {year} {2000})\BibitemShut{NoStop}%
\bibitem{Bruno:2005}%
  \BibitemOpen
  \bibfield{author}{%
  \bibinfo {author} {\bibfnamefont{R.}~\bibnamefont{{Bruno}}}\ and\ \bibinfo
  {author} {\bibfnamefont{V.}~\bibnamefont{{Carbone}}},\ }%
  \bibfield{title}{%
  \enquote{\bibinfo {title} {{The Solar Wind as a Turbulence Laboratory}},}\ }%
  \bibfield{journal}{%
  \bibinfo {journal} {Living Reviews in Solar Physics}\ }%
  \textbf{\bibinfo {volume} {2}},\ \bibinfo {pages} {4} (\bibinfo {month}
  {Sep.}\ \bibinfo {year} {2005})\BibitemShut{NoStop}%
\bibitem{Podesta:2007}%
  \BibitemOpen
  \bibfield{author}{%
  \bibinfo {author} {\bibfnamefont{J.~J.}\ \bibnamefont{{Podesta}}}, \bibinfo
  {author} {\bibfnamefont{D.~A.}\ \bibnamefont{{Roberts}}},\ and\ \bibinfo
  {author} {\bibfnamefont{M.~L.}\ \bibnamefont{{Goldstein}}},\ }%
  \bibfield{title}{%
  \enquote{\bibinfo {title} {{Spectral Exponents of Kinetic and Magnetic Energy
  Spectra in Solar Wind Turbulence}},}\ }%
  \bibfield{journal}{%
  \Doi{10.1086/519211}{\bibinfo {journal} {Astrophys.~J.}}\ }%
  \textbf{\bibinfo {volume} {664}},\ \bibinfo {pages} {543--548} (\bibinfo
  {month} {Jul.}\ \bibinfo {year} {2007})\BibitemShut{NoStop}%
\bibitem{Salem:2009}%
  \BibitemOpen
  \bibfield{author}{%
  \bibinfo {author} {\bibfnamefont{C.~S.}\ \bibnamefont{{Salem}}}, \bibinfo
  {author} {\bibfnamefont{D.~J.}\ \bibnamefont{{Sundkvist}}},\ and\ \bibinfo
  {author} {\bibfnamefont{S.}~\bibnamefont{{Bale}}},\ }%
  \bibfield{title}{%
  \enquote{\bibinfo {title} {{Wavemode identification in the
  dissipation/dispersion range of solar wind turbulence: Kinetic Alfv\'en Waves
  and/or Whistlers? (Invited)}},}\ }%
  \bibfield{journal}{%
  \bibinfo {journal} {AGU Fall Meeting Abstracts},\ \bibinfo {pages} {A4+}}%
   (\bibinfo {month} {Dec.}\ \bibinfo {year} {2009})\BibitemShut{NoStop}%
\bibitem{Boldyrev:2009b}%
  \BibitemOpen
  \bibfield{author}{%
  \bibinfo {author} {\bibfnamefont{S.}~\bibnamefont{{Boldyrev}}}\ and\ \bibinfo
  {author} {\bibfnamefont{J.~C.}\ \bibnamefont{{Perez}}},\ }%
  \bibfield{title}{%
  \enquote{\bibinfo {title} {{Spectrum of Weak Magnetohydrodynamic
  Turbulence}},}\ }%
  \bibfield{journal}{%
  \Doi{10.1103/PhysRevLett.103.225001}{\bibinfo {journal} {Phys.~Rev.~Lett.}}\
  }%
  \textbf{\bibinfo {volume} {103}},\ \bibinfo {eid} {225001} (\bibinfo {month}
  {Nov.}\ \bibinfo {year} {2009}),\
  \Eprint{http://arxiv.org/abs/0907.4475}{arXiv:0907.4475
  [astro-ph.GA]}\BibitemShut{NoStop}%
\bibitem{Muller:2005}%
  \BibitemOpen
  \bibfield{author}{%
  \bibinfo {author} {\bibfnamefont{{W.-C.}}\ \bibnamefont{{M{\"u}ller}}}\ and\
  \bibinfo {author} {\bibfnamefont{R.}~\bibnamefont{{Grappin}}},\ }%
  \bibfield{title}{%
  \enquote{\bibinfo {title} {{Spectral Energy Dynamics in Magnetohydrodynamic
  Turbulence}},}\ }%
  \bibfield{journal}{%
  \Doi{10.1103/PhysRevLett.95.114502}{\bibinfo {journal} {Phys.~Rev.~Lett.}}\
  }%
  \textbf{\bibinfo {volume} {95}},\ \bibinfo {pages} {114502--+} (\bibinfo
  {month} {Sep.}\ \bibinfo {year} {2005})\BibitemShut{NoStop}%
\bibitem{Wang:2011}%
  \BibitemOpen
  \bibfield{author}{%
  \bibinfo {author} {\bibfnamefont{Y.}~\bibnamefont{{Wang}}}, \bibinfo {author}
  {\bibfnamefont{S.}~\bibnamefont{{Boldyrev}}},\ and\ \bibinfo {author}
  {\bibfnamefont{J.~C.}\ \bibnamefont{{Perez}}},\ }%
  \bibfield{title}{%
  \enquote{\bibinfo {title} {{Residual Energy in Magnetohydrodynamic
  Turbulence}},}\ }%
  \bibfield{journal}{%
  \Doi{10.1088/2041-8205/740/2/L36}{\bibinfo {journal} {Astrophys.~J.~Lett.}}\
  }%
  \textbf{\bibinfo {volume} {740}},\ \bibinfo {eid} {L36} (\bibinfo {month}
  {Oct.}\ \bibinfo {year} {2011}),\
  \Eprint{http://arxiv.org/abs/1106.2238}{arXiv:1106.2238
  [astro-ph.GA]}\BibitemShut{NoStop}%
\bibitem{Tronko:2012}%
  \BibitemOpen
  \bibfield{author}{%
  \bibinfo {author} {\bibfnamefont{N.}~\bibnamefont{{Tronko}}}, \bibinfo
  {author} {\bibfnamefont{S.~V.}\ \bibnamefont{{Nazarenko}}},\ and\ \bibinfo
  {author} {\bibfnamefont{S.}~\bibnamefont{{Galtier}}},\ }%
  \bibfield{title}{%
  \enquote{\bibinfo {title} {{Weak turbulence in two-dimensional
  magnetohydrodynamics}},}\ }%
  \bibfield{journal}{%
  \bibinfo {journal} {ArXiv e-prints}}%
   (\bibinfo {month} {Dec.}\ \bibinfo {year} {2012}),\
  \Eprint{http://arxiv.org/abs/1212.0769}{arXiv:1212.0769
  [physics.plasm-ph]}\BibitemShut{NoStop}%
\bibitem{Dmitruk:2009}%
  \BibitemOpen
  \bibfield{author}{%
  \bibinfo {author} {\bibfnamefont{P.}~\bibnamefont{{Dmitruk}}}\ and\ \bibinfo
  {author} {\bibfnamefont{W.~H.}\ \bibnamefont{{Matthaeus}}},\ }%
  \bibfield{title}{%
  \enquote{\bibinfo {title} {{Waves and turbulence in magnetohydrodynamic
  direct numerical simulations}},}\ }%
  \bibfield{journal}{%
  \Doi{10.1063/1.3148335}{\bibinfo {journal} {Phys.~Plasmas}}\ }%
  \textbf{\bibinfo {volume} {16}},\ \bibinfo {pages} {062304--+} (\bibinfo
  {month} {Jun.}\ \bibinfo {year} {2009})\BibitemShut{NoStop}%
\bibitem{Boldyrev:2009}%
  \BibitemOpen
  \bibfield{author}{%
  \bibinfo {author} {\bibfnamefont{S.}~\bibnamefont{{Boldyrev}}}, \bibinfo
  {author} {\bibfnamefont{J.}~\bibnamefont{{Mason}}},\ and\ \bibinfo {author}
  {\bibfnamefont{F.}~\bibnamefont{{Cattaneo}}},\ }%
  \bibfield{title}{%
  \enquote{\bibinfo {title} {{Dynamic Alignment and Exact Scaling Laws in
  Magnetohydrodynamic Turbulence}},}\ }%
  \bibfield{journal}{%
  \Doi{10.1088/0004-637X/699/1/L39}{\bibinfo {journal} {Astrophys.~J.~Lett.}}\
  }%
  \textbf{\bibinfo {volume} {699}},\ \bibinfo {pages} {L39--L42} (\bibinfo
  {month} {Jul.}\ \bibinfo {year} {2009})\BibitemShut{NoStop}%
\bibitem{Parashar:2009}%
  \BibitemOpen
  \bibfield{author}{%
  \bibinfo {author} {\bibfnamefont{T.~N.}\ \bibnamefont{{Parashar}}}, \bibinfo
  {author} {\bibfnamefont{M.~A.}\ \bibnamefont{{Shay}}}, \bibinfo {author}
  {\bibfnamefont{P.~A.}\ \bibnamefont{{Cassak}}},\ and\ \bibinfo {author}
  {\bibfnamefont{W.~H.}\ \bibnamefont{{Matthaeus}}},\ }%
  \bibfield{title}{%
  \enquote{\bibinfo {title} {{Kinetic dissipation and anisotropic heating in a
  turbulent collisionless plasma}},}\ }%
  \bibfield{journal}{%
  \Doi{10.1063/1.3094062}{\bibinfo {journal} {Phys.~Plasmas}}\ }%
  \textbf{\bibinfo {volume} {16}},\ \bibinfo {pages} {032310--+} (\bibinfo
  {month} {Mar.}\ \bibinfo {year} {2009})\BibitemShut{NoStop}%
\bibitem{Servidio:2009}%
  \BibitemOpen
  \bibfield{author}{%
  \bibinfo {author} {\bibfnamefont{S.}~\bibnamefont{{Servidio}}}, \bibinfo
  {author} {\bibfnamefont{W.~H.}\ \bibnamefont{{Matthaeus}}}, \bibinfo {author}
  {\bibfnamefont{M.~A.}\ \bibnamefont{{Shay}}}, \bibinfo {author}
  {\bibfnamefont{P.~A.}\ \bibnamefont{{Cassak}}},\ and\ \bibinfo {author}
  {\bibfnamefont{P.}~\bibnamefont{{Dmitruk}}},\ }%
  \bibfield{title}{%
  \enquote{\bibinfo {title} {{Magnetic Reconnection in Two-Dimensional
  Magnetohydrodynamic Turbulence}},}\ }%
  \bibfield{journal}{%
  \Doi{10.1103/PhysRevLett.102.115003}{\bibinfo {journal} {Phys.~Rev.~Lett.}}\
  }%
  \textbf{\bibinfo {volume} {102}},\ \bibinfo {eid} {115003} (\bibinfo {month}
  {Mar.}\ \bibinfo {year} {2009})\BibitemShut{NoStop}%
\bibitem{Parashar:2010}%
  \BibitemOpen
  \bibfield{author}{%
  \bibinfo {author} {\bibfnamefont{T.~N.}\ \bibnamefont{{Parashar}}}, \bibinfo
  {author} {\bibfnamefont{S.}~\bibnamefont{{Servidio}}}, \bibinfo {author}
  {\bibfnamefont{B.}~\bibnamefont{{Breech}}}, \bibinfo {author}
  {\bibfnamefont{M.~A.}\ \bibnamefont{{Shay}}},\ and\ \bibinfo {author}
  {\bibfnamefont{W.~H.}\ \bibnamefont{{Matthaeus}}},\ }%
  \bibfield{title}{%
  \enquote{\bibinfo {title} {{Kinetic driven turbulence: Structure in space and
  time}},}\ }%
  \bibfield{journal}{%
  \Doi{10.1063/1.3486537}{\bibinfo {journal} {Phys.~Plasmas}}\ }%
  \textbf{\bibinfo {volume} {17}},\ \bibinfo {pages} {102304--+} (\bibinfo
  {month} {Oct.}\ \bibinfo {year} {2010})\BibitemShut{NoStop}%
\bibitem{Servidio:2010}%
  \BibitemOpen
  \bibfield{author}{%
  \bibinfo {author} {\bibfnamefont{S.}~\bibnamefont{{Servidio}}}, \bibinfo
  {author} {\bibfnamefont{W.~H.}\ \bibnamefont{{Matthaeus}}}, \bibinfo {author}
  {\bibfnamefont{M.~A.}\ \bibnamefont{{Shay}}}, \bibinfo {author}
  {\bibfnamefont{P.}~\bibnamefont{{Dmitruk}}}, \bibinfo {author}
  {\bibfnamefont{P.~A.}\ \bibnamefont{{Cassak}}},\ and\ \bibinfo {author}
  {\bibfnamefont{M.}~\bibnamefont{{Wan}}},\ }%
  \bibfield{title}{%
  \enquote{\bibinfo {title} {{Statistics of magnetic reconnection in
  two-dimensional magnetohydrodynamic turbulence}},}\ }%
  \bibfield{journal}{%
  \Doi{10.1063/1.3368798}{\bibinfo {journal} {Phys.~Plasmas}}\ }%
  \textbf{\bibinfo {volume} {17}},\ \bibinfo {pages} {032315} (\bibinfo {month}
  {Mar.}\ \bibinfo {year} {2010})\BibitemShut{NoStop}%
\bibitem{Servidio:2011}%
  \BibitemOpen
  \bibfield{author}{%
  \bibinfo {author} {\bibfnamefont{S.}~\bibnamefont{{Servidio}}}, \bibinfo
  {author} {\bibfnamefont{P.}~\bibnamefont{{Dmitruk}}}, \bibinfo {author}
  {\bibfnamefont{A.}~\bibnamefont{{Greco}}}, \bibinfo {author}
  {\bibfnamefont{M.}~\bibnamefont{{Wan}}}, \bibinfo {author}
  {\bibfnamefont{S.}~\bibnamefont{{Donato}}}, \bibinfo {author}
  {\bibfnamefont{P.~A.}\ \bibnamefont{{Cassak}}}, \bibinfo {author}
  {\bibfnamefont{M.~A.}\ \bibnamefont{{Shay}}}, \bibinfo {author}
  {\bibfnamefont{V.}~\bibnamefont{{Carbone}}},\ and\ \bibinfo {author}
  {\bibfnamefont{W.~H.}\ \bibnamefont{{Matthaeus}}},\ }%
  \bibfield{title}{%
  \enquote{\bibinfo {title} {{Magnetic reconnection as an element of
  turbulence}},}\ }%
  \bibfield{journal}{%
  \Doi{10.5194/npg-18-675-2011}{\bibinfo {journal} {Nonlin.~Proc.~Geophys.}}\
  }%
  \textbf{\bibinfo {volume} {18}},\ \bibinfo {pages} {675--695} (\bibinfo
  {month} {Oct.}\ \bibinfo {year} {2011})\BibitemShut{NoStop}%
\bibitem{Markovskii:2011}%
  \BibitemOpen
  \bibfield{author}{%
  \bibinfo {author} {\bibfnamefont{S.~A.}\ \bibnamefont{{Markovskii}}}\ and\
  \bibinfo {author} {\bibfnamefont{B.~J.}\ \bibnamefont{{Vasquez}}},\ }%
  \bibfield{title}{%
  \enquote{\bibinfo {title} {{A Short-timescale Channel of Dissipation of the
  Strong Solar Wind Turbulence}},}\ }%
  \bibfield{journal}{%
  \Doi{10.1088/0004-637X/739/1/22}{\bibinfo {journal} {Astrophys.~J.}}\ }%
  \textbf{\bibinfo {volume} {739}},\ \bibinfo {eid} {22} (\bibinfo {month}
  {Sep.}\ \bibinfo {year} {2011})\BibitemShut{NoStop}%
\bibitem{Parashar:2011}%
  \BibitemOpen
  \bibfield{author}{%
  \bibinfo {author} {\bibfnamefont{T.~N.}\ \bibnamefont{{Parashar}}}, \bibinfo
  {author} {\bibfnamefont{S.}~\bibnamefont{{Servidio}}}, \bibinfo {author}
  {\bibfnamefont{M.~A.}\ \bibnamefont{{Shay}}}, \bibinfo {author}
  {\bibfnamefont{B.}~\bibnamefont{{Breech}}},\ and\ \bibinfo {author}
  {\bibfnamefont{W.~H.}\ \bibnamefont{{Matthaeus}}},\ }%
  \bibfield{title}{%
  \enquote{\bibinfo {title} {{Effect of driving frequency on excitation of
  turbulence in a kinetic plasma}},}\ }%
  \bibfield{journal}{%
  \Doi{10.1063/1.3630926}{\bibinfo {journal} {Physics of Plasmas}}\ }%
  \textbf{\bibinfo {volume} {18}},\ \bibinfo {pages} {092302} (\bibinfo {month}
  {Sep.}\ \bibinfo {year} {2011})\BibitemShut{NoStop}%
\bibitem{Servidio:2012}%
  \BibitemOpen
  \bibfield{author}{%
  \bibinfo {author} {\bibfnamefont{S.}~\bibnamefont{{Servidio}}}, \bibinfo
  {author} {\bibfnamefont{F.}~\bibnamefont{{Valentini}}}, \bibinfo {author}
  {\bibfnamefont{F.}~\bibnamefont{{Califano}}},\ and\ \bibinfo {author}
  {\bibfnamefont{P.}~\bibnamefont{{Veltri}}},\ }%
  \bibfield{title}{%
  \enquote{\bibinfo {title} {{Local Kinetic Effects in Two-Dimensional Plasma
  Turbulence}},}\ }%
  \bibfield{journal}{%
  \Doi{10.1103/PhysRevLett.108.045001}{\bibinfo {journal} {Phys.~Rev.~Lett.}}\
  }%
  \textbf{\bibinfo {volume} {108}},\ \bibinfo {eid} {045001} (\bibinfo {month}
  {Jan.}\ \bibinfo {year} {2012})\BibitemShut{NoStop}%
\bibitem{Vasquez:2012}%
  \BibitemOpen
  \bibfield{author}{%
  \bibinfo {author} {\bibfnamefont{B.~J.}\ \bibnamefont{{Vasquez}}}\ and\
  \bibinfo {author} {\bibfnamefont{S.~A.}\ \bibnamefont{{Markovskii}}},\ }%
  \bibfield{title}{%
  \enquote{\bibinfo {title} {{Velocity Power Spectra from Cross-field
  Turbulence in the Proton Kinetic Regime}},}\ }%
  \bibfield{journal}{%
  \Doi{10.1088/0004-637X/747/1/19}{\bibinfo {journal} {Astrophys.~J.}}\ }%
  \textbf{\bibinfo {volume} {747}},\ \bibinfo {eid} {19} (\bibinfo {month}
  {Mar.}\ \bibinfo {year} {2012})\BibitemShut{NoStop}%
\bibitem{Svidzinski:2009}%
  \BibitemOpen
  \bibfield{author}{%
  \bibinfo {author} {\bibfnamefont{V.~A.}\ \bibnamefont{{Svidzinski}}},
  \bibinfo {author} {\bibfnamefont{H.}~\bibnamefont{{Li}}}, \bibinfo {author}
  {\bibfnamefont{H.~A.}\ \bibnamefont{{Rose}}}, \bibinfo {author}
  {\bibfnamefont{B.~J.}\ \bibnamefont{{Albright}}},\ and\ \bibinfo {author}
  {\bibfnamefont{K.~J.}\ \bibnamefont{{Bowers}}},\ }%
  \bibfield{title}{%
  \enquote{\bibinfo {title} {{Particle in cell simulations of fast magnetosonic
  wave turbulence in the ion cyclotron frequency range}},}\ }%
  \bibfield{journal}{%
  \Doi{10.1063/1.3274559}{\bibinfo {journal} {Phys.~Plasmas}}\ }%
  \textbf{\bibinfo {volume} {16}},\ \bibinfo {pages} {122310--+} (\bibinfo
  {month} {Dec.}\ \bibinfo {year} {2009})\BibitemShut{NoStop}%
\bibitem{Hunana:2011}%
  \BibitemOpen
  \bibfield{author}{%
  \bibinfo {author} {\bibfnamefont{P.}~\bibnamefont{{Hunana}}}, \bibinfo
  {author} {\bibfnamefont{D.}~\bibnamefont{{Laveder}}}, \bibinfo {author}
  {\bibfnamefont{T.}~\bibnamefont{{Passot}}}, \bibinfo {author}
  {\bibfnamefont{P.~L.}\ \bibnamefont{{Sulem}}},\ and\ \bibinfo {author}
  {\bibfnamefont{D.}~\bibnamefont{{Borgogno}}},\ }%
  \bibfield{title}{%
  \enquote{\bibinfo {title} {{Reduction of Compressibility and Parallel
  Transfer by Landau Damping in Turbulent Magnetized Plasmas}},}\ }%
  \bibfield{journal}{%
  \Doi{10.1088/0004-637X/743/2/128}{\bibinfo {journal} {Astrophys.~J.}}\ }%
  \textbf{\bibinfo {volume} {743}},\ \bibinfo {eid} {128} (\bibinfo {month}
  {Dec.}\ \bibinfo {year} {2011}),\
  \Eprint{http://arxiv.org/abs/1109.2636}{arXiv:1109.2636
  [physics.plasm-ph]}\BibitemShut{NoStop}%
\bibitem{Salem:2012}%
  \BibitemOpen
  \bibfield{author}{%
  \bibinfo {author} {\bibfnamefont{C.~S.}\ \bibnamefont{{Salem}}}, \bibinfo
  {author} {\bibfnamefont{G.~G.}\ \bibnamefont{{Howes}}}, \bibinfo {author}
  {\bibfnamefont{D.}~\bibnamefont{{Sundkvist}}}, \bibinfo {author}
  {\bibfnamefont{S.~D.}\ \bibnamefont{{Bale}}}, \bibinfo {author}
  {\bibfnamefont{C.~C.}\ \bibnamefont{{Chaston}}}, \bibinfo {author}
  {\bibfnamefont{C.~H.~K.}\ \bibnamefont{{Chen}}},\ and\ \bibinfo {author}
  {\bibfnamefont{F.~S.}\ \bibnamefont{{Mozer}}},\ }%
  \bibfield{title}{%
  \enquote{\bibinfo {title} {{Identification of Kinetic Alfv{\'e}n Wave
  Turbulence in the Solar Wind}},}\ }%
  \bibfield{journal}{%
  \Doi{10.1088/2041-8205/745/1/L9}{\bibinfo {journal} {Astrophys.~J.~Lett.}}\
  }%
  \textbf{\bibinfo {volume} {745}},\ \bibinfo {eid} {L9} (\bibinfo {year}
  {2012})\BibitemShut{NoStop}%
\bibitem{Klein:2012}%
  \BibitemOpen
  \bibfield{author}{%
  \bibinfo {author} {\bibfnamefont{K.~G.}\ \bibnamefont{{Klein}}}, \bibinfo
  {author} {\bibfnamefont{G.~G.}\ \bibnamefont{{Howes}}}, \bibinfo {author}
  {\bibfnamefont{J.~M.}\ \bibnamefont{{TenBarge}}}, \bibinfo {author}
  {\bibfnamefont{S.~D.}\ \bibnamefont{{Bale}}}, \bibinfo {author}
  {\bibfnamefont{C.~H.~K.}\ \bibnamefont{{Chen}}},\ and\ \bibinfo {author}
  {\bibfnamefont{C.~S.}\ \bibnamefont{{Salem}}},\ }%
  \bibfield{title}{%
  \enquote{\bibinfo {title} {{Using Synthetic Spacecraft Data to Interpret
  Compressible Fluctuations in Solar Wind Turbulence}},}\ }%
  \bibfield{journal}{%
  \Doi{10.1088/0004-637X/755/2/159}{\bibinfo {journal} {Astrophys.~J.}}\ }%
  \textbf{\bibinfo {volume} {755}},\ \bibinfo {eid} {159} (\bibinfo {year}
  {2012}),\ \Eprint{http://arxiv.org/abs/1206.6564}{arXiv:1206.6564
  [physics.space-ph]}\BibitemShut{NoStop}%
\bibitem{Howes:2012b}%
  \BibitemOpen
  \bibfield{author}{%
  \bibinfo {author} {\bibfnamefont{G.~G.}\ \bibnamefont{{Howes}}}, \bibinfo
  {author} {\bibfnamefont{D.~J.}\ \bibnamefont{{Drake}}}, \bibinfo {author}
  {\bibfnamefont{K.~D.}\ \bibnamefont{{Nielson}}}, \bibinfo {author}
  {\bibfnamefont{T.~A.}\ \bibnamefont{{Carter}}}, \bibinfo {author}
  {\bibfnamefont{C.~A.}\ \bibnamefont{{Kletzing}}},\ and\ \bibinfo {author}
  {\bibfnamefont{F.}~\bibnamefont{{Skiff}}},\ }%
  \bibfield{title}{%
  \enquote{\bibinfo {title} {{Toward Astrophysical Turbulence in the
  Laboratory}},}\ }%
  \bibfield{journal}{%
  \Doi{10.1103/PhysRevLett.109.255001}{\bibinfo {journal} {Phys.~Rev.~Lett.}}\
  }%
  \textbf{\bibinfo {volume} {109}},\ \bibinfo {eid} {255001} (\bibinfo {month}
  {Dec.}\ \bibinfo {year} {2012}),\
  \Eprint{http://arxiv.org/abs/1210.4568}{arXiv:1210.4568
  [physics.plasm-ph]}\BibitemShut{NoStop}%
\bibitem{Howes:2013b}%
  \BibitemOpen
  \bibfield{author}{%
  \bibinfo {author} {\bibfnamefont{G.~G.}\ \bibnamefont{{Howes}}}, \bibinfo
  {author} {\bibfnamefont{K.~D.}\ \bibnamefont{{Nielson}}}, \bibinfo {author}
  {\bibfnamefont{D.~J.}\ \bibnamefont{{Drake}}}, \bibinfo {author}
  {\bibfnamefont{J.~W.~R.}\ \bibnamefont{{Schroeder}}}, \bibinfo {author}
  {\bibfnamefont{C.~A.}\ \bibnamefont{{Skiff}}, \bibfnamefont{F.~{Kletzing}}},\
  and\ \bibinfo {author} {\bibfnamefont{T.~A.}\ \bibnamefont{{Carter}}},\ }%
  \bibfield{title}{%
  \enquote{\bibinfo {title} {{Alfven Wave Collisions, The Fundamental Building
  Block of Plasma Turbulence III: Theory and Simulations for Experimental
  Design}},}\ }%
  \bibfield{journal}{%
  \bibinfo {journal} {Phys.~Plasmas}}%
   (\bibinfo {year} {2012}),\ \bibinfo {note} {in
  preparation}\BibitemShut{NoStop}%
\bibitem{Drake:2013}%
  \BibitemOpen
  \bibfield{author}{%
  \bibinfo {author} {\bibfnamefont{D.~J.}\ \bibnamefont{{Drake}}}, \bibinfo
  {author} {\bibfnamefont{J.~W.~R.}\ \bibnamefont{{Schroeder}}}, \bibinfo
  {author} {\bibfnamefont{G.~G.}\ \bibnamefont{{Howes}}}, \bibinfo {author}
  {\bibfnamefont{C.~A.}\ \bibnamefont{{Skiff}}, \bibfnamefont{F.~{Kletzing}}},
  \bibinfo {author} {\bibfnamefont{T.~A.}\ \bibnamefont{{Carter}}},\ and\
  \bibinfo {author} {\bibfnamefont{D.~W.}\ \bibnamefont{{Auerbach}}},\ }%
  \bibfield{title}{%
  \enquote{\bibinfo {title} {{Alfven Wave Collisions, The Fundamental Building
  Block of Plasma Turbulence IV: Experiment}},}\ }%
  \bibfield{journal}{%
  \bibinfo {journal} {Phys.~Plasmas}}%
   (\bibinfo {year} {2012}),\ \bibinfo {note} {in
  preparation}\BibitemShut{NoStop}%
\end{thebibliography}

%

\end{document}